\documentclass[12pt,letterpaper]{JHEP3}
\usepackage{amsmath,amssymb,amsfonts,xspace}
\numberwithin{equation}{section}

\def\varpi{t}

\def\det{\,{\rm det}\, }

\def\Im{\,{\rm Im}\,}
\def\Re{\,{\rm Re}\,}

\def\({\left(}
\def\){\right)}
\def\[{\left[}
\def\]{\right]}
\def\<{\left\langle}
\def\>{\right\rangle}
\def\hf{{1\over 2}}

\renewcommand{\d}{\mathrm{d}}
\newcommand{\de}{\mathrm{d}}

\newcommand{\I}{\mathrm{i}}
\newcommand{\e}{\mathrm{e}}
\newcommand{\cL}{\mathcal{L}}
\newcommand{\cD}{\mathcal{D}}

\newcommand{\p}{\partial}

\newcommand{\half}{\frac{1}{2}}

\newcommand{\cF}{\mathcal{F}}
\newcommand{\cV}{\mathcal{V}}

\newcommand{\cS}{\mathcal{S}}
\newcommand{\cG}{\mathcal{G}}
\newcommand{\cK}{\mathcal{K}}
\newcommand{\cM}{\mathcal{M}}

\newcommand{\cN}{\mathcal{N}}
\newcommand{\cE}{\mathcal{E}}
\newcommand{\cX}{\mathcal{X}}

\newcommand{\cJ}{\mathcal{J}}

\DeclareSymbolFont{AMSa}{U}{msa}{m}{n}
\DeclareSymbolFont{AMSb}{U}{msb}{m}{n}
\DeclareMathSymbol{\fieldR}{\mathalpha}{AMSb}{"52}

\newcommand{\N}{{\mathcal N}}
\newcommand{\kahler}{{K\"ahler}\xspace}
\newcommand{\hk}{{hyperk\"ahler}\xspace}
\newcommand{\qk}{{quaternion-K\"ahler}\xspace}

\newcommand{\R}{{\mathbb R}}
\newcommand{\Z}{{\mathbb Z}}
\newcommand{\cZ}{\mathcal{Z}}
\newcommand{\cI}{\mathcal{I}}
\newcommand{\cO}{\mathcal{O}}

\newcommand{\cU}{\mathcal{U}}

\newcommand{\pa}{\partial}
\newcommand{\nn}{\nonumber}

\newcommand{\eps}{\epsilon}
\newcommand{\veps}{\varepsilon}
\newcommand{\IR}{\mathbb{R}}

\newcommand{\IZ}{\mathbb{Z}}

\newcommand{\Tr}{\mbox{Tr}}

\newcommand{\sgn}{\mbox{sgn}}
\newcommand{\tzeta}{\tilde\zeta}
\newcommand{\tsigma}{\tilde\sigma}

\newcommand{\txi}{\tilde\xi}

\newcommand{\CP}{\mathbb{P}^1}

\def\bea{\begin{eqnarray}}
\def\eea{\end{eqnarray}}
\def\be{\begin{equation}}
\def\ee{\end{equation}}
\def\ba{\begin{align}}
\def\ea{\end{align}}
\def\bse{\begin{subequations}}
\def\ese{\end{subequations}}

\def\bi{\bar \imath}

\def\ba{\bar a}

\def\bz{\bar z}
\def\bY{\bar Y}
\def\bZ{\bar Z}

\def\bF{\bar F}

\newcommand{\CD}{{\cal{D}}}

\def\Hp{H_{\scriptscriptstyle{\smash{(1)}}}}

\def\ui#1{^{[#1]}}

\def\txii#1{{\tilde\xi}^{[#1]}}

\def\ai#1{{\alpha}^{[#1]}}

\def\xii#1{\xi_{[#1]}}

\def\Hij#1{H^{[#1]}}

\newcommand{\Li}{{\rm Li}}

\def\XXint#1#2#3{{\setbox0=\hbox{$#1{#2#3}{\int}$}
\vcenter{\hbox{$#2#3$}}\kern-.5\wd0}}

\def\hHij#1{H^{[#1]}}

\newcommand{\hCX}{\mathcal{X}}

\def\cij#1{c}
\def\ci#1{c}

\def\ellg#1{\ell_{#1}}

\def\Fcl{F^{\rm cl}}
\def\bFcl{\bF^{\rm cl}}

\newcommand{\expe}[1]{{\bf E}\!\left( #1\right)}
\def\lvol{{\rm cl}}
\def\ws{{\rm w.s.}}

\DeclareMathOperator{\Td}{Td}
\DeclareMathOperator{\ch}{ch}

\DeclareMathOperator{\Erf}{Erf}
\DeclareMathOperator{\Erfc}{Erfc}

\def\hcU{{\cal U}}

\newcommand{\beq}{\begin{eqnarray}}
\newcommand{\eeq}{\end{eqnarray}}
\def\tleta{\tilde\eta}

\def\cla{\tilde c_a}
\def\cl0{\tilde c_0}

\newcommand{\bfb}{{\boldsymbol b}}
\newcommand{\bfc}{{\boldsymbol c}}
\newcommand{\bfd}{{\boldsymbol d}}
\newcommand{\bfk}{{\boldsymbol k}}
\newcommand{\bfp}{{\boldsymbol p}}
\newcommand{\bfq}{{\boldsymbol q}}
\newcommand{\bfr}{{\boldsymbol r}}
\newcommand{\bft}{{\boldsymbol t}}
\newcommand{\bfz}{{\boldsymbol z}}
\newcommand{\bfx}{{\boldsymbol x}}
\newcommand{\bfy}{{\boldsymbol y}}
\newcommand{\bfeps}{{\boldsymbol \epsilon}}
\newcommand{\bfmu}{{\boldsymbol \mu}}
\newcommand{\bfnu}{{\boldsymbol \nu}}
\newcommand{\bfch}{{\bf{\boldsymbol c}_2}}
\newcommand{\bfcl}{\tilde{\boldsymbol c}}
\newcommand{\bftxi}{\tilde{\boldsymbol\xi}}
\newcommand{\bftxip}{\smash{\tilde{\boldsymbol\xi}}\vphantom{\txi}'}
\newcommand{\bfxi}{{\boldsymbol \xi}}
\newcommand{\bfDb}{\overline{\boldsymbol D}}
\newcommand{\bfD}{{\boldsymbol D}}
\newcommand{\Db}{\bar D}

\newcommand{\gammap}{\gamma}
\newcommand{\gammam}{-\gamma}
\newcommand{\gammapm}{\pm\gamma}

\newcommand{\antipod}{\varsigma}

\def\trans{g}
\def\hgamma{\hat\gamma}

\def\CY{\mathfrak{Y}}
\def\CYm{\mathfrak{\hat Y}}

\def\ZTheta{\Xi}
\def\BTheta{\Upsilon}

\def\kbp{(k+b)_+}

\def\signkp{(-1)^{\bfk\cdot\bfp}}

\def\wh{\mathfrak{h}}
\def\bwh{\bar\wh}

\title{D3-instantons, Mock Theta Series  and Twistors}

\preprint{L2C:12-090\\ Bonn-TH-2012-13\\CERN-PH-TH/2012-157\\arXiv:1207.1109v3}

\author{Sergei Alexandrov$^{1,2}$, Jan Manschot$^{3,4}$, Boris Pioline$^{5,6}$
\\
$^1$ {\it Universit\'e Montpellier 2, Laboratoire Charles Coulomb UMR 5221, \\ F-34095,
Montpellier, France}\\

$^2$ {\it CNRS, Laboratoire Charles Coulomb UMR 5221, F-34095,
Montpellier, France}\\

$^3$ {\it Bethe Center for Theoretical Physics, Physikalisches Institut, Universit\"at Bonn, \\
53115 Bonn, Germany}\\

$^4$ {\it Max Planck Institute for Mathematics, 53111 Bonn, Germany} \\

$^5$ {\it CERN PH-TH,
Case C01600, CERN, CH-1211 Geneva 23, Switzerland}\\

$^6$ {\it Laboratoire de Physique Th\'eorique et Hautes
Energies, CNRS UMR 7589, \\
Universit\'e Pierre et Marie Curie,
4 place Jussieu, 75252 Paris cedex 05, France} \\

\vspace*{2mm}
{%\tt e-mail:
\email{salexand@univ-montp2.fr},\ \
\email{manschot@uni-bonn.de},\ \
\email{boris.pioline@cern.ch}
}

\vspace*{-3mm}

}

\abstract{The D-instanton corrected hypermultiplet moduli space of type II string theory
compactified on a Calabi-Yau threefold is known  in the type IIA picture to be determined
in terms of the generalized Donaldson-Thomas invariants, through a twistorial construction.
At the same time, in the mirror type IIB picture, and in the limit where only
D3-D1-D(-1)-instanton corrections are retained, it should carry an isometric action of the S-duality
group $SL(2,\IZ)$. We prove that this is the case in the one-instanton
approximation, by constructing a holomorphic action of $SL(2,\IZ)$ on the
linearized twistor space. Using the modular invariance of the D4-D2-D0 black hole
partition function, we show that the standard Darboux coordinates in twistor space have modular anomalies
controlled by period integrals of a
Siegel-Narain theta series, which   can be canceled by a contact transformation
generated by a holomorphic mock theta series.
}

\begin{document}

%%%%%%%%%%%%%%%%%%%%%%%%%%%%%%%%%%%%%%%%%%%%%%%%%%%%%%%%%%%%%%%%%%%%
\section{Introduction}
%%%%%%%%%%%%%%%%%%%%%%%%%%%%%%%%%%%%%%%%%%%%%%%%%%%%%%%%%%%%%%%%%%%%

In type II string vacua with $N=2$ supersymmetry in four dimensions, the moduli
space spanned by massless scalar fields famously factorizes as the product
$\cM_V \times \cM_H$ of the vector multiplet (VM) and hypermultiplet (HM) moduli spaces,
respectively \cite{Bagger:1983tt,deWit:1984px}.  A complete understanding of the former was
achieved by the end of the last millenium, and has led to key developments in black hole
physics and algebraic geometry. By contrast,  until recently our understanding of the latter
was confined to the tree-level \cite{Cecotti:1988qn,Bodner:1989cg,Ferrara:1989ik}
and one-loop \cite{Antoniadis:1997eg,Gunther:1998sc,Antoniadis:2003sw,Robles-Llana:2006ez,Alexandrov:2007ec}
approximations to the metric, together with some general expectations about the form of instanton
corrections \cite{Becker:1995kb,Becker:1999pb}.
While it has been clear to many that an exact solution
for the \qk (QK) metric on $\cM_H$ would have considerable impact  both for physics and
geometry, progress was hindered mainly by the lack of a convenient parametrization of this
class of metrics.

The situation has improved dramatically in recent years, as projective superspace and
twistorial techniques \cite{MR664330,Karlhede:1984vr,Hitchin:1986ea,MR1001707, MR1096180,deWit:2001dj,Lindstrom:2008gs}
were brought to bear on this problem.
The key idea behind these methods is that the QK metric on $\cM_H$ is encoded in
the complex contact structure on the twistor space $\cZ$, a $\CP$ bundle over $\cM_H$ \cite{MR1001707}.
The latter is in turn specified by a set of holomorphic generating
functions for complex contact transformations between local Darboux coordinate systems,
subject to certain global consistency and reality requirements \cite{Alexandrov:2008nk}.
These gluing conditions can be converted into a system of integral equations,
which can often be solved by iteration. The QK metric on $\cM_H$ is then obtained by
expanding the resulting Darboux coordinates around any fixed section of $\cZ$,
similar to the usual twistorial construction of HK manifolds \cite{Hitchin:1986ea,Alexandrov:2008ds}.

Using these techniques, combining earlier results on the twistorial description
of the perturbative metric and of D1-D(-1)-instantons
\cite{RoblesLlana:2006is,Alexandrov:2006hx,RoblesLlana:2007ae,Saueressig:2007dr}
and taking inspiration from a similar construction  in the context of gauge
theories \cite{Gaiotto:2008cd}, a general construction
of the D-instanton corrected HM moduli space was laid out in \cite{Alexandrov:2008gh,Alexandrov:2009zh},
in terms of the generalized Donaldson-Thomas (DT) invariants
$\Omega(\gamma;\bfz)$ which count D-instantons. For reasons that will become clear below,
we refer to the construction of \cite{Alexandrov:2008gh,Alexandrov:2009zh} as the
`type IIA' construction.
In type IIA string theory compactified on a Calabi-Yau (CY) threefold $\CY$,
D-instantons come from  D2-branes wrapping special Lagrangian cycles,
while in type IIB string theory compactified on the mirror threefold $\CYm$,
they come from D5-D3-D1-D(-1)-branes wrapping even-dimensional cycles,
or more generally coherent sheaves on $\CYm$. Although the DT invariants typically depend on
the complex structure moduli (in type IIA) or  \kahler moduli (in type IIB),
they are piecewise constant away from walls of marginal stability,
and their discontinuity is such that the
resulting contact structure on $\cZ$ (and hence, the QK metric on $\cM_H$) is
smooth \cite{Gaiotto:2008cd}. In fact, the type IIA construction of
\cite{Alexandrov:2008gh,Alexandrov:2009zh} is essentially dictated by consistency with wall-crossing,
and is isomorphic to the gauge theory construction of  \cite{Gaiotto:2008cd} under the QK/HK
correspondence \cite{Haydys, Alexandrov:2011ac}.

In addition to these D-instantons, one also expects contributions
from NS5-brane instantons wrapped on the CY threefold. Those effects are in principle
determined from the known D5-instanton corrections, by requiring that the QK metric
should carry an isometric action of  $SL(2,\IZ)$, originating from  S-duality in
ten-dimensional type IIB string theory. NS5-brane corrections
were constructed at linear order in \cite{Alexandrov:2010ca}
(see also \cite{Pioline:2009qt,Bao:2009fg,Alexandrov:2010np}),
but an understanding of these effects at non-linear level is still missing.

In this work, we focus on a subset of the allowed instanton corrections,
namely those corresponding to D3-D1-D(-1)-instantons on the type IIB side.
Although these effects are exponentially suppressed compared to
D1-D(-1)-instantons in the large volume limit,
they are still exponentially larger than D5 and NS5-brane instantons. Thus, a natural
question is whether the twistorial construction of \cite{Alexandrov:2008gh,Alexandrov:2009zh}
produces a modular invariant \qk metric, in the large volume limit where D5-NS5-instantons
are ignored but the effects of D3-D1-D(-1)-instantons with non-zero D3-brane
charge are still retained.\footnote{For zero D3-brane charges, i.e. in the limit where  only D1-D(-1)-instantons
are retained, the QK metric and twistor space are known to be modular invariant
\cite{RoblesLlana:2006is,Alexandrov:2009qq,Alexandrov:2012bu},
although this is not apparent in the usual
type IIA formulation.} In this work, we shall
answer this question in the affirmative, at least in the one-instanton approximation. As will
become clear shortly, even in this simplified setting, modular invariance
is achieved in a very non-trivial way, and depends on the sophisticated machinery of
Eichler integrals and indefinite theta series. We expect an even richer structure beyond the one-instanton
approximation, but we shall hardly touch upon it in this work.

The reason why this problem is non-trivial is that modular invariance is far from manifest
in the `type IIA'  twistorial construction \cite{Alexandrov:2008gh,Alexandrov:2009zh}. Indeed,
when lifted to a holomorphic action on twistor space, the action of S-duality involves a
$SU(2)$ rotation along the fiber, and therefore relates Darboux coordinate systems on different patches.
The only exception are the patches around the points $t=\pm \I$, where $t$ is the usual
stereographic coordinate on the fiber, which are mapped onto
themselves under S-duality. But even in these patches the Darboux coordinates need only
be modular covariant up to a local complex contact transformation.
Finally, a technical but serious difficulty is that the natural type IIA coordinates
parametrizing the twistor fiber need not have simple transformation properties
under S-duality, and it is a priori unclear how to construct a basis of
type IIB fields which would transform covariantly.
Put differently, there is no general method to construct the quantum mirror map
(see however \cite{Alexandrov:2012bu} for recent progress in a closely related set-up).

Fortunately, we find that this picture drastically simplifies in the large volume
limit $\cV\to\infty$, provided
one simultaneously scales $z\sim(t\pm\I)\to 0$ keeping the product $z \cV^{1/3}$
fixed. In this case all BPS rays
(the contours on $\CP$ associated to D-instantons \cite{Gaiotto:2008cd,Alexandrov:2008gh})
coincide with the real axis in the complex $z$ plane, shifted
slightly above or below the origin $z=0$ depending on the D1-brane charge.
The corrections to Darboux coordinates are then given by  integrals of certain
Siegel-Narain theta series along these contours, corresponding to the sum over
D1-brane charges at fixed D3-brane charge. These Gaussian theta series
are far simpler than the non-Gaussian sums which
would otherwise arise at finite volume.

Another simplification of the large volume limit is that the moduli-dependent DT invariants
can be traded for another set of invariants, which we call the MSW invariants, which are independent
of the \kahler moduli in type IIB (or complex structure moduli in type IIA), and
have simple modular properties. The idea is that D3-D1-D(-1)-instantons are governed
by the same DT invariants which count D4-D2-D0 black holes in type IIA string theory
compactified on the same CY threefold $\CYm$. At large volume, the latter can be described
by micro-states of the MSW superconformal field theory (SCFT)
\cite{Maldacena:1997de}, and bound states thereof
\cite{Manschot:2009ia, Manschot:2010xp}. We define
the multi-instanton expansion in the large volume limit in such a way that only elementary MSW states
contribute in the one-instanton approximation. With this definition, the Gaussian theta
series alluded to in the previous paragraph reconstruct  the elliptic genus of the SCFT (or modular derivatives thereof),
and the corrections to Darboux coordinates are now identified as Eichler integral of the
MSW elliptic genus. For illustration, we display the result for
the correction to the Darboux coordinates $\bfxi$ conjugate to the D1-brane charges $\bfq$, evaluated
for simplicity at the point $z=0$ on the twistor fiber,
\be
\delta\bfxi(0) =- \frac{\bfp}{4\pi}\,e^{-2\pi S_{\rm cl}} \sum_{\bfmu\in\Lambda^*/\Lambda}
h_{\bfp,\bfmu}(\tau)\,
\int_{\bar\tau}^{-\I \infty}
\frac{\overline{\tilde\theta_{\bfp,\bfmu}(w,\bar\tau,\bft,\bfb,\bfc)} \,\de \bar w}{\sqrt{\I(\bar w-\tau)}},
\ee
where $S_{\rm cl}$ is the classical D3-instanton action \eqref{clactinst}. It involves
the Siegel-Narain type theta series  $\tilde\theta_{\bfp,\bfmu}(w,\bar\tau,\bft,\bfb,\bfc)$
\eqref{defSieg1} of  weight $(\tfrac32,\tfrac{b_2-1}{2})$,
where $b_2\equiv b_2(\CYm)$, analytically continued away from
the slice $w=\tau$, and the partition function of the MSW invariants $h_{\bfp,\bfmu}(\tau)$,
which is a vector-valued modular form of weight $(-\tfrac{b_2}{2}-1,0)$.
As a consequence, the Darboux coordinates around $t=\pm \I$ are not modular covariant, but transform with a specific
modular anomaly given by the period integral of the MSW elliptic genus. In the example above, this implies
\be
\delta\bfxi(0) \mapsto (c\tau+d)^{-1} \left(
\delta\bfxi(0) + \frac{\bfp}{4\pi} \, e^{-2\pi S_{\rm cl}} \sum_{\bfmu\in\Lambda^*/\Lambda}
h_{\bfp,\bfmu}(\tau)\,
\int_{-d/c}^{\I\infty} \frac{\overline{\tilde\theta_{\bfp,\bfmu}(w,\bar\tau,\bft,\bfb,\bfc)}\,
\de \bar w }{\sqrt{\I(\bar w-\tau)}}\right) .
\ee

The final step in the proof is to show that the modular anomalies in
the Darboux coordinates can be absorbed by a complex contact
transformation, which will then ensure that the complex contact structure on $\cZ$ transforms covariantly
under S-duality. For this purpose, we observe that
the Eichler integral which determines the corrections to the Darboux coordinates is in fact
the modular completion of the holomorphic mock theta
series  introduced in \cite{Zwegers-thesis},
\be
\label{eq:indefintro}
\overline{\Theta}_{\bfp,\bfmu}(\tau,\bft,\bft',\bfb,\bfc)
=\sum_{\bfk \in \Lambda+\bfmu+\smash{\tfrac12} \bfp}
\signkp
\[\sgn( (\bfk+\bfb)\cdot \bft )-\sgn( (\bfk+\bfb)\cdot \bft' )\]
e^{-\pi\I\tau(\bfk+\bfb)^2+ 2\pi\I\bfc\cdot (\bfk+\tfrac12\bfb)},
\ee
where $\bft$ are physical \kahler moduli and $\bft'$ is an arbitrary reference point on the
boundary of the \kahler cone.
In an amusing role reversal, we find that
this holomorphic mock theta series provides the generating function of a complex contact
transformation which cancels the modular anomaly in the Darboux coordinates. This
establishes the modular invariance of the QK metric in the large volume, one-instanton approximation.

Clearly, an important challenge for the future is to establish modularity
at the non-linear level. One possible strategy would be to recast the type IIA twistorial
construction into the manifestly $SL(2,\IZ)$-invariant type IIB formalism presented
in \cite{Alexandrov:2012bu}. As a first step in this direction, one may try to use the Poincar\'e
series representation (also known as Farey tail \cite{Dijkgraaf:2000fq,Manschot:2007ha})
of the generating functions of the MSW invariants to represent the section of $H^1(\cZ,\cO(2))$
describing D3-instanton corrections at linear order as a sum $H=\sum_{m,n} G_{m,n}$, such that
S-duality acts by permuting these contributions. Unfortunately, this manipulation is formal since
Poincar\'e series with negative weight are divergent without a suitable regularization.
Even though this naive idea fails, it nevertheless provides very useful guidance for determining
the quantum mirror map, as we explain in Appendices B and C.

The remainder of this work is organized as follows. In \S\ref{sec_rev}, we review the general
properties of the HM moduli space in type IIB string theory compactified on a CY threefold,
with particular emphasis on the discrete symmetries under S-duality and large gauge
transformations, and recall  the type IIA twistorial construction of D-instantons.
In \S\ref{sec_moddt}, we recall the geometric description of D3-brane instantons in terms of
coherent sheaves, the definition of the generalized DT invariants which count them and
their relation to the MSW invariants, which are independent of the moduli and have simple
modular properties. We define a multi-instanton expansion which is best suited for studying modularity,
and give a heuristic explanation of modular invariance in twistor space.
In \S\ref{sec_d3}, we analyze the instanton corrections to the Darboux coordinates
in the large volume, one-instanton limit, zooming in near  the S-duality
invariant points $t=\pm\I$. We express these instanton corrections in terms of
Eichler integrals of Siegel-Narain theta series, and construct their modular invariant completion.
Some open problems are discussed in \S \ref{sec_discuss}.
The relevant properties of indefinite theta series and Eichler integrals are reviewed
in Appendix \ref{app_mock}, while Appendix \ref{sec_inslve} contains details about the
derivation of instanton corrections and quantum mirror map
in the large volume limit. Finally, Appendix \ref{sec_poinca}
gives a preliminary attempt to recast the type IIA twistorial construction of D3-instantons
in the type IIB framework of \cite{Alexandrov:2012bu}.

%%%%%%%%%%%%%%%%%%%%%%%%%%%%%%%%%%%%%%%%%%%%%%%%%%%%%%%%%%%%%%%%%%%%
\section{HM moduli space in type IIB CY vacua\label{sec_rev}}
%%%%%%%%%%%%%%%%%%%%%%%%%%%%%%%%%%%%%%%%%%%%%%%%%%%%%%%%%%%%%%%%%%%%

In this section, we briefly recall some basic facts about the hypermultiplet moduli space
in type IIB  CY vacua. We mainly follow
\cite{Alexandrov:2008gh}, incorporating further improvements introduced
in \cite{Alexandrov:2009zh,Alexandrov:2010ca,Alexandrov:2011ac}, as well
as some new insights on the action of S-duality on the tree level geometry.

%%%%%%%%%%%%%%%%%%%%%%%%%%%%%%%%%%
\subsection{Tree-level metric}
%%%%%%%%%%%%%%%%%%%%%%%%%%%%%%%%%%

Recall that the hypermultiplet moduli space  in type IIB string theory  compactified on a CY
threefold $\CYm$ is a \qk manifold $\cM_H$ of dimension $4b_2+4$, where $b_2\equiv b_2(\CYm)$, which
describes the dynamics of
\begin{enumerate}
\item the ten-dimensional dilaton $g_s$;
\item the K\"ahler moduli $b^a + \I t^a\equiv \int_{\gamma^a} \mathcal{J}$
($a=1,\dots, b_2$)
where $\mathcal{J} \equiv B+\I\, J$ is the complexified \kahler form on  $\CYm$
and $\gamma^a$ is a basis of  $H_2(\CYm,\IZ)$;
\item the  Ramond-Ramond (RR) scalars $c^0,c^a,\cla,\cl0$, corresponding to periods of the RR 0-form,
2-form, 4-form and 6-form on a basis of $H_{\rm even}(\CYm,\IZ)$;
\item the NS axion $\psi$, dual to the NS 2-form $B$ in four dimensions.
\end{enumerate}
The ten-dimensional
string coupling $\tau_2\equiv 1/g_s$ and the RR axion $\tau_1\equiv c^0$
combine into the ten-dimensional axio-dilaton field $\tau = \tau_1+\I \tau_2$.
The resulting field basis $\tau,t^a,b^a, c^a,\cla,\cl0,\psi$ is adapted to the action
of S-duality in ten dimensions, as will become apparent below.

At tree level in type IIB string perturbation theory,
the metric on $\cM_H$ is obtained from the moduli space $\cM_{\cS\cK}$ of
complexified K\"ahler deformations
(which also describes the vector multiplet moduli space in Type IIA string theory
compactified on the {\it same} CY threefold $\CYm$) via the
$c$-map construction \cite{Cecotti:1988qn,Ferrara:1989ik}.
In the large volume limit, the special \kahler manifold
$\cM_{\cS\cK}$ is characterized by the holomorphic prepotential
\be
\label{Flv}
F^\lvol (X^\Lambda)=-\kappa_{abc} \,\frac{X^a X^b X^c}{6 X^0} +
\frac12 A_{\Lambda\Sigma} X^\Lambda X^\Sigma,
\ee
where $X^\Lambda$ are homogeneous complex coordinates on $\cM_{\cS\cK}$
such that $X^\Lambda/X^0=z^\Lambda$ (with $z^0=1$),
$ \kappa_{abc}$ is the triple intersection product in $H_4(\CYm,\IZ)$, and
$A_{\Lambda\Sigma}$ is a constant, real symmetric
matrix  which satisfies the quantization conditions
\cite{Hosono:1993qy,Alexandrov:2010ca}\footnote{For simplicity we set $A_{00}=0$ in this paper,
as this can always be achieved by an integer symplectic transformation. The constants $c_{2,a}$
are the components of the second Chern class of $\CYm$ along the basis of
$H_4(\CYm,\IZ)$.}
\be
A_{00}\in \IZ ,
\qquad
A_{0a} = \frac{c_{2,a}}{24} + \IZ ,
\qquad
A_{ab} p^p - \frac12 \kappa_{abc} p^b p^c\in \IZ \quad \text{for}\ \forall p^a\in\IZ.
\ee
The $c$-map construction \cite{Ferrara:1989ik} produces the QK metric
\be
\begin{split}
ds_{\cM_H}^2=&\frac{\de r^2}{r^2}
+4 \de s^2_{\cS\cK}
-\frac{1}{2r}\left(\de\tzeta_\Lambda - \bar\cN_{\Lambda\Sigma} \de\zeta^\Sigma\right)
\Im \cN^{\Lambda\Lambda'}\left(\de\tzeta_{\Lambda'} - \cN_{\Lambda'\Sigma'} \de\zeta^{\Sigma'}\right)
\\
&+ \frac{1}{16 r^2} \left(\de \sigma + \tzeta_\Lambda \de \zeta^\Lambda -\zeta^\Lambda \de \tzeta_\Lambda\right)^2,
\end{split}
\label{hypmettree}
\ee
where $\de s^2_{\cS\cK}$ is the metric on $\cM_{\cS\cK}$ with \kahler
potential $\cK=-\log[ \I ( \bar X^\Lambda F^\lvol_\Lambda - X ^\Lambda \bar F^\lvol_\Lambda)]$,
\be
\label{defcN}
\cN_{\Lambda\Lambda'} =\bar\tau_{\Lambda\Lambda'} +
2\I \frac{ [\Im\tau \cdot X]_\Lambda
[\Im \tau \cdot X]_{\Lambda'}}
{X^\Sigma \, \Im\tau_{\Sigma\Sigma'}X^{\Sigma'}}
\ee
is the Weil period matrix,
$\tau_{\Lambda\Sigma}\equiv \partial_{X^\Lambda}\partial_{X^\Sigma} F^\lvol(X)$ the Griffiths period matrix, and
\be
\label{defrphi}
r =   \frac{\tau_2^2}{2} \, \cV  ,
\qquad
\cV =\frac16\int_\CYm J\wedge J\wedge J
= \frac16 \,\kappa_{abc}t^a t^b t^c.
\ee
Here $\cV $ is the volume of the threefold $\CYm$ in string units, and the variable $r$ is related to the
four-dimensional string coupling by $r=1/g_4^2$.
The variables $z^a,\zeta^\Lambda, \tzeta_\Lambda,\sigma$ appearing in \eqref{hypmettree}
are the natural variables in type IIA string theory compactified on the threefold $\CY$
mirror to $\CYm$, and are related to the type IIB fields $\tau_1,t^a,b^a,c^a,\cla,\cl0,\psi$ by
the `classical mirror map' \cite{Bohm:1999uk}
\be
\label{symptobd}
\begin{split}
z^a & =b^a+\I t^a\, ,
\qquad
\zeta^0=\tau_1\, ,
\qquad
\zeta^a = - (c^a - \tau_1 b^a)\, ,
\\
\tzeta'_a &=  \cla+ \frac{1}{2}\, \kappa_{abc} \,b^b (c^c - \tau_1 b^c)\, ,
\qquad
\tzeta'_0 = \cl0-\frac{1}{6}\, \kappa_{abc} \,b^a b^b (c^c-\tau_1 b^c)\, ,
\\
\sigma &= -2 (\psi+\frac12  \tau_1 \cl0) + \cla (c^a - \tau_1 b^a)
-\frac{1}{6}\,\kappa_{abc} \, b^a c^b (c^c - \tau_1 b^c)\, .
\end{split}
\ee
Here the primed coordinates $\tzeta'_\Lambda\equiv \tzeta_\Lambda-A_{\Lambda\Sigma}
\zeta^\Lambda$ are the RR fields in a non-integer homology basis on the type IIA side, which
is best suited for comparison with type IIB \cite{Alexandrov:2010ca}.
The electric charges $q'_\Lambda$ in this basis
are related to the electric charges $q_\Lambda$ in the integer basis
by $q'_\Lambda=q_\Lambda-A_{\Lambda\Sigma} p^\Sigma$, and therefore
satisfy the quantization properties
\be
\label{fractionalshiftsD5}
q'_a \in \IZ - \frac{p^0}{24} c_{2,a} - \frac12 \kappa_{abc} p^b p^c ,
\qquad
q'_0\in \IZ-\frac{1}{24} p^a c_{2,a} .
\ee
The prepotential in the non-integer
basis is given by \eqref{Flv} with $A_{\Lambda\Sigma}=0$, which we henceforth
denote by $F'^\lvol(X^\Lambda)$.

%%%%%%%%%%%%%%%%%%%%%%%%%%%%%%%%%%
\subsection{Continuous symmetries at tree-level}
%%%%%%%%%%%%%%%%%%%%%%%%%%%%%%%%%%
\label{subsec_sym}

The tree-level metric \eqref{hypmettree} is invariant under a large group of
continuous isometries which is generated by three subgroups:
\begin{itemize}
\item continuous shifts of the RR and NS axions
%, often known as `Heisenberg shifts',
%\footnote{The constant shifts
%in the variation of $\sigma$ in \eqref{heis0}, of $\phi^a$ in \eqref{bjacr} and  of $\cla$ in \eqref{SL2Z}
%below are not determined by the invariance of the metric at the classical level.
%However, they become important once these isometries are broken by quantum effects to their discrete subgroups
%and are further discussed in \S\ref{sec_Sinst}.}
\be
\label{heis0}
T_{(\eta^\Lambda,\tleta_\Lambda,\kappa)}:\
\bigl(\zeta^\Lambda,\tzeta_\Lambda,\sigma\bigr)\mapsto
\bigl(\zeta^\Lambda + \eta^\Lambda ,\
\tzeta_\Lambda+ \tleta_\Lambda ,\
\sigma + 2 \kappa- \tleta_\Lambda \zeta^\Lambda
+ \eta^\Lambda \tzeta_\Lambda  \bigr),
%\sigma + 2 \kappa- \tleta_\Lambda (\zeta^\Lambda -2 \theta^\Lambda)
%+ \eta^\Lambda (\tzeta_\Lambda -2 \phi_\Lambda)
%- \eta^\Lambda \tleta_\Lambda \bigr),
\ee
%where $\theta^\Lambda,\phi_\Lambda$ are constant characteristics; as observed
%in \cite{Alexandrov:2010ca}, and as we shall
%confirm in \S\ref{sec_twilin}, S-duality requires $\theta^\Lambda=0$, which we
%assume in the sequel.
\item
monodromies around the large volume point
\be
\label{bjacr}
\begin{split}
M_{\epsilon^a}\ :\ b^a&\mapsto b^a+\eps^a\, ,
\qquad
\zeta^a\mapsto \zeta^a + \epsilon^a \zeta^0\, ,
\qquad
\tzeta'_a\mapsto \tzeta'_a -\kappa_{abc}\zeta^b \epsilon^c
-\frac12\,\kappa_{abc} \epsilon^b \epsilon^c \zeta^0\, ,
\\
\tzeta'_0&\mapsto \tzeta'_0 -\tzeta'_a \epsilon^a+\frac12\, \kappa_{abc}\zeta^a \epsilon^b \epsilon^c
+\frac16\,\kappa_{abc} \epsilon^a \epsilon^b \epsilon^c \zeta^0\, ;
%\qquad
%\phi_a \mapsto \phi_a +\frac12 \kappa_{aab} \epsilon^b\ ;
\qquad
\end{split}
\ee

\item
continuous $SL(2,\IR)$ transformations given in the type IIB field basis by
\be\label{SL2Z}
\begin{split}
&\quad \tau \mapsto \frac{a \tau +b}{c \tau + d} \, ,
\qquad
t^a \mapsto t^a |c\tau+d| \, ,
\qquad
\cla\mapsto \cla \, ,% - c_{2,a}\, \varepsilon(\trans)\, ,
\\
&
\begin{pmatrix} c^a \\ b^a \end{pmatrix} \mapsto
\begin{pmatrix} a & b \\ c & d  \end{pmatrix}
\begin{pmatrix} c^a \\ b^a \end{pmatrix}\, ,
\qquad
\begin{pmatrix} \cl0 \\ \psi \end{pmatrix} \mapsto
\begin{pmatrix} d & -c \\ -b & a  \end{pmatrix}
\begin{pmatrix} \cl0 \\ \psi \end{pmatrix},
\end{split}
\ee
with $ad-bc=1$ \cite{Gunther:1998sc,Bohm:1999uk}.
%The constant shift in $\cla$ is proportional
%to the  multiplier system $\varepsilon(\trans)$ of the Dedekind
%eta function, defined by
%\be
%\label{multeta}
%\eta\left(\frac{a\tau+b}{c\tau+d}\right)\slash \eta(\tau)
%=e^{2\pi\I \eps(\trans)}(c\tau+d)^{1/2}\, .
%\ee
\end{itemize}

The algebra generated by these
isometries is the semi-direct product $SL(2,\IR)\ltimes N$, where $N=N^{(1)}\oplus
N^{(2)}\oplus N^{(3)}$ is a nilpotent algebra of dimension $3b_2+ 2$,
satisfying
\be
\label{Ncomm}
[N^{(1)},N^{(1)}] \subset N^{(2)} ,
\qquad
[N^{(p)},N^{(q)}] = 0 \quad \mbox{if} \ p+q\geq 3 \, ,
\ee
with the generators in $N^{(1)}, N^{(2)}, N^{(3)}$ transforming as $b_2$ doublets,
$b_2$ singlets and one doublet under $SL(2,\IR)$, respectively. The relation of this algebra decomposition
to the symmetries presented above is the following:
\begin{itemize}
\item
the group elements $T^{(1)}_{(\epsilon^a,\eta^a)}$ obtained by exponentiating
$N^{(1)}$ (with a suitable admixture of higher level generators) consist of the monodromies \eqref{bjacr} and
the Heisenberg shifts \eqref{heis0} with non-vanishing $\eta^a$;
\item
the group elements $T^{(2)}_{\tilde\eta_a}$ obtained by exponentiating
$N^{(2)}$ correspond to the Heisenberg shifts with non-vanishing $\tilde\eta_a$;
\item
the group elements $T^{(3)}_{(\tilde\eta_0,\kappa)}$
associated to $N^{(3)}$ coincide with Heisenberg shifts with non-vanishing $\tilde\eta_0$ and $\kappa$;
\item
unlike the other Heisenberg shifts which belong to $N$, the Heisenberg shift $\zeta^0\mapsto\zeta^0+\eta^0$
belongs to $SL(2,\IR)$ and in fact coincides
with the $SL(2,\R)$ transformation $\tau\mapsto\tau+\eta^0$.
\end{itemize}
In the presence of quantum corrections, all these continuous isometries are lifted,
but a discrete subgroup of these symmetries is conjectured to remain unbroken,
as will be discussed in \S\ref{sec_Sinst}.

%%%%%%%%%%%%%%%%%%%%%%%%%%%%%%%%%%
\subsection{Twistorial description of the tree-level metric}
%%%%%%%%%%%%%%%%%%%%%%%%%%%%%%%%%%

In studying instanton corrections to hypermultiplet moduli spaces, it has proven
very useful to encode the metric on a QK manifold $\cM$ in terms of the complex
contact structure on its twistor space $\cZ$, the total space of the $\CP$-bundle over
$\cM$ twisted with the projectivized $SU(2)$ connection
(see \cite{Alexandrov:2011va} for a review of this approach).
This contact structure is represented by a (twisted) holomorphic one-form $\cX$,
which locally can always be expressed in terms of complex Darboux coordinates as
\be
\cX^{[i]}=  \de\ai{i}+ \txii{i}_\Lambda \de \xii{i}^\Lambda\, ,
\label{contform}
\ee
where the index $\scriptsize [i]$ labels the
patches $\cU_i$ of an open covering\footnote{In principle, since $\cZ$ is
a non-trivial $\CP$-bundle over $\cM$, one should be using an open covering of $\cZ$.
However, for QK manifolds with one quaternionic isometry, as is the case in this paper, the QK/HK
correspondence gives a natural trivialization of an open subset of $\cZ$, which allows to
view $\cZ$ as a $\cM$-bundle over $\CP$.}
of $\CP$. The contact structure on $\cZ$
(hence, the QK metric on $\cM$)  is then encoded in holomorphic
generating functions $\Hij{ij}(\xii{i}^\Lambda,\txii{j}_\Lambda,\ai{j})$
for the contact transformations between Darboux coordinate systems on the overlap
$\cU_i \cap \cU_j$. For QK manifolds with one quaternionic isometry, which in our case is realized as constant
shifts of the NS axion $\sigma$, the transition functions  $\Hij{ij}$ should be independent of $\ai{j}$.
In this case, the corresponding contact transformation takes the following form
\be
\begin{split}
\xii{j}^\Lambda =&\,  \xii{i}^\Lambda -\p_{\txii{j}_\Lambda }\Hij{ij},
\qquad
\txii{j}_\Lambda =  \txii{i}_\Lambda + \p_{\xii{i}^\Lambda } \Hij{ij},
\\
&\
\ai{j} =  \ai{i}+\Hij{ij}- \xii{i}^\Lambda \p_{\xii{i}^\Lambda}\Hij{ij}
\end{split}
\label{QKgluing}
\ee
and reduces to a complex symplectomorphism in the subspace spanned by $(\xi^\Lambda,\txi_\Lambda)$ variables.
In addition to the set of transition functions $\Hij{ij}$, one must also specify
a real constant $c_\alpha$, known as the anomalous dimension, which characterizes
the singular behavior of $\alpha$ near the poles of $\CP$.
The QK metric is then obtained
by `parametrizing the twistor lines', i.e.  solving the gluing conditions \eqref{QKgluing}
for the Darboux coordinates as functions of coordinates on
the base $\cM$ and the $\CP$  coordinate $\varpi$, and expanding the contact one-form
in the vicinity of any fixed point on $\CP$. A useful construct in this procedure
is the contact potential $e^\Phi$, which determines a K\"ahler potential on $\cZ$
(see \cite{Alexandrov:2008nk} for more details on this procedure).

As explained in \cite{Alexandrov:2008nk,Alexandrov:2008gh}, the QK metric \eqref{hypmettree}
can be cast in this twistorial framework by choosing a covering of $\CP$ consisting of
two patches $\cU_+$, $\cU_-$ around the north and south poles,
$\varpi=0$ and $\varpi=\infty$,
and a third patch $\cU_0$ which covers the equator. The transition functions
between complex Darboux coordinates on each patch are given by
\be
\label{symp-cmap}
\hHij{+0}= F^\lvol(\xi^\Lambda) ,
\qquad
\hHij{-0}=\bF^\lvol(\xi^\Lambda) ,
\ee
whereas the corresponding Darboux coordinates in the patch $\cU_0$ read as \cite{Neitzke:2007ke,Alexandrov:2008nk}
\be
\label{gentwi}
\begin{array}{rcl}
\xi^\Lambda &=& \zeta^\Lambda + \frac{\tau_2}{2}
\left( \varpi^{-1} z^{\Lambda} -\varpi \,\bz^{\Lambda}  \right) ,
\\
\txi_\Lambda &=& \tzeta_\Lambda + \frac{\tau_2}{2}
\left( \varpi^{-1} F^\lvol_\Lambda-\varpi \,\bF^\lvol_\Lambda \right) ,
\\
\tilde\alpha&=& \sigma + \frac{\tau_2}{2}
\left(\varpi^{-1} W^\lvol-\varpi \,\bar W^\lvol \right),
\end{array}
\ee
where $W^\lvol$ denotes the `superpotential'
\be
\label{defW}
W^\lvol(z) \equiv  F^\lvol_\Lambda(z) \, \zeta^\Lambda - z^\Lambda \tzeta_\Lambda
\ee
and  we have traded the Darboux coordinate $\alpha$
for $\tilde \alpha \equiv -2\alpha - \txi_\Lambda \xi^\Lambda$.
Moreover, the contact potential $e^\Phi$ turns out to coincide with the four-dimensional string coupling
$r$ given in \eqref{defrphi}.

A very useful property of this twistorial framework is that isometries
of $\cM$ can always be lifted to a holomorphic contact transformation on the twistor
space $\cZ$, by a suitable choice of action on the $\CP$ fiber. Thus,
the nilpotent generators \eqref{heis0} and  \eqref{bjacr} (supplemented with the
trivial action on $\CP$) lift to the holomorphic contact transformations
\be
\label{heisalgz}
T_{(\eta^\Lambda,\tleta_\Lambda,\kappa)}:  \
\bigl(\xi^\Lambda,\txi_\Lambda,\tilde\alpha\bigr)\mapsto
\bigl(\xi^\Lambda + \eta^\Lambda ,\ \txi_\Lambda+ \tleta_\Lambda,\
\tilde\alpha +2\kappa-  \tleta_\Lambda \xi^\Lambda+
\eta^\Lambda \txi_\Lambda
\bigr),
%\tilde\alpha +2\kappa-  \tleta_\Lambda (\xi^\Lambda-2 \theta^\Lambda) +
%\eta^\Lambda (\txi_\Lambda-2\phi_\Lambda)
%- \eta^\Lambda \tleta_\Lambda
%\bigr),
\ee
\be
\label{bjacz}
\begin{split}
M_{\epsilon^a}\ :\ \xi^0\mapsto & \xi^0\, ,
\qquad
\xi^a\mapsto \xi^a + \epsilon^a \xi^0\, ,
\qquad
\txi'_a\mapsto \txi'_a -\kappa_{abc}\xi^b \epsilon^c-\frac12\,\kappa_{abc} \epsilon^b \epsilon^c \xi^0\, ,
\\
&\txi'_0\mapsto \txi'_0 -\txi_a \epsilon^a+\frac12\, \kappa_{abc}\xi^a \epsilon^b \epsilon^c
+\frac16\,\kappa_{abc} \epsilon^a \epsilon^b \epsilon^c \xi^0\, ,
\qquad
\tilde\alpha\mapsto\tilde\alpha\, ,
\end{split}
\ee
respectively, where we defined the Darboux coordinates in the non-integer symplectic basis
in the same way as in \eqref{symptobd},
\be
\txi'_\Lambda=\txi_\Lambda-A_{\Lambda\Sigma} \xi^\Sigma,
\qquad
\alpha'=\alpha-\frac12\, A_{\Lambda\Sigma} \xi^\Lambda \xi^\Sigma.
\ee

Similarly, the isometric $SL(2,\R)$ action \eqref{SL2Z} lifts to the following
holomorphic action on $\cZ$ (here expressed in terms of the Darboux coordinates
in the patch $\hcU_0$):
\be
\label{SL2Zxi}
\begin{split}
\xi^0 &\mapsto \frac{a \xi^0 +b}{c \xi^0 + d} \, ,
\qquad
\xi^a \mapsto \frac{\xi^a}{c\xi^0+d} \, ,
\qquad
\txi'_a \mapsto \txi'_a +  \frac{c}{2(c \xi^0+d)} \kappa_{abc} \xi^b \xi^c\, ,
%- c_{2,a}\, \varepsilon(\trans)\, ,
\\
\begin{pmatrix} \txi'_0 \\ \alpha' \end{pmatrix} &\mapsto
\begin{pmatrix} d & -c \\ -b & a  \end{pmatrix}
\begin{pmatrix} \txi'_0 \\ \alpha' \end{pmatrix}
+ \frac{1}{6}\, \kappa_{abc} \xi^a\xi^b\xi^c
\begin{pmatrix}
c^2/(c \xi^0+d)\\
-[ c^2 (a\xi^0 + b)+2 c] / (c \xi^0+d)^2
\end{pmatrix}\, .
\end{split}
\ee
This action agrees with \eqref{SL2Z} provided the latter is supplemented with
a $SU(2)$ rotation along the fiber,
\be
\label{ztrans}
z\mapsto \frac{c\bar\tau+d}{|c\tau+d|}\, z  ,
\ee
where the coordinate $z$ is related to the coordinate $t$ appearing in \eqref{gentwi} by
the Cayley transformation:
\be
\label{Cayley}
z=\frac{t+\I}{t-\I},
\qquad
t=-\I \,\frac{1+z}{1-z}  .
\ee
Under this  action,  the complex contact one-form \eqref{contform} transforms by
an overall holomorphic factor $\hCX\ui{0}\mapsto \hCX\ui{0}/(c\xi^0+d)$, leaving the complex
contact structure invariant, while the contact potential
$e^\Phi$ transforms with modular weight $(-\frac12,-\frac12)$,
\be
\label{SL2phi}
e^\Phi \mapsto \frac{e^\Phi}{|c\tau+d|}\, .
\ee
It will be key to note that the points $z=0,\infty$ (corresponding to $t=\pm\I$)
are invariant under $SL(2,\R)$. As a result, the leading Taylor coefficients of the Darboux
coordinates around these points have simple modular transformations, e.g.
$\xi^a = \tau b^a-c^a+\cO(z)$ transforms with modular weight $(-1,0)$.

%%%%%%%%%%%%%%%%%%%%%%%%%%%%%%%%%%
\subsection{Quantum corrections and S-duality \label{sec_Sinst}}
%%%%%%%%%%%%%%%%%%%%%%%%%%%%%%%%%%

Away from the large volume, weak coupling limit, the continuous $SL(2,\IR)$
isometric action is broken
by several types of quantum corrections. Firstly, due to worldsheet instantons
 the holomorphic prepotential \eqref{Flv} entering the  $c$-map metric \eqref{hypmettree}
is corrected into $F=F^{\lvol}+F^{\ws}$ where
\be
\label{Fws}
F^{\ws}(X^\Lambda)=
\chi_\CYm\,
\frac{\zeta(3)(X^0)^2}{2(2\pi\I)^3}
-\frac{(X^0)^2}{(2\pi\I)^3}{\sum_{k_a\gamma^a\in H_2^+(\CYm)}} n_{k_a}^{(0)}\,
\Li_3\left( e^{2\pi \I  k_a X^a/X^0}\right).
\ee
The first term, proportional to the  Euler number $\chi_\CYm$ of $\CYm$,
corresponds to a perturbative correction in the $\alpha'$-expansion around large volume.
The second term corresponds to
a sum of worldsheet instantons, labeled by their effective\footnote{Effective
means $k_a\in \IZ^+$ for all $a$,
not all $k_a$'s vanishing simultaneously.}
 homology class
$k_a\gamma^a\in H_2^+(\CYm,\IZ)$, and weighted by the genus zero
Gopakumar-Vafa invariants $n_{k_a}^{(0)}\in \IZ$.
These instantons contribute through the tri-logarithm function
$\Li_3(x)=\sum_{m=1}^\infty m^{-3}x^m$, which takes into account
multi-covering effects. Note that the two terms
in \eqref{Fws} may be combined by including the zero class
$k_a=0$ in the sum and setting $n_0^{(0)}=-\chi_\CYm/2$. In the presence of these
worldsheet instantons, the twistor space is
still described by three patches related  by the transition functions \eqref{symp-cmap},
where $F^\lvol(\xi^\Lambda)$ is replaced by $F(\xi^\Lambda)$.

Secondly, the HM moduli space receives corrections controlled by the string coupling $g_s$.
At the perturbative level, it is believed \cite{Gunther:1998sc,Robles-Llana:2006ez,Alexandrov:2008nk}
that the only corrections occurs at one-loop, and is proportional to the Euler number
$\chi_\CYm$ \cite{Antoniadis:1997eg,Gunther:1998sc}. Its effect on the metric was explicitly
calculated in \cite{Robles-Llana:2006ez,Alexandrov:2007ec}. At the level of the twistor space,
the one-loop correction is simply incorporated by allowing a logarithmic singularity in
the Darboux coordinate $\tilde\alpha$ with prescribed coefficient,
\be
\label{alp1loop}
\tilde\alpha= \sigma + \frac{\tau_2}{2}
\left(\varpi^{-1} W-\varpi \,\bar W \right)
-\frac{\I\chi_\CYm}{24\pi} \,\log \varpi \, .
\ee
Including both worldsheet instantons and the one-loop correction, the contact potential
(equivalently the one-loop corrected four-dimensional
dilaton) is now given by
\be
\label{phipertB}
r=e^{\Phi}= \frac{\tau_2^2}{2} \,\cV -\frac{\zeta(3)\chi_\CYm}{8(2\pi)^3}\,\tau_2^2
+ e^{\Phi_{\rm ws}}- \frac{\chi_\CYm}{192\pi},
\ee
where
\be
\label{phiws}
e^{\Phi_{\rm ws}} =\frac{\tau_2^2}{4(2\pi)^3}\sum_{k_a\gamma^a\in H_2^+(\CYm)} n_{k_a}^{(0)}
\Re\left[ \Li_3 \left( e^{2\pi \I k_a z^a} \right) + 2\pi k_a t^a\,
\Li_2 \left( e^{2\pi \I k_a z^a} \right)  \right].
\ee

In addition to these perturbative corrections, one expects instanton corrections from Euclidean D(-1), D1, D3, D5-branes
wrapping supersymmetric 0,2,4,6-cycles on $\CYm$, as well as from Euclidean NS5-branes
wrapping $\CYm$. Lacking a non-perturbative definition of string theory, an effective way
of computing these effects is to assume that a certain discrete subgroup of the classical
symmetry group $SL(2,\IR)\times N$ stays unbroken at the quantum level,
and investigate how this symmetry constrains quantum contributions.

For what concerns the nilpotent algebra of isometries $N$, it is natural to assume
that quantum corrections will be invariant under the unipotent group $N(\IZ)$
consisting of integer monodromies around the large volume point, and large gauge
transformations of the RR and NS axions. The former consists of monodromies
$M_{\epsilon^a}$ in \eqref{bjacr} with $\epsilon^a$ integer.
The latter were analyzed in \cite{Alexandrov:2010np,Alexandrov:2010ca} using properties of the five-brane
partition function. It was found that large gauge transformations act by integer shifts
of the RR axions $\zeta^\Lambda,\tzeta_\Lambda$, but require an additional shift of the NS-axion,
\be
T_{(\eta^\Lambda,\tleta_\Lambda,\kappa)}:\ \sigma \mapsto
\sigma + 2 \kappa- \tleta_\Lambda (\zeta^\Lambda -2 \theta^\Lambda)
+ \eta^\Lambda (\tzeta_\Lambda -2 \phi_\Lambda)
- \eta^\Lambda \tleta_\Lambda\, ,
\ee
where  $\theta^\Lambda,\phi_\Lambda$ are the characteristics entering
the fivebrane partition function, and $\kappa$ is an arbitrary integer.
Clearly, the same correction must also affect the
transformation of the Darboux coordinate $\tilde\alpha$ in \eqref{heisalgz}.
%\be
%\tilde\alpha \mapsto \tilde\alpha +2\kappa-  \tleta_\Lambda (\xi^\Lambda-2 \theta^\Lambda) +
%\eta^\Lambda (\txi_\Lambda-2\phi_\Lambda)
%- \eta^\Lambda \tleta_\Lambda \ .
%\ee

Finally, since the $SL(2,\IR)$ action \eqref{SL2Z}
descends from the S-duality of ten-dimensional type IIB string theory (or, in the
T-dual picture, from the diffeomorphism symmetry of M-theory compactified on $\CYm\times
T^2$), it is natural to assume that the modular subgroup $SL(2,\IZ)$ remains a symmetry at the quantum level.
It was noted in \cite{Alexandrov:2010ca}, and will be further confirmed in this
paper, that the action of S-duality must however differ from the naive action \eqref{SL2Z} by
an additional shift in the RR axion dual to D3-branes
\be
\cla\mapsto \cla \, - c_{2,a}\, \varepsilon(g)\, ,
\ee
where $\varepsilon(g) \in \left(-\frac{1}{2},\frac{1}{2}\right]$ is defined in terms of the multiplier system of the Dedekind
eta function $\eta(\tau)$:
\be
\label{multeta}
e^{2\pi\I\, \varepsilon(g) }=\frac{\eta\left(\frac{a\tau+b}{c\tau+d}\right)}{(c\tau+d)^{1/2}\,
\eta(\tau)},  \qquad g={\left(\begin{array}{cc} a
    & b\\ c & d \end{array}
\right)}\, .
\ee
Of course, the same correction must affect the Darboux coordinate $\txi'_a$,
\be
\txi'_a \mapsto \txi'_a +  \frac{c}{2(c \xi^0+d)} \kappa_{abc} \xi^b \xi^c\,
- c_{2,a}\, \varepsilon(\trans)\, .
\ee
These additional shifts are closely related to the
quantization condition \eqref{fractionalshiftsD5} for the charge $q'_0$, and
ensure the coincidence of the Heisenberg shift $\zeta^0\mapsto\zeta^0+\eta^0$
with the $SL(2,\Z)$ transformation $\tau\mapsto\tau+\eta^0$, which was noticed at
the classical level.
Moreover, it was observed that the characteristics $\theta^\Lambda$ must vanish,
and that $\phi_a$ must transform under monodromies $M_{\epsilon^a}$ as
$\phi_a \mapsto \phi_a +\frac12 \kappa_{aab} \epsilon^b$.

Based on these considerations, it is therefore natural to expect that the instanton corrected
moduli space wil be invariant under the action of\footnote{Further enhancements of
S-duality to  $SL(3,\IZ)$ or $SU(2,1,\IZ)$ were proposed in \cite{Pioline:2009qt,Bao:2009fg}.}
 $SL(2,\IZ)\ltimes N(\IZ)$, where the
action of the unipotent group $N(\IZ)$ in the type IIB field basis is summarized in the following table:

\vspace{0.5cm}
\noindent
\begin{tabular}{|c|c|c|c|c|c|}
\hline $\vphantom{\frac{A^{A^A}}{A_{A_A}}}$
& $\delta b^a$ & $\delta c^a$ & $\delta\cla$ &$\delta\cl0$ & $\delta\psi$
\\
\hline $\vphantom{\frac{A^{A^A}}{A_{A_A}}}$
$T^{(1)}_{(\epsilon^a,0)}$ &   $\epsilon^a$    &   $0$  & $\frac12\, \kappa_{abc} \epsilon^b c^c$
& $-\epsilon^a \cla-\frac16\, \kappa_{abc} \epsilon^a (b^b+2\epsilon^b)c^c$
& $\frac16\, \kappa_{abc}\epsilon^a c^b c^c$
\\
\hline $\vphantom{\frac{A^{A^A}}{A_{A_A}}}$
$T^{(1)}_{(0,\eta^a)}$&  $0$    &    $\eta^a$  & $-\frac12\, \kappa_{abc} \eta^b b^c+A_{ab} \eta^b$
&  $\frac16\, \kappa_{abc} \eta^a b^b b^c + A_{0a} \eta^a $
&  $\begin{array}{c}\eta^a (\cla-\phi_a)+\frac12\, A_{ab} \eta^a \eta^b\\
-\frac16\, \kappa_{abc}\eta^a b^b (c^c+2\eta^c)\end{array}$
\\
\hline $\vphantom{\frac{A^{A^A}}{A_{A_A}}}$
$T^{(2)}_{\tilde\eta_a}$
&  $0$  & $0$ & $\tilde\eta_a$
& $0$  & $ 0$
\\
\hline $\vphantom{\frac{A^{A^A}}{A_{A_A}}}$
$T^{(3)}_{(\tilde\eta_0,\kappa)}$ &  $0$  & $0$ & $0$
&  $\tilde\eta_0$ & $\kappa$
\\
\hline
\end{tabular}
\vspace{0.5cm}

In the remainder of this paper, we shall ignore the action of this discrete symmetry
on the axions $\tilde c_0$ and $\psi$, since they couple only to D5-brane and NS5-branes.
In this sector, the
generators $T^{(3)}_{\tilde\eta_0,\kappa}$ act trivially, and the remaining
generators  satisfy the Heisenberg-type relations
 (not to be confused
with the Heisenberg algebra satisfied by the generators \eqref{heis0})
\be
T^{(1)}_{(0,-\eta^a)}\cdot
T^{(1)}_{(-\epsilon^a,0)}\cdot
T^{(1)}_{(0,\eta^a)}\cdot T^{(1)}_{(\epsilon^a,0)}
= T^{(2)}_{\kappa_{abc} \eps^b \eta^c}\, .
\ee
It is interesting to note that this
Heisenberg-type algebra is isomorphic to the one which appears in the Coulomb branch
of five-dimensional gauge theories compactified on a torus \cite{Haghighat:2011xx}. We comment
on the relation between the string theory and field theory set-ups in \S\ref{sec_discuss}.

%%%%%%%%%%%%%%%%%%%%%%%%%%%%%%%%%%
\subsection{Twistorial construction of D-instanton corrections \label{sec_twiA}}
%%%%%%%%%%%%%%%%%%%%%%%%%%%%%%%%%%

Using the twistorial methods reviewed above and the symmetry restrictions
discussed in the previous subsection, in \cite{Alexandrov:2008gh,Alexandrov:2009zh}
it was shown how to incorporate all D-brane multi-instanton corrections.
For convenience we shall phrase
this construction in the language of type IIA string theory compactified on the mirror
threefold $\CY$, where  D-instantons correspond to Euclidean D2-branes
wrapping special Lagrangian cycles with homology class
$\gamma=(p^\Lambda,q_\Lambda)\in H_3(\CY,\IZ)$
(or more generally, elements in the Fukaya category of $\CY$).

To specify the twistor space corrected by these D2-brane contributions,
one should replace the patch $\hcU_0$, appearing in the above perturbative construction,
by a (in general infinite) set of angular sectors which extend between two consecutive BPS rays
\be
\ell_\gamma = \{ \varpi\in \CP\ :\ \ Z_\gamma(z^a)/\varpi\in \I \IR^-\} ,
\ee
where $Z_\gamma(z^a)$ is the central charge
\be
Z_\gamma(z^a) = q_\Lambda z^\Lambda - p^\Lambda F_\Lambda(z^a) ,
\ee
evaluated at a fixed point $z^a$ in $\cM_{\cS\cK}$.
Across each  BPS ray $\ell_\gamma$, the Darboux coordinates are required to jump
by the following complex contact transformation
\be
\begin{split}
\Delta \xi^\Lambda =& \frac{\Omega(\gamma;z^a)}{2\pi\I} \, p^\Lambda\, \log\left[1-\sigma_\gamma \cX_{\gamma}\right]
\\
\Delta \txi_\Lambda =& \frac{\Omega(\gamma;z^a)}{2\pi\I} \, q_\Lambda\, \log\left[1-\sigma_\gamma \cX_\gamma\right]
\\
\Delta\tilde\alpha =& \frac{\Omega(\gamma;z^a)}{2\pi^2} \,
\left( {\rm Li}_2\left(\sigma_\gamma \cX_\gamma \right)
-\hf\, \log\cX_\gamma\,  \log\left[1-\sigma_\gamma \cX_\gamma\right]\right),
\end{split}
\label{elemks}
\ee
where $\cX_\gamma = \expe{p^\Lambda \txi_\Lambda-q_\Lambda \xi^\Lambda}$,
$\Omega(\gamma;z^a)$ are the generalized Donaldson-Thomas (DT)
invariants\footnote{As further discussed in \S\ref{sec_dtmod},  the DT invariants
are piecewise constant functions of the moduli $z^a$.
Thus, despite appearances \eqref{elemks} is still a holomorphic complex transformation.
Moreover, as in the gauge theory setting \cite{Gaiotto:2008cd},
the consistency of the construction for different values of the moduli  $z^a$
requires that the jump of the DT invariants across lines of marginal
stability (where two BPS rays collide) satisfies the Kontsevich-Soibelman wall-crossing formula \cite{ks}.},
$\sigma_\gamma$ is the quadratic refinement
\be
\sigma_\gamma = \expe{-\frac12 q_\Lambda p^\Lambda + q_\Lambda \theta^\Lambda -
p^\Lambda \phi_\Lambda} ,
\ee
and we introduced the notation $\expe{x}=e^{2\pi\I x}$ which will be extensively used below.
In \cite{Alexandrov:2009zh} it was shown how to integrate the contact transformation \eqref{elemks}
to a generating function $\Hij{ij}\equiv H^{[\gamma]}$.
 While the result is somewhat cumbersome,  we will only require its limit in the
 one-instanton approximation,
\be
H^{[\gamma]}(\xi,\txi)=\frac{\Omega(\gamma;z^a)}{(2\pi)^2}\,
\Li_2\left(\sigma_\gamma \cX_\gamma \right).
\label{prepH}
\ee

Using the fact that the Darboux coordinates must approach the classical answer \eqref{gentwi}, \eqref{alp1loop}
exponentially fast at $\varpi=0,\ \varpi=\infty$, the gluing conditions \eqref{elemks}
can be rewritten as integral equations for the Fourier modes $\cX_\gamma$,
having the form of the Thermodynamical Bethe Ansatz equations \cite{Gaiotto:2008cd,Alexandrov:2010pp},
\be
\cX_\gamma = \cX^{\rm sf}_\gamma\, \expe{
\frac{1}{8\pi^2}
\sum_{\gamma'} \Omega(\gamma';z^a)\, \langle \gamma ,\gamma'\rangle
\int_{\ell_{\gamma'} }\frac{\de t'}{t'}\, \frac{t'+t}{t'-t}\,
\log\left[1-\sigma_{\gamma'} \, \cX_{\gamma'} \right] },
\label{eqTBA}
\ee
where
\be
\cX^{\rm sf}_\gamma=\expe{\frac{\tau_2}{2}\(\bZ_\gamma\,t-\frac{Z_\gamma}{t}\)
+p^\Lambda \tzeta_\Lambda- q_\Lambda \zeta^\Lambda}
\ee
are the Fourier modes constructed from the
perturbative Darboux coordinates \eqref{gentwi}. Eq. \eqref{eqTBA}
uniquely determines the Darboux coordinates in each angular sector in terms of
the DT invariants $\Omega(\gamma;z^a)$. In the weak coupling limit, it
can be solved iteratively, leading to a (formal) multi-instanton series.
The one-instanton approximation is obtained by substituting $\cX_\gamma\to\cX^{\rm sf}_{\gamma'}$
on the r.h.s. and evaluating the integral over $t'$. The  latter is
dominated by a saddle point  at
\be
\label{gensaddle}
t'_{\gamma} = \I\, \frac{Z_{\gamma}}{|Z_\gamma|},
\qquad
S_{\gamma} = \tau_2|Z_\gamma| - \I ( p^\Lambda \tzeta_\Lambda
- q_\Lambda \zeta^\Lambda),
\ee
leading to a correction of order $\cO(e^{-2\pi S_{\gamma}})$ consistent
with the expected semi-classical action for single-centered D-instantons.
Having solved these equations, the contact potential is then obtained from the Penrose-type integral
\be
e^{\Phi} = \frac{\tau_2^2}{16}\, e^{-\cK(z,\bz)} +\frac{\chi_\CY}{192\pi}
-\frac{ \I\tau_2}{64\pi^2}\sum\limits_{\gamma} \Omega(\gamma;z^a)
\int_{\ellg{\gamma}}\frac{\d \varpi}{\varpi}\,
\( \varpi^{-1} Z_\gamma -\varpi\bZ_\gamma\)
\log\left[1-\sigma_\gamma \cX_\gamma  \right].
\label{phiinstmany}
\ee

Since the HM moduli spaces of type IIA and type IIB string theories compactified on mirror
Calabi-Yau threefolds must be identical, the above construction also defines in principle
the D-instanton corrections in the type IIB formulation. In practice however, this is complicated by
the fact that the relation between the coordinates $z^a,\zeta^\Lambda, \tzeta_\Lambda,\sigma$
appearing in the boundary conditions \eqref{gentwi} and the physical type IIB fields\footnote{In contrast,
it is reasonable to assume that the coordinates $z^a,\zeta^\Lambda, \tzeta_\Lambda,\sigma$
are identified with the physical type IIA fields, as they transform in the required way
under symplectic transformations.}
is a priori unknown
beyond the classical mirror map \eqref{symptobd}. One way to specify this relation is to require
that the physical type IIB fields $\tau,c^a,t^a,b^a,\tilde c_a,\tilde c_0,\psi$ should satisfy the same
transformation rules \eqref{SL2Z} as at tree-level. But then, one faces the second shortcoming
of the  construction above, namely that of the lack of manifest S-duality invariance. This
problem was solved in the D1-D(-1) sector in \cite{Alexandrov:2009qq,Alexandrov:2012bu}
where both, a manifestly S-duality invariant twistor construction and the quantum corrected mirror map were found.
In this paper we address the more complicated case of D3-instantons and
uncover their S-duality invariant description in the linearized approximation.

%%%%%%%%%%%%%%%%%%%%%%%%%%%%%%%%%%%%%%%%%%%%%%%%%%%%%%%%%%%%%%%%%%%%
\section{Modularity of DT invariants for D3-instantons \label{sec_moddt}}
%%%%%%%%%%%%%%%%%%%%%%%%%%%%%%%%%%%%%%%%%%%%%%%%%%%%%%%%%%%%%%%%%%%%

In this section, we establish the modular properties satisfied by the
generalized DT invariants which count D3-instantons. The arguments are
very similar to those which allow to establish the modularity of the partition function of
D4-D2-D0 black holes in type IIA/$\CYm$, or equivalently of M5-branes in M-theory
on $\CYm$ \cite{Maldacena:1997de,Gaiotto:2006ns,Gaiotto:2006wm,deBoer:2006vg,Denef:2007vg}.
This is not a surprise, since T-duality relates  the HM moduli space in type IIB on
$\CYm\times S_1$ to the VM moduli space in type IIA on $\CYm\times S_1$, and maps
D3-brane instantons wrapping a divisor $\cD$ in type IIB to a D4-brane instanton
wrapping $\cD \times S^1$ in type IIA.
In \S \ref{sec_dtmod},  we recall the geometric representation of D3-instantons as coherent
sheaves on divisors, and recall the main properties of the generalized
DT invariants which count them. This subsection
may be skipped by the reader uninterested in mathematical aspects of D3-instantons.
In \S \ref{sec_msw}, we express the DT invariants in the large volume limit in terms
of the MSW invariants, which have simple modularity properties, and
define a novel multi-instanton expansion which is well suited
for modularity. In \S \ref{sec_twilin}, we
 give a heuristic description of D3-instanton corrections
in twistor space, formally summing up the transition functions while ignoring the
fact that each of them is associated to a different contour. A more rigorous treatment
of modularity in twistor space is deferred till \S\ref{sec_d3}.

%%%%%%%%%%%%%%%%%%%%%%%%%%%%%%%%%%
\subsection{D3-instantons, coherent sheaves and DT invariants\label{sec_dtmod}}
%%%%%%%%%%%%%%%%%%%%%%%%%%%%%%%%%%

First, we recall the description of D3-brane
instantons in type IIB string theory compactified on a Calabi-Yau
threefold $\CYm$ in the large volume limit.
Similarly to their D4-brane cousins \cite{Douglas:2000gi,Aspinwall:2004jr,
Dabholkar:2005dt,Denef:2007vg}, D3-instantons are described  by pure,
dimension-two, semi-stable sheaves $\cE$  supported on a divisor $\cD\subset \CYm$,
i.e. a complex hypersurface in $\CYm$ (more generally, objects in the derived category
of coherent sheaves \cite{Sharpe:1999qz,Douglas:2000gi}.).
Let the 4-cycles $\gamma_a$ (Poincar\'e dual to 2-forms $\omega_a$)
denote an integer basis of $ H_4(\CYm,\IZ)\equiv \Lambda$,
and the 2-cycles $\gamma^a$ (Poincar\'e dual to 4-forms $\omega^a$)
denote an integer basis of $H_2(\CYm,\IZ)\equiv \Lambda^*$, such that
\be
\omega_a \wedge\omega_b = \kappa_{abc}\, \omega^c\, ,
\qquad
\omega_a \wedge \omega^b = \delta_a^b \, \omega_\CYm\, ,
\qquad
 \int_{\gamma^a}\omega_b= \int_{\gamma_b}\omega^a= \delta^a_b\, ,
\ee
where $\omega_\CYm$ is the volume form, normalized to $\int_\CYm \omega_\CYm =1$,
and $ \kappa_{abc}=\int_\CYm \omega_a \omega_b \omega_c =\langle
\gamma_a, \gamma_b, \gamma_c\rangle$.
The homology class of the divisor $\cD$ may be expanded on the
basis of 4-cycles as $\cD= d^a \gamma_a$, or equivalently its Poincar\'e dual may be
expanded on a basis of 2-forms, $[\cD]=d^a \omega_a$.
We denote by $\bfd$ the vector $(d^1,\dots,d^{b_2})$. We assume that $\cD$ is an ample divisor, i.e.
that $[\cD]$ belongs to the \kahler cone,
\be
\bfd^3=\kappa_{abc} d^a d^b d^c >0,
\qquad
\bfr\cdot \bfd^2=\kappa_{abc} d^a d^b r^c >0,
\qquad
\bfq\cdot \bfd = q_a d^a >0,
\ee
for all effective divisors $r^a \gamma_a \in H_4(\CYm,\IZ)$ and effective curves $q_a \gamma^a \in H_2(\CYm,\IZ)$.
The intersection matrix of 2-cycles of an ample divisor $\cD$ provides a natural quadratic form $\kappa_{abc} d^c$ on
$\Lambda$, which has signature $(1, b_2-1)$. This also provides the quadratic form $\kappa^{ab}=(\kappa_{abc}d^c)^{-1}$ of $\Lambda^*$.
In the following, we shall use this quadratic form
to identify $\Lambda$ as a sublattice of $\Lambda^*$,
$k^a \mapsto k_a\equiv \kappa_{abc} p^b k^c$.

Given any coherent sheaf $\cE$ on $\CYm$, the D-brane charges are given by the
components of the generalized Mukai charge vector $\gamma'$ on a basis of $H^{\rm even}(\CYm,\IZ)$,
\be
\gamma'= \ch \cE \, \sqrt{\Td \CYm}
= p^0 + p^a \omega_a - q'_a \omega^a + q_0'\, \omega_{\CYm}\, ,
\ee
and satisfy the quantization conditions \eqref{fractionalshiftsD5}. D3-brane instantons
correspond to dimension-two sheaves with $p^0=0$. Such a sheaf
$\cE$ can be considered as the push forward $\cE=\iota_{*}(\cF)$ of a coherent
sheaf $\cF$ on $\cD$, where $\iota: \cD\to \CYm$ is the embedding of
the surface $\cD$ in the Calabi-Yau. The Chern character $\ch \cE$ is related to the Chern character
of the sheaf $\cF$ on $\CYm$ by the Grothendieck-Riemann-Roch
formula \cite{Minasian:1997mm,Douglas:2006jp}
\be
\label{grr}
\ch( \cE) = \iota_{*}\left( \ch (\cF)\, \Td (\cD)  \right) / \Td(\CYm)\, ,
\ee
where $\iota_{*}$ is the push-forward in cohomology, obtained by first Poincar\'e dualizing on
$\cD$, then pushing-forward in homology, then Poincar\'e dualizing on
$\CYm$ (thereby increasing the form degree by 2). The Chern character
$\ch (\cF)$ is given in terms of the rank $r$ of $\cF$ and Chern
classes $c_i(\cF)$ by: $\ch (\cF)=r+c_1(\cF)+\half c_1(\cF)^2-c_2(\cF)$.
We recall that the first Chern class of $\cF$ must  satisfy
$c_1(\cF)\in H^2(\cD,\IZ)+ \frac12 c_1(\cN)$, where $\cN$ is the normal bundle of
$\cD$ in $\CYm$.
The components of the charge vector $\gamma'$ are then
\cite{Dabholkar:2005dt,Diaconescu:2007bf}
\be
\begin{split}
p^0=&\,0,
\qquad
p^a=r\, d^a,
\qquad
q'_a=- \int_{\cD}\,  \iota^{*}(\omega_a) \Big(c_1(\cF)+\frac{r}{2}\, c_1(\cD)\Big) ,
\\
q'_0=&\,\int_{\cD}   \frac{1}{2r} \left( c_1(\cF)+\frac{r}{2} c_1(\cD)\right)^2
+ \frac{r}{24}\, \chi(\cD) - r \, \Delta(\cF),
\end{split}
\label{d5d3d1c}
\ee
where $\iota^{*}$ denotes the pull-back map in cohomology, and
$c_1(\cD)=-c_1(\cN)=-[\cD]$. The Euler number of $\cD$ is
\be
\chi(\cD)=[\cD]^3+c_2(\CYm)\cdot [\cD]=\bfd^3 + {\bfch}\cdot \bfd,
\ee
while $\Delta(\cF)$ is the Bogomolov discriminant, invariant under tensoring by line bundles,
\be
\label{defbog}
\Delta(\cF)=\frac{1}{r}\int_{\cD} \(c_2(\cF)-\frac{(r-1)}{2r}\, c_1^2(\cF)\).
\ee

Similarly to BPS D4-brane black holes, D3-instantons with the minimal number
of fermionic zero modes (namely, four) correspond to semi-stable sheaves. Recall
that a sheaf $\cF$ is semi-stable if all  subsheaves $\cF'\subset \cF$ satisfy
$\mu(\cF')\leq \mu(\cF)$. The (generalized)
slope\footnote{The relevant stability condition in string theory is $\Pi$-stability,
but this reduces in the large volume limit to (generalized) slope stability
(see \cite{Douglas:2000gi} for further details). }
for D3-instantons is defined by
\be
\label{eq:slope}
\mu(\cF(\gamma))=\frac{(\bfq'+\bfb)\cdot \bft}{\bfp \cdot \bft^2}\, .
\ee
In particular the Bogomolov discriminant \eqref{defbog} of a
semi-stable sheaf is positive, which can be stated physically as a bound
\be
\label{boundqhat0}
\hat q_0 \leq
\frac{r}{24}\, \chi(\cD)= \frac{1}{24} \left( \frac{1}{r^2}\, \bfp^3+
  \bfch\cdot \bfp \right) ,
\ee
on the `invariant D(-1) charge'
\be
\label{defqhat0}
\hat q_0 \equiv
q_0' -\frac12\, \bfq'^2
=\frac{r}{24}\, \chi(\cD) - r \, \Delta(\cF)\, ,
\ee
where $\bfq'^2=\kappa^{ab}q_a'q_b'$. A useful property of $\hat q_0$ is
that it is, like the Bogomolov discriminant, invariant under tensoring of $\cF$ by a
line bundle $\cL$ on $\cD$. If $c_1(\cL)=-\bfeps\in\Lambda$,
the corresponding change of the Chern character $\mathrm{ch(\cE)}$ is the
`spectral flow'
\be
\label{flow}
\bfp \mapsto \bfp ,
\qquad
\bfq'\mapsto \bfq' - \bfeps ,
\qquad
q'_0\mapsto q'_0- \bfeps\cdot \bfq' + \frac12\, \bfp \cdot \bfeps^2 .
\ee
Note that the residue class of $\bfq'-\tfrac12 \bfp \in\Lambda^*$ modulo $\Lambda$ is unaffected by this flow.
This operation however does not leave
invariant the stability condition for sheaves $\cE$ on $\CYm$,
unless one also changes the moduli $\bfz$ accordingly: $\bfz \mapsto
\bfz+\bfeps$ \cite{Manschot:2009ia}.

The moduli space of semi-stable dimension-two sheaves $\cE$ on $\CYm$ is a
complex space of finite dimension. The generalized Donaldson-Thomas invariant
$\Omega(0,\bfp,\bfq',q'_0; \bfz)$ is an integer given, informally, by the Euler characteristics
of this moduli space. It is also useful to consider the rational DT invariant \cite{ks,Joyce:2008pc,Manschot:2010qz},
\be
\label{defntilde}
\bar\Omega(\gamma; \bfz) = \sum_{d|\gamma}  \frac{1}{d^2}\,
\Omega(\gamma/d; \bfz)\, ,
\ee
which coincides with the integer $\Omega(\gamma; \bfz)$ whenever $\gamma$ is a primitive
vector, but in general is a rational number. Both $\Omega$ and $\bar\Omega$ are
piecewise constant as a function of the K\"ahler moduli $\bfz$, but are discontinuous across walls of
marginal stability where the sheaf becomes unstable. This moduli dependence persists
even in the large volume limit, with the exception of CY threefolds
with $b_2=1$, where the invariants $\Omega(\bfp,\bfq',q'_0;\bfz)$ are
independent of $\bfz$ at large volume.

%%%%%%%%%%%%%%%%%%%%%%%%%%%%%%%%%%
\subsection{MSW invariants and modularity \label{sec_msw}}
%%%%%%%%%%%%%%%%%%%%%%%%%%%%%%%%%%

The moduli dependence of the DT invariants makes it difficult to state their
modular properties directly.
Fortunately, it is possible to express the DT invariants  in the large volume limit
in terms of a different set of invariants $\bar\Omega_{\bfp}(\bfq,q_0')$, which we call the MSW
invariants. They  are moduli independent and given by Fourier coefficients of a certain
Jacobi form, namely the elliptic genus of the MSW superconformal field theory
which describes D4-D2-D0 black holes \cite{Maldacena:1997de,Gaiotto:2006ns,Gaiotto:2006wm,deBoer:2006vg,Denef:2007vg},
\begin{eqnarray}
\label{eq:ellgenus}
\mathcal{Z}_{\bfp}(\tau,\bfy) &=& \Tr' (2J_3)^2 (-1)^{2J_3}\,
\expe{\left( L_0 - \frac{c_L}{24}\right)\tau -
\left( \bar L_0 - \frac{c_R}{24}\right)\bar\tau + \bfq'\cdot \bfy} \nn \\
&=& \sum_{\bfq', q'_0} \, (-1)^{\bfp\cdot \bfq'}\, \bar\Omega_\bfp(\bfq',q'_0)\,
\expe{-\hat q_0 \tau  - {1\over 2} \,\bfq'^2_- \tau - {1\over 2}\,\bfq'^2_+ \bar \tau
  +\bfq'\cdot \bfy},
\end{eqnarray}
Here, the summation runs over
\be
\bfq'  \in \IZ^{b_2} + \tfrac12\, \bfp,
\qquad
q'_0 \in \IZ-\tfrac{1}{24}\, {\bfch}\cdot \bfp,
\ee
where $(\bfp)_a\equiv \kappa_{abc} p^b p^c$, subject to the
bound \eqref{boundqhat0}. For a general vector
$\bfk\in\Lambda$,
the vectors $\bfk_\pm\in \Lambda\otimes \mathbb{R}$ are projections of $\bfk$ onto the positive
and negative definite subspaces of $\Lambda\otimes \mathbb{R}$ defined by the magnetic
charge vector $\bfp$ and the \kahler moduli\footnote{This result is usually stated for
$\bft\propto \bfp$, but the SCFT has a moduli space of marginal deformations which allows
to go away from this point \cite{deBoer:2008ss}.}
$\bft$,
\be
\label{defkpm1}
\bfk_+=\frac{\bfk\cdot \bft }{\bfp \cdot \bft^2}\, \bft,
\qquad
\bfk_-=\bfk-\bfk_+,
\qquad
\bfk^2 = \bfk_+^2+\bfk_-^2,
\ee
which satisfy
\be
\bfk_+^2>0,\qquad \bfk_-^2<0,
\qquad \forall\, \bfk \neq \bf 0.
\ee
We also use the notation $k_+=\bfk\cdot \bft /\sqrt{\bfp \cdot \bft^2}$, so that
$|k_+|=\sqrt{\bfk_+\cdot \bfk_+}$.
It follows from general properties of the CFT that the elliptic genus  \eqref{eq:ellgenus}
is a Jacobi form of
weight $(-\tfrac32,\frac12)$ under $SL(2,\IZ)$,
with multiplier system $M_{\cZ}=\expe{\varepsilon(\trans)\,{\bfch}\cdot \bfp}$,
where $\varepsilon(\trans)$ is the multiplier system \eqref{multeta}
of the Dedekind eta function.

The MSW invariants $\bar\Omega_\bfp(\bfq',q'_0)$ coincide with the generalized DT invariants
at the so-called `large volume attractor point' \cite{deBoer:2008fk}
\be
\bar\Omega_\bfp(\bfq',q'_0) = \bar\Omega\(0,\bfp,\bfq',q'_0; \bfz_{\infty}(\gamma) \),
\qquad
\bfz_{\infty}(\gamma)=
\lim_{\lambda\to +\infty}\(\bfb_{}(\gamma)+\I\lambda \, \bft_{}(\gamma)\),
\ee
where $\bfz(\gamma)=\bfb(\gamma)+\I \bft(\gamma)$ is the
standard attractor point. In particular, $\bar\Omega_\bfp(\bfq',q'_0)$ is invariant under tensoring
by a line bundle \eqref{flow}, which corresponds to the spectral flow symmetry in the SCFT.
Decomposing
\be
\bfq' = \bfmu + \bfeps +\tfrac12\,\bfp ,
\ee
where  $\bfmu \in \Lambda^*/\Lambda$ is the residue
class\footnote{The quotient $\Lambda^*/\Lambda$  is
a finite group of order $|\det\kappa_{ab}|$, sometimes known as the discriminant group.}
of $\bfq'-\tfrac12\bfp$ modulo $ \bfeps$, it follows that the MSW invariant
$\bar \Omega_\bfp(\bfq',q'_0)\equiv  \bar\Omega_{\bfp,\bfmu}(\hat q_0)$
depends only on $\bfp$, $\bfmu$ and $\hat q_0$. As a result,
the elliptic genus has a theta function decomposition:
\be
\label{thetaeg}
\mathcal{Z}_\bfp (\tau,\bfy,\bft) = \sum_{\bfmu \in \Lambda^*/\Lambda}
h_{\bfp,\bfmu} (\tau)\, \overline{\theta_{\bfp,\bfmu} (\tau,\bfy,\bft)}\, ,
\ee
where $\theta_{\bfp,\bfmu}$
is the Siegel-Narain theta series\footnote{Note that  \eqref{defth} coincides
with the theta series \eqref{thSieg} with $\bfy_+=\bfb_+ \tau-\bfc_+$, $\bfy_-=\bfb_- \bar\tau-\bfc_-$,
up to a phase $\expe{\frac12 (\bfb_+^2 \tau+ \bfb_-^2 \bar\tau - \bfb\cdot \bfc )}$.\label{foophase}}
 associated to the lattice $\Lambda$ equipped
with the  quadratic form $\kappa_{ab}$ of signature $(1,b_2-1)$,
\be
\label{defth}
\theta_{\bfp,\bfmu} (\tau,\bfy,\bft) =
\sum_{\bfk\in \Lambda+\bfmu+ \tfrac12 \bfp}
 \signkp\,\expe{ \frac12\, \bfk_+^2 \tau+\frac12\, \bfk_-^2 \bar\tau
+ \bfk\cdot \bfy},
\ee
and the $\bfy$-independent coefficients are given by
\be
\label{defchimu}
h_{\bfp,\bfmu}(\tau) = \sum_{\hat q_0 \leq r \chi(\cD)/24}
\bar\Omega_{\bfp,\bfmu}(\hat q_0)\,
\expe{-\hat q_0 \tau }.
\ee
Since the theta series \eqref{defth} is a vector valued Jacobi form of
modular weight $(\tfrac12,\tfrac{b_2-1}{2})$ and multiplier system
$M_\theta$ under $SL(2,\IZ)$, it follows that the function $h_{\bfp,\bfmu}$
must transform as a vector-valued
holomorphic\footnote{\label{foot:mock}Non-compact directions in the target space of the
CFT could potentially lead to mock modular forms and thus holomorphic anomalies \cite{Troost:2010ud}. For
local Calabi-Yau manifolds, e.g. $\cO(-K_{\mathbb{P}^2})\to
\mathbb{P}^2$, it is known that the holomorphic generating series of DT-invariants
for sheaves with rank $>1$ requires a non-holomorphic addition in order to transform as a
modular form \cite{Vafa:1994tf}. We assume that this issue does not arise here.}
modular form of negative weight $(-\frac{b_2}{2}-1,0)$
and  multiplier system $M(\trans)=M_{\cZ}\times \overline{M_\theta}^{-1}$
under the full modular group $SL(2,\IZ)$. This is equivalent to
$M(\trans)=M_{\cZ}\times M_\theta$ since $M_\theta$ is unitary.
The Fourier coefficients of $h_{\bfp,\bfmu}$ with $\hat q_0>0$ are known as the polar degeneracies \cite{Denef:2007vg},
and determine the remaining MSW invariants by the usual Rademacher
representation.\footnote{The most polar term $q^{-r\chi(\CD)/24}$, arises in
the CFT from the fact that $L_0\geq 0$ and that $c_L=r\chi(\cD)$.
Application of the Cardy formula shows that the asymptotic
growth of states is dominated by $r=1$ for given magnetic charge
$\bfp$, therefore the contribution of non-Abelian D3-instantons
is exponentially suppressed compared to Abelian instantons.}

Away from the large volume attractor point $\bfz_{\infty}(\gamma)$ (but still at large
volume), the DT invariant $\bar\Omega(0,\bfp,\bfq',q'_0; \bfz)$ differs from the MSW
invariant $\bar\Omega_{\bfp,\bfmu}(\hat q_0)$ by terms of higher order in the MSW invariants
 \cite{Manschot:2009ia, Manschot:2010xp}:
\be
\begin{split}
\bar \Omega(0,\bfp,\bfq',q'_0;\bfz)=&\,\bar\Omega_{\bfp,\bfmu}( \hat q_0)
\\
&\, +\sum_{\gamma_1+\gamma_2=(0,\bfp,\bfq',q_{0}')\atop\bfp_i> {\bf 0}}
\frac{1}{4}\left(\sgn\left(\,\tfrac{(\bfq_1'+\bfb)\cdot\bft}{\bfp_1\cdot \bft^2}
-\tfrac{(\bfq_2'+\bfb)\cdot\bft}{\bfp_2\cdot \bft^2}\,\right)
-\sgn(\left<\gamma_1,\gamma_2 \right>)\right)
      \\
& \quad \times\, \left<\gamma_1,\gamma_2 \right> (-1)^{\left<\gamma_1,\gamma_2 \right>}\,
\bar\Omega_{\bfp_1,\bfmu_1}(\hat q_{0,1})\,\bar\Omega_{{\bfp_2},\bfmu_2}(\hat q_{0,2})
\\
&\,+\dots
\end{split}
\label{multiMSW}
\ee
where $\gamma_i=(0,\bfp_i,\bfq'_i,q_{0,i}')$, and therefore
$\left<\gamma_1,\gamma_2\right>=\bfp_2\cdot \bfq'_1-\bfp_1\cdot \bfq'_2$.
Moreover one recognizes on the second line the difference of
the slopes (\ref{eq:slope}) of the constituents. The higher order terms
can be thought of as describing bound states of the MSW constituents,
which exist away from the large volume attractor point $\bfz_{\infty}(\gamma)$.
This decomposition is analogous to the
decomposition of the index in terms of the multi-centered black hole bound states, although
the elementary MSW constituents may themselves arise
as bound states of more basic constituents such as D6-anti D6 bound
states \cite{Denef:2007vg}, which may decay across walls in the interior of the K\"ahler cone.
The decomposition \eqref{multiMSW} hinges on the fact
that walls for constituents with $\bfp_1=0$, $\bfp_2\neq 0$ lie on the
boundary or outside the K\"ahler cone, such that the only relevant
walls are those between D3-instantons, i.e. constituents with charges $\bfp_i>0$ \cite{Manschot:2009ia}.
As shown in \cite{Manschot:2009ia}, the `two-centered' contribution in \eqref{multiMSW}
(after smoothing out the sign function into a error function as in Eq. \eqref{signtoErf})
leads to a modular invariant partition function with the same modular properties as the
elliptic genus, and it is expected that modularity persists to all orders. It is worthwhile
to note that the expansion \eqref{multiMSW} has a finite number of terms in any chamber
related by a finite number of wall-crossings from the large volume attractor chamber.

Having expressed the generalized DT invariants in terms of the moduli-independent
MSW invariants, which have simple modular properties, it is thus natural to reorganize
the multi-instanton series expansion of the integral equations \eqref{eqTBA} by first
expressing $\Omega(\gamma;\bfz)$ in terms of  $\bar\Omega_{\bfp,\bfmu}(\hat q_0)$
using \eqref{multiMSW},
and then expanding in powers of the MSW invariants $\bar\Omega_{\bfp,\bfmu}$. In terms
of the usual multi-instanton expansion of  \eqref{eqTBA}, the
$n$-th order in this reorganized multi-instanton expansion will contain the usual
$n$-instanton contribution weighted by products of MSW degeneracies
$\bar\Omega_{\bfp,\bfmu}$,
together with terms with $n'<n$ instantons weighted by terms of order $n-n'$ in the
expansion \eqref{multiMSW}.
In the remainder of this work, we shall be concerned with only the first term in this expansion, which we
dub the `one-instanton approximation'.

%%%%%%%%%%%%%%%%%%%%%%%%%%%%%%%%%%
\subsection{One-instanton approximation: heuristics \label{sec_twilin}}
%%%%%%%%%%%%%%%%%%%%%%%%%%%%%%%%%%

In the one-instanton approximation introduced in the previous subsection, quantum corrections
to the perturbative metric are effectively encoded by the set of holomorphic
functions\footnote{Indeed, the multi-covering contributions induced by the dilogarithm function in
\eqref{prepH} are described just as well by replacing
${\rm Li_2}(\sigma_{\gamma}\cX_{\gamma})\mapsto \sigma_{\gamma} \cX_{\gamma}$
and using the  rational DT invariants \eqref{defntilde}
in place of the integer DT invariants.
Due to \eqref{multiMSW}, the rational DT invariants can then be replaced by $\bar\Omega_{\bfp,\bfmu}$.}
\be
H_\gamma(\xi^0,\bfxi,\bftxi)= \frac{1}{(2\pi)^2}\,\bar\Omega_{\bfp,\bfmu}(\hat q_0)\,\sigma_\gamma\cX_\gamma\, ,
\label{prepHnew}
\ee
which generate the infinitesimal complex contact
 transformation across the BPS ray $\ell_\gamma$.
The set of all $(\ell_\gamma,H_{\gamma})$ can be viewed as a Cech representative
of $H^1(\cZ,\cO(2))$, which is known to classify the linear deformations of QK manifolds.
In this subsection, we take a heuristic approach and consider the formal
sum $H=\sum_\gamma H_\gamma$, ignoring the important fact
that the functions $H_{\gamma}$ are attached to different contours. This will allow
us to highlight the main mechanism which ensures invariance under S-duality, deferring
a rigorous treatment to \S\ref{sec_lv} onward.

Ignoring the contributions of  D5 and NS5-instantons, which are exponentially suppressed
at large volume with respect to D3-D1-D(-1)-instantons, and restricting to a
fixed  D3-brane charge $\bfp$, the D3-D1-D(-1) one-instanton corrections are thus described
by the formal sum\footnote{The symbol $\sum\limits^!$ indicates a divergent sum.}
\be
\label{eqHp}
\Hp= \frac{1}{(2\pi)^2}\sum_{\bfq', q'_0}^! \, (-1)^{\bfp\cdot \bfq'}\,
\bar\Omega_{\bfp,\bfmu}(\hat q_0)\,
\expe{ \bfp\cdot\bftxip - \bfq' \cdot \bfxi - q'_0 \xi^0},
\ee
where $\bar\Omega_{\bfp,\bfmu}(\hat q_0)$ are the MSW invariants introduced in the previous
subsection. In writing \eqref{eqHp}, we identified the quadratic refinement $\sigma_\gamma$
with the phase $(-1)^{\bfp\cdot\bfq'}$ by setting the characteristics $\theta^\Lambda,\phi_\Lambda$ to zero.
We also omit the phase $\expe{\tfrac12
A_{\Lambda\Sigma}p^\Lambda p^\Sigma}$ which is
unimportant for fixed magnetic charges $p^\Lambda$.
The assumption $\phi_\Lambda=0$ is  for simplicity, since $\phi_\Lambda$ couples only
to magnetic charges  which are kept constant in our set-up. In contrast, the condition $\theta^\Lambda=0$
is necessary for modular invariance, which requires the phase factor
$(-1)^{\bfp\cdot \bfq'}$ in the sum \eqref{eqHp}. The vanishing of $\theta^\Lambda$
was also seen to be necessary in the linear analysis of NS5-instanton corrections
in \cite{Alexandrov:2010ca}.

Using the same steps as before, we can therefore
trade the sum over $\bfq',q'_0$ in \eqref{eqHp} for a sum over $\bfmu,\hat q_0$ and
$\bfk\equiv \boldsymbol{\epsilon}+\bfmu+ \tfrac12\, \bfp$, and represent \eqref{eqHp} as
a sum of theta series analogous to \eqref{thetaeg},
\be
\label{Hdec}
\Hp = \frac{1}{(2\pi)^2}\,
\expe{ \bfp\cdot \bftxip }
\sum_{\bfmu\in \Lambda^*/\Lambda}
h_{\bfp,\bfmu} (\xi^0)\, \ZTheta_{\bfp,\bfmu} (\xi^0,\bfxi)\, ,
\ee
where
\be
\label{defTheta}
\ZTheta_{\bfp,\bfmu} (\xi^0,\bfxi) =
\sum_{\bfk \in\Lambda + \bfmu+ \tfrac12 \bfp}^!
\expe{-\frac12\, \bfk\cdot \bfp^2
-\bfk\cdot  \bfxi- \frac12\, \bfk^2 \xi^0}
\ee
and $h_{\bfp,\bfmu}(\xi^0)$ is the same modular function which was defined in \eqref{defchimu},
evaluated at $\tau=\xi^0$. It therefore transforms as a  vector-valued holomorphic modular form
of modular weight $-\tfrac{b_2}{2}-1$ and multiplier system
$M(\trans)=M_{\cZ}\times M_\theta$ under the  S-duality action
\eqref{SL2Zxi}. On the other hand, $\ZTheta_{\bfp,\bfmu} (\xi^0,\bfxi)$
is formally a holomorphic theta series for
the lattice $\Lambda$ of rank $b_2$ with quadratic form $-\kappa_{ab}$,
with modular parameter $\tau=\xi^0$ and elliptic parameters $\xi^a$.
Thus, under the action \eqref{SL2Zxi},
it is expected to transform as a holomorphic Jacobi form of weight
$\tfrac{b_2}{2}$, multiplier system $M_\theta^{-1}$ and index $m_{ab}=-\frac12 \kappa_{ab}$.
Finally, thanks to the term proportional to $c_{2,a}$ in the transformation of $\bftxip$,
the exponential prefactor in \eqref{Hdec} transforms as the automorphy
factor of a multi-variable holomorphic Jacobi theta series with the index
$m_{ab}=\frac12 \kappa_{ab}$ and multiplier system $M_{\cZ}^{-1}$. We thus
conclude that, under the action of S-duality in twistor space, the formal sum \eqref{eqHp} transforms as a
holomorphic Jacobi form of weight $-1$ and trivial multiplier system.
In \cite{Alexandrov:2012bu} it was shown that this is precisely the condition ensuring
the presence of the $SL(2,\IZ)$ isometry group. Thus, $\Hp$
is expected to generate an S-duality invariant deformation of the perturbative HM metric.

However, this analysis overlooks the important fact that the quadratic form $\kappa_{ab}$ has
indefinite signature $(1,b_2-1)$, and therefore the theta series \eqref{defTheta}
is formally divergent. Fortunately, the theta series \eqref{defTheta} never arises
as such in the computation of the metric, rather each of the terms in \eqref{defTheta} must
be integrated along a different contour, which renders the resulting series convergent.
Another fortunate circumstance is that there does exist a natural holomorphic `mock' theta
series of signature $(1,b_2-1)$, the convergence of which is ensured by
restricting the sum to lattice vectors lying in a certain cone where the quadratic form
is definite positive \cite{Zwegers-thesis}. While each of these two series  transform with
modular anomalies, their sum is in fact modular invariant, ensuring the modular
invariance  of the D3-instanton corrected metric, as the next section will
aim to demonstrate.

%%%%%%%%%%%%%%%%%%%%%%%%%%%%%%%%%%%%%%%%%%%%%%%%%%%%%%%%%%%%%%%%%%%%
\section{D3-instantons, period integrals and mock theta series\label{sec_d3}}
%%%%%%%%%%%%%%%%%%%%%%%%%%%%%%%%%%%%%%%%%%%%%%%%%%%%%%%%%%%%%%%%%%%%

In this section, we discuss how S-duality is realized in the presence of D3-D1-D(-1)
instantons, but in the absence of D5 and NS5-instantons. This
situation arises in the large volume limit, where D5 and NS5-brane instantons
are suppressed compared to D3-D1-D(-1) and worldsheet
instantons. In \S \ref{sec_lv} we
introduce a large volume limit which retains modular invariance but reduces the
D3-instanton sums to (generalized) Gaussian theta series, and  obtain
the instanton corrections to the Darboux coordinates and contact potential in this limit.
We establish their modularity (or lack thereof) in \S \ref{sec_modphi}, \S\ref{sec_dperiod}.
Finally, in \S \ref{sec_mock} we show how modular anomalies can be canceled
by a contact transformation generated by a mock theta series.

%%%%%%%%%%%%%%%%%%%%%%%%%%%%
\subsection{Instanton corrections in the large volume limit \label{sec_lv}}
%%%%%%%%%%%%%%%%%%%%%%%%%%%%

In order to analyze the action of S-duality on the D-instanton corrected twistor space,
it is useful to focus on the  Darboux coordinates around the point $t=-\I$, or equivalently
$z=0$, which  is invariant under the S-duality action \eqref{ztrans} on the $\CP$
fiber. In the large volume limit $t^a\to\infty$,
it may be checked that the saddle points \eqref{gensaddle} dominating
the integral equations \eqref{eqTBA},
for D3-D1-D(-1) charges $\gamma=(0,\bfp,\bfq,q_0)$ such that $\bfp\cdot \bft^2>0$,
accumulate near the same point
\be
t'_{\gamma}\approx -\I - 2 \frac{\kbp}{\sqrt{\bfp\cdot \bft^2}} + \dots\, ,
\qquad
z'_{\gamma}\approx -\I \frac{\kbp}{\sqrt{\bfp\cdot \bft^2}} +\dots\, .
\label{saddlep}
\ee
Thus, it is natural to expand the Darboux coordinates around the point $z=0$, $t^a\to\infty$,
keeping the product $z t^a$ fixed. Retaining all the terms from the perturbative result \eqref{gentwi}
which are finite or divergent in this limit,
we find
\bea
\label{darblv}
\xi^0 &=& \tau ,
\qquad\nn\\
\xi^a &=& \tau b^a-c^a +2\tau_2 t^a z + \delta\xi^a ,
\nn\\
\txi'_a &=&
\frac{\I}{2} \tau_2 \kappa_{abc} t^b t^c +
\cla+\frac12 \kappa_{abc} b^b(c^c-\tau b^c)
-2\tau_2\,\kappa_{abc} t^b b^c\,z +\I\tau_2\,\kappa_{abc} t^b t^c\, z^2
 + \delta\txi'_a,
\nn\\
\txi'_0
&=&\frac16 \kappa_{abc} \tau_2 t^a t^b  (2 t^c z+3 b^c ) +
 \cl0-\frac16 \kappa_{abc} b^a b^b (c^c-\tau b^c)
+ \tau_2  \, \kappa_{abc} t^a b^b (b^c\, z - \I t^c z^2)
+\delta\txi'_0,
\nn\\
 \alpha'
&=&-\frac16 \kappa_{abc} \tau_2 t^a t^b  (2 t^c z \bar\tau +3 c^c)+
 \psi+\frac16 \kappa_{abc} b^a(b^b\tau-c^b)(b^c\tau-2c^c)
 \\
 &&
+\tau_2 z \kappa_{abc} t^a \left( b^b (b^c\tau-2 c^c)  + t^b(2 z b^c \tau_2 + \I z^2 c^c) \right)
+\delta\alpha' ,
\nn
\eea
where $\delta\xi^a$, etc, denote the instanton corrections which include two types of contributions:
from the integral terms in \eqref{eqTBA} and from instanton corrections to the classical mirror map \eqref{symptobd}.
Note that the former contributions can be split into two parts.
The first corresponds to D-branes with charge $\gamma$ with $\bfp \cdot \bft^2>0$, while
the second comes from anti D-branes with opposite
charge $-\gamma$. In the former case, the integrals are dominated  by a saddle point at
$t'_\gamma$, which approaches $t'=-\I$ in the large volume limit, while in the
latter case, they are dominated by a saddle point at the antipodal
point $t'_{-\gamma}=-1/\bar t'_\gamma$, which approaches $t'=\I$ or $z'=\infty$ in the large volume limit.
This suggests to keep fixed inside these integrals a different combination, $t^a/z'$.

Besides the Darboux coordinates, it is important to consider also the contact potential $e^\Phi$.
In our approximation, it is dominated by the classical term in \eqref{phipertB}, so we can write it as
\be
\label{philv}
e^\Phi = \frac12 \tau_2^2 \, \cV+ \delta e^\Phi.
\ee

Our task will be to compute the corrections $\delta\xi^a$, etc. in the one-instanton
and large volume approximation, and to show that the resulting
instanton corrected Darboux coordinates \eqref{darblv}
and contact potential \eqref{philv} continue to transform in the same way \eqref{SL2Zxi}, \eqref{SL2phi}
as their perturbative counterpart, possibly up to a contact transformation.
For what concerns the coordinates $\xi^a$ and the contact potential $e^\Phi$, this requires showing
\be
\label{delphiSL2}
\delta\xi^a \mapsto \frac{\delta\xi^a}{c\tau+d},
\qquad
\delta e^\Phi \mapsto \frac{\e^\Phi}{|c\tau+d|}.
\ee
For what concerns the other coordinates $\txi'_a, \txi'_0,\alpha'$,
the corresponding constraints are best expressed by forming suitable
linear combinations\footnote{These linear combinations are equal to
the large volume limit of the linear combinations in Eq. (B.6) of  \cite{Alexandrov:2012bu}, up
to a divergent term which also transforms as in \eqref{hatdelSL2} and can therefore be
dropped.}
\be
\label{defhctJa}
\begin{split}
\hat\delta\txi'_a=&\,\delta\txi'_a+\kappa_{abc} \(b^b-\I zt^b\)\delta\xi^a,
\\
\hat\delta_+ \alpha'=&\,
\delta\alpha' + \tau\delta\txi'_0
+\kappa_{abc} (b^a-\I z t^a )(c^b-\tau b^b-2\tau_2 z t^a) \delta\xi^c,
\\
\hat\delta_-\alpha'=&\,
\delta\alpha'+ \bar\tau\delta\txi'_0 +
\kappa_{abc} (b^a-\I z t^a )(c^b-\tau_1 b^b-\tau_2 z t^a) \delta\xi^c.
\end{split}
\ee
These linear combinations are designed to transform canonically under S-duality as
\be
\label{hatdelSL2}
\hat\delta\txi'_a \mapsto \hat\delta\txi'_a,
\qquad
\hat\delta_+ \alpha'\mapsto \frac{\hat\delta_+\alpha}{c\tau+d},
\qquad
\hat\delta_- \alpha' \mapsto \frac{\hat\delta_-\alpha'}{c\bar\tau+d}
\ee
if and only if the corrected Darboux coordinates \eqref{darblv}  transform according
to \eqref{SL2Zxi}.

As was mentioned above, the instanton corrections acquire two types of contributions.
The first is computed by replacing  $\cX_\gamma$ by  $\cX^{\rm sf}_\gamma$ on the r.h.s. of the integral equations
\eqref{eqTBA} and \eqref{phiinstmany}, taking the large volume limit, and
evaluating the integrals in
terms of the type IIA coordinates $\tau_2,z^a,\zeta^\Lambda,\tzeta_\Lambda,\sigma$.
The latter can then be replaced by the type IIB coordinates $\tau,t^a,b^a,c^a,\cla,\cl0,\psi$,
adapted for S-duality, using the classical mirror map \eqref{symptobd}.
However, there are further contributions
stemming from instanton corrections to this mirror map and coming from the tree level terms on the type IIA side.
In Appendix \ref{sec_inslve} we extend the procedure of \cite{Alexandrov:2012bu} and compute these
corrections to the mirror map in our approximation. The result is presented explicitly in \eqref{inst-mp}.
In the same appendix it is shown that, adding up the two types of contributions, the instanton corrections to
the Darboux coordinates and the contact potential can be written as
\bea
\label{delxi}
\delta\bfxi &=&\,
2\pi\I\, \bfp\, \cJ_\bfp,
\nn\\
\hat\delta\bftxip &=&- \bfD \cdot \cJ_\bfp,
\vphantom{\frac{A}{A}}\nn\\
\hat\delta_+\alpha'
&=&
\[\frac{\tau_2}{\pi} \bfD \cdot \bfDb  -(b_2+1)\] \cdot \cJ_\bfp,
\label{cov-DC}
\\
\hat\delta_-\alpha'
&=&-  \left[4\pi \tau_2 \cD_{-1}+(\bfc-\bar\tau \bfb)\cdot \bfD \right]\cdot \cJ_\bfp
+\I \tau_2 z(p\cdot t^2) \sum_{q_\Lambda}\int_{\ell_{\gammam}}
\frac {\de z'}{(z')^2}\,H_{\gammam},
\nn\\
\delta e^\Phi& =&\,
-\frac{\tau_2}{4}\sum_{q_\Lambda}\int_{\ell_{\gammap}} {\de z'}\[
\hat q_0+\hf\,
(\bfk+\bfb-\I z' \bft)\cdot(\bfk+\bfb-3\I z' \bft)
\] H_{\gammap}
+{\rm c.c.}
\nn
\eea
where
\be
\cJ_\bfp(z)=\sum_{q_\Lambda} \int_{\ell_{\gammap}} \frac{\de z'}{2\pi\I(z'-z)}\,H_{\gammap} ,
\label{defcJ}
\ee
and $\cD_{-1},\bfD,\bfDb$ are the covariant derivative operators\footnote{Acting on
theta functions of weight $(\wh,\bwh)$, the operators
$\cD_{\wh}, \bfD, \bfDb$ raise the modular weight by (2,0), (1,0), (0,1), respectively. To
avoid cluttering, we abuse notation and declare that  $\cD_{\wh}$ annihilates the classical,
modular invariant contribution $S_{\rm cl}$ in
$H_{\gamma}$.\label{fooD}}
\be
\label{defDw}
\cD_{\wh} = \frac{1}{2\pi \I}\(\partial_\tau+\frac{\wh}{2\I\tau_2}\),
\qquad
\bfD =  -\frac{\I}{2\tau_2} \left[
\pa_{\bfb} + \bar\tau \pa_{\bfc} - \I\pi (\bfc - \bar\tau \bfb) \right],
\qquad
\bfDb = (\bfD)^\star.
\ee
Moreover,
$H_{\pm\gamma}$ is the large volume limit of \eqref{prepHnew} where one keeps fixed $z^{\pm 1} t^a$, respectively.
Explicitly, these two functions are given by
\be
\begin{split}
H_{\gammap} =& \frac{(-1)^{\bfp\cdot\bfk}}{(2\pi)^2}\, \bar\Omega_{\bfp,\bfmu}( \hat q_0) \,
\expe{\I S_{\rm cl} -\frac12  (\bfk+ \bfb)_+^2 \,
\bar\tau - \(\hat q_0+ \frac12 (\bfk+ \bfb)_-^2\)\tau+\bfc \cdot(\bfk +\frac12 \bfb)
+\I Q_{\gammap}(z')},
\\
H_{-\gammap} =& \frac{(-1)^{\bfp\cdot\bfk}}{(2\pi)^2}\, \bar\Omega_{\bfp,\bfmu}( \hat q_0)\,
\expe{\I \bar S_{\rm cl} +\frac12  (\bfk+ \bfb)_+^2 \,\tau
+ \(\hat q_0+ \frac12 (\bfk+ \bfb)_-^2\)\bar\tau-\bfc \cdot(\bfk +\frac12 \bfb)
+\I Q_{-\gammap}(z')},
\end{split}
\ee
where $S_{\rm cl}$ is the leading part of the Euclidean D3-brane action
in the large volume limit, and  $Q_{\pm\gamma}(z)$ is the only part which depends on $z$,
\be
S_{\rm cl} = \frac{\tau_2}{2}\, \bfp\cdot \bft ^2 - \I\, \bfcl\cdot \bfp ,
\qquad
Q_{\pm\gammap}(z')=\tau_2 \, \bfp\cdot \bft ^2\,
\left((z')^{\pm 1} \pm \I\, \frac{ \kbp}{\sqrt{\bfp\cdot \bft ^2}}\right)^2.
\label{clactinst}
\ee
The notation $\int_{\ell_{\gammap}}$ signifies that each term
must be integrated on a contour which extends
from $z'=-\infty$ to $z'=+\infty$ and goes through the saddle point $z'_{\gamma}=-\I\kbp/\sqrt{\bfp\cdot \bft ^2}$.
This corresponds to the BPS ray $\ell_{\gammap}$ in the large volume limit.

As we shall see momentarily, the corrections \eqref{delxi} do not quite transform
as required in \eqref{hatdelSL2}, but the modular anomaly can be absorbed by a
contact transformation. This issue however does not arise in the case of the contact
potential, which we consider first.

%%%%%%%%%%%%%%%%%%%%%%%%%%%%%%%%%%
\subsection{Modular invariance of the contact potential \label{sec_modphi}}
%%%%%%%%%%%%%%%%%%%%%%%%%%%%%%%%%%

The one-instanton correction to the contact potential in \eqref{delxi} involves a Gaussian integral
over $z'$ (renamed as $z$) which is easily calculated. Rewriting the integrand
\be
\label{truephi}
\begin{split}
\delta e^\Phi
=&-\frac{\tau_2}{4}\sum_{q_\Lambda}\int_{\ell_{\gammap}} {\de z}\[
\hat q_0+\frac12 (\bfk+\bfb)_-^2 -\frac{3}{8\pi\tau_2} + \frac{1}{4\pi\tau_2} \pa_{z}
\left( z - \frac{1}{8\pi\tau_2 \, \bfp\cdot \bft^2} \pa_z \right)
\] H_{\gammap}+{\rm c.c.}
\end{split}
\ee
and dropping the total derivative, we arrive at
\be
\label{truephi2}
\begin{split}
\delta e^\Phi
=&-\frac{\tau_2}{16\pi^2\sqrt{2\tau_2\, \bfp\cdot \bft^2}}
\sum_{q_\Lambda} \signkp\, \bar\Omega_{\bfp,\bfmu} \,
\[\hat q_0+\frac12 (\bfk+\bfb)_-^2 -\frac{3}{8\pi\tau_2} \]
\\
 & \times\,
\expe{\I S_{\rm cl} -\frac12  (\bfk+ \bfb)_+^2 \, \bar\tau
- \(\hat q_0+ \frac12 (\bfk+ \bfb)_-^2\)\,\tau+\bfc\cdot (\bfk +\frac12 \bfb)}+{\rm c.c.}
\end{split}
\ee
The sum over $q_\Lambda$ is recognized as the partition function \eqref{thetaeg},
except for the insertion in the square bracket.
The latter in turn arises by acting with the covariant derivative
$\cD_{-\frac{3}{2}}$, obtaining
\be
\begin{split}
\delta e^\Phi = &
\frac{\tau_2\, e^{-2\pi S_{\rm cl}}}{16\pi^2\sqrt{2\tau_2\, \bfp\cdot \bft^2}}\,
\cD_{-\frac{3}{2}}\, \sum_{\bfmu\in\Lambda^*/\Lambda}
h_{\bfp,\bfmu}(\tau)\,\overline{\theta_{\bfp,\bfmu}(\tau,\bft,\bfb,\bfc)} +{\rm c.c.}
\end{split}
\label{resPhi}
\ee
The action of $\cD_{-\frac{3}{2}}$ on $h_{\bfp,\bfmu}(\tau) \overline{\theta_{\bfp,\bfmu} (\tau,\bft,\bfb,\bfc)}$
raises the modular weight from $( -\tfrac32,\half)$ to $(\half,\half)$, while the overall factor of
$\tau_2$ reduces this to $(-\half,-\half)$ (recall that the combination $\tau_2\, \bfp\cdot \bft^2$
is modular invariant). Therefore, the correction $\delta e^{\Phi}$ transforms as required in  \eqref{delphiSL2}.

%%%%%%%%%%%%%%%%%%%%%%%%%%%%%%%%%%
\subsection{Instanton corrections to Darboux coordinates and period integrals
%%%%%%%%%%%%%%%%%%%%%%%%%%%%%%%%%%
\label{sec_dperiod}}

We now turn to the instanton corrections to the Darboux coordinates. Unlike
the  contact potential, which is modular covariant, it will turn out that the Darboux coordinates
have anomalous modular transformations, due to the fact that the contour $\ell_\gamma$
is not invariant. However, we shall be able to give a precise characterization of this modular anomaly,
by rewriting the Penrose-type integral over $z'$ as an Eichler integral.

For simplicity we start from the last term in $\hat\delta_-\alpha'$ \eqref{cov-DC}.
It is given by a Gaussian integral
in $1/z'$ which can be easily evaluated. The result is
\be
\I \tau_2 z(p\cdot t^2) \sum_{q_\Lambda}\int_{\ell_{\gammam}}
\frac {\de z'}{(z')^2}\,H_{\gammam}=
-\frac{\I z}{8\pi^2}\,\sqrt{2\tau_2\,\bfp\cdot \bft^2}\,
e^{-2\pi \bar S_{\rm cl}}
\sum_{\bfmu\in\Lambda^*/\Lambda}
\overline{h_{\bfp,\bfmu}(\tau)}\,{\theta_{\bfp,\bfmu}(\tau,\bft,\bfb,\bfc)}.
\label{antiterm}
\ee
It transforms as a modular function of weight $(0,-1)$ in full agreement with
the required transformation properties \eqref{hatdelSL2}.

All the other instanton corrections to Darboux coordinates in the large volume limit are determined by the function
$\cJ_\bfp(z)$ \eqref{defcJ}. Therefore, our prime interest is to understand its modular properties.
To this aim, let us consider the integral appearing in this  function, namely
\be
\label{defIz}
\cI (z)= \int_{\ell_{\gammap}} \frac{\de z'}{z'-z}\, \expe{\I Q_{\gammap}(z')}.
\ee
By shifting the contour, this may be rewritten as
\be
\label{defI}
\cI (z)= \int_{-\infty}^{+\infty} \frac{\de z'}{z'-z-\I\alpha}\,e^{-\beta^2 z'^2}\, ,
\qquad
\alpha=\frac{ \kbp}{\sqrt{\bfp\cdot \bft ^2}}\, ,
\quad
\beta=\sqrt{2\pi\tau_2 \, \bfp\cdot \bft ^2}\, .
\ee
Since $z$ can be absorbed by a shift of $\alpha$, we shall first consider the value of $\cI$
at $z=0$. Using the identity
\be
\label{Intab}
\int_{-\infty}^{\infty} \frac{\de z'}{z'-\I \alpha}\, e^{-\beta^2 z'^2}
=\I\pi\, \sgn(\Re(\alpha))\,e^{\alpha^2\beta^2} \Erfc\Bigl(\sgn(\Re(\alpha\beta))\,\alpha\beta\Bigr),
\ee
valid for $\alpha, \beta \in \mathbb{C}$ with $\Re(\beta^2)>0$ and
$\Re(\alpha)\neq 0$, the integral representation
\be
\Erfc( \sqrt{\pi}\, x ) = \int_{x^2}^{\infty}\de u\, u^{-1/2}\, e^{-\pi
  u},\qquad \Re(x)\geq 0,
\ee
and changing variable from $u$ to $\bar w=\tau-\I u/(\bfk+\bfb)^2_+$,
valued in the lower half plane, one may cast $\cI(0)$ in the form of
an Eichler integral \eqref{PhiEich}\footnote{
In fact, the integral over $z'$ \eqref{Intab}  can be rewritten as an integral over $\bar w$
without ever evaluating it in terms of the error function, upon representing both
factors in the integrand in terms of their Fourier transforms
\bea
\frac{1}{z'-\I\alpha} &=& \I \, \sgn(\alpha)\, \int_{-\infty}^{\infty} \, \de\omega\,
e^{-\alpha\, \omega-\I\,\omega\,z'}\, \eta(\sgn(\alpha)\,\omega)
\nn\\
e^{-\beta^2 z'^2} &=& \frac{1}{2|\beta|\sqrt{\pi}}
\int_{-\infty}^{\infty} \, \de\omega'\, e^{-\I\,\omega'z'-\omega'^2/(4\beta^2)},
\eea
where $\eta$ is the Heaviside step function, and carrying out the integral over $z',\omega,\omega'$.
Thus, the integration variable $\bar w$ in \eqref{eqI0per} is Fourier dual to the twistor
coordinate $z'$ in \eqref{defIz}.
}
\be
\label{eqI0per}
\cI(0) = -\pi
\int_{\bar\tau}^{-\I \infty}
\frac{ \de \bar w}{\sqrt{\I(\bar w-\tau)}} \, \kbp\,
\, \expe{\frac12 (\bfk+\bfb)^2_+ (\bar \tau-\bar w)} .
\ee

Inserting the representation \eqref{eqI0per} into \eqref{defcJ}, restricting
to $z=0$ for simplicity, and carrying out the sum over $q_\Lambda$ in the same fashion
as in \S\ref{sec_msw}, we arrive at
\be
\label{xia0period}
\begin{split}
\cJ_\bfp(0) =& \, \frac{\I\, e^{-2\pi S_{\rm cl}}}{8\pi^2} \sum_{\bfmu\in\Lambda^*/\Lambda}
h_{\bfp,\bfmu}(\tau)\,
\int_{\bar\tau}^{-\I \infty}
\frac{ \de \bar w}{\sqrt{\I(\bar w-\tau)}}
\\
& \times
\sum_{\bfk \in \Lambda+\bfmu+\frac12 \bfp}\, (-1)^{\bfk\cdot\bfp}\,
\kbp\,
\expe{- \tfrac12(\bfk+\bfb)_+^2 \bar w -\tfrac12 (\bfk+\bfb)_-^2 \tau+\bfc \cdot (\bfk+\tfrac12\bfb)} .
\end{split}
\ee
The second line in this expression is recognized as the (complex conjugate of the)
Siegel-Narain theta series \eqref{defSieg1} of  weight $(\frac32,\tfrac12\,(b_2-1))$,
analytically continued away from the slice $\bar w=\bar\tau$. Eq. \eqref{xia0period}
is thus an Eichler integral of weight $(\frac{b_2}{2},0)$, multiplied by the vector-valued
modular form $h_{\bfp, \bfmu}$ of weight $(-1-\frac{b_2}{2},0)$.
Thus, $\cJ_\bfp(z=0)$ transforms with modular weight $(-1,0)$, except for a modular
anomaly of the form \eqref{Eichmod} given by the period integral of the theta series
\eqref{defSieg1}.

More generally, using this type of manipulations, one may rewrite the full function $\cJ_\bfp(z)$
away from $z=0$  as
\be
\label{xiazperiod}
\cJ_\bfp(z)=\frac{\I\, e^{-2\pi S_{\rm cl}}}{8\pi^2}
\sum_{\bfmu\in\Lambda^*/\Lambda}
h_{\bfp,\bfmu}(\tau)\,
\int_{\bar\tau}^{-\I \infty}
\frac{\overline{\BTheta_{\bfmu} (w,\bar\tau;\bz)}\, \de \bar w}{\sqrt{\I(\bar w-\tau)}}
\ee
where
\be
\begin{split}
\overline{\BTheta_{\bfmu} (w,\bar\tau;\bz)}
=&\, \expe{\I\tau_2\, \bfp\cdot \bft^2 \frac{\bar \tau-\bar w}{\tau-\bar w}\, z^2}\,
\,\sum_{\bfk \in \Lambda+\bfmu
+\frac12 \bfp}\, (-1)^{\bfk\cdot\bfp}\,
\(\kbp +  \frac{2\tau_2 \sqrt{ \bfp\cdot \bft^2} z}{\bar w-\tau}\)
\\
&\,\times
\expe{- \tfrac12(\bfk+\bfb)_+^2 \bar w -\tfrac12 (\bfk+\bfb)_-^2 \tau+\bfc \cdot (\bfk+\tfrac12\bfb)} .
\end{split}
\ee
It is easy to see that the integral is identical to the complex conjugate of
the generating function $\cG^\Phi_{\bfp,\bfmu}(y)$ \eqref{genf-Phi} evaluated at
$y=2\tau_2\sqrt{\bfp\cdot \bft^2}\bz$, i.e.
\be
\cJ_\bfp(z)=\frac{e^{-2\pi S_{\rm cl}}}{8\pi^2}
\sum_{\bfmu\in\Lambda^*/\Lambda}
h_{\bfp,\bfmu}(\tau)\,\overline{\cG^\Phi_{\bfp,\bfmu}(2\tau_2\sqrt{\bfp\cdot \bft^2}\bz)}.
\label{JGG}
\ee
In Appendix \ref{app_mock} it is shown that provided $y$ transforms with the modular weight $(0,-1)$,
which is indeed the case for our identification between $y$ and $z$, the generating function $\cG^\Phi_{\bfp,\bfmu}(y)$
transforms as an Eichler integral of weight $(0,b_2/2)$.
As a result, the full function \eqref{JGG}
possesses the following transformation property
\be
\label{modcJ}
\cJ_\bfp(z) \mapsto (c\tau+d)^{-1} \left(
\cJ_\bfp(z) -\frac{\I\,e^{-2\pi S_{\rm cl}}}{8\pi^2} \sum_{\bfmu\in\Lambda^*/\Lambda}
h_{\bfp,\bfmu}(\tau)\,
\int_{-d/c}^{\I\infty} \frac{\overline{\BTheta_{\bfmu} (w,\bar\tau;\bz)}\,
\de \bar w }{[\I(\bar w-\tau)]^{1/2}}\right) ,
\ee
i.e. transforms with modular weight $(-1,0)$ and an anomaly given by a period integral.
Recalling the properties of the covariant derivative operators $\cD_{\wh},\bfD,\bfDb$ mentioned in
footnote \ref{fooD}, this allows to conclude that the corrections
$\delta\bfxi, \hat\delta\bftxip, \hat\delta_\pm\alpha'$
transform as required in \eqref{hatdelSL2},  except for the modular anomaly \eqref{modcJ},
acted upon by the corresponding covariant derivatives.

%%%%%%%%%%%%%%%%%%%%%%%%%%%%%%%%%%
\subsection{Anomaly cancelation from mock theta series \label{sec_mock}}
%%%%%%%%%%%%%%%%%%%%%%%%%%%%%%%%%%

Having expressed the corrections to the Darboux coordinates as Eichler integrals,
or modular derivative thereof, it remains to show how the modular anomalies can be
canceled by a suitable contact transformation.

As explained in Appendix \ref{app_mock}, the modular anomaly of an Eichler integral
can be canceled by adding to it a certain mock theta series. Typically it appears as
a theta series with insertion of a difference of two sign functions defined by
two \kahler moduli $\bft$ and $\bft'$, with one of them chosen to lie on the boundary of the \kahler cone, i.e.
$\bfp \cdot \bft'^2=0$. The simplest example is provided by \eqref{eq:completedPhi} and
\eqref{eq:indefTheta}, respectively, which describe the instanton corrections to $\cJ_\bfp(0)$.

Similarly, the modular anomaly in \eqref{modcJ} can be canceled by replacing the
generating function $\cG^\Phi_{\bfp,\bfmu}$ in \eqref{JGG} by its modular completion
$\cG^\Phi_{\bfp,\bfmu}-\cG^\Theta_{\bfp,\bfmu}$, where the second term is the generating
function defined in \eqref{defcG}, evaluated at $y=2\tau_2\sqrt{\bfp\cdot \bft^2}\bz$.
In this way, we find that the  modular completion of $\cJ_\bfp(z)$ is given by
\be
\hat\cJ_{\bfp} = \cJ_{\bfp} - H_{\rm anom}\,  ,
\ee
where
\be
H_{\rm anom}=
\frac{e^{-2\pi S_{\rm cl}}}{8\pi^2}
\sum_{\bfmu\in\Lambda^*/\Lambda}
h_{\bfp,\bfmu}(\tau)\,\overline{\cG^\Theta_{\bfp,\bfmu}(2\tau_2\sqrt{\bfp\cdot \bft^2}\bz)}\ .
\ee
Using \eqref{genf-Theta}, this can be rewritten as
\bea
H_{\rm anom}&=&
\frac{1}{8\pi^2} \sum_{\bfmu\in\Lambda^*/\Lambda}
h_{\bfp,\bfmu}(\xi^0)
\label{Mockanom}
\\
& \times&
\sum_{\bfk \in \Lambda+\bfmu+\frac12 \bfp}(-1)^{\bfk\cdot\bfp}
\left[ \sgn\((\bfk+\bfb)\cdot \bft\) - \sgn\((\bfk+\bfb)\cdot \bft'\)\right]
\expe{\bfp\cdot \bftxi-\bfk\cdot \bfxi-\frac12 \,\xi^0\, (\bfk)^2 }\ ,
\nn
\eea
which shows that miraculously, $H_{\rm anom}$ is a holomorphic function of the Darboux coordinates. In fact, $H_{\rm anom}$
is proportional to the holomorphic mock theta series
\eqref{eq:indefTheta} evaluated at $\tau=\xi^0, \bfy=\bfxi$ (upon taking into account
the identification between $\bfy$ and $\bfb,\bfc$ mentioned in footnote \ref{foophase}).

Moreover, one can show that the corresponding modular completions of
the Darboux coordinates $\delta\bftxi, \delta\txi'_0,\delta\alpha'$, following from \eqref{defhctJa},
\eqref{cov-DC} and obtained by substitution of $\cJ_{\bfp}$ by $\hat\cJ_{\bfp}$, have the following form
\be
\delta\bfxi- \partial_{\bftxi} H_{\rm anom},
\qquad
\delta\txi'_\Lambda+ \partial_{\xi^\Lambda} H_{\rm anom},
\qquad
\delta\alpha'+ (1- \xi^\Lambda \partial_{\xi^\Lambda} )\, H_{\rm anom}
\label{modcompl}
\ee
and transform according to \eqref{SL2Zxi}, without any further anomaly.
Moreover, the modular completion \eqref{modcompl} is recognized as a holomorphic contact transformation \eqref{QKgluing}.
It is interesting to note that the theta series \eqref{Mockanom} can be written as
\be
H_{\rm anom}=\hf\,\sum_{q_\Lambda}\left[ \sgn\((\bfk+\bfb)\cdot \bft\) - \sgn\((\bfk+\bfb)\cdot \bft'\)\right] H_\gamma
\label{HHHH}
\ee
and differs from the formal sum \eqref{eqHp} only by an insertion of
the two signs which makes the sum over charges convergent.
As a result, the compensating contact transformation is generated
by a proper subset of the original transition functions,
and the modular covariant Darboux coordinates differ from the original type IIA Darboux
coordinates
by a sequence of symplectomorphisms associated to the BPS states
for which the two signs are  different.
This mimicks the situation in the D1-D(-1) sector \cite{Alexandrov:2012bu}.

Thus, we have demonstrated that the modular anomaly is canceled
by performing the transformation \eqref{QKgluing} generated by the mock
theta series \eqref{Mockanom}. In fact, we have
a continuous family of modular completions labeled by the parameter $\bft'$ on the
boundary of the \kahler cone. It would be interesting if this ambiguity was fixed by
some physical principle.
This concludes the proof of the modular invariance of the HM moduli space corrected by D3-instantons in the large
volume, one-instanton approximation.

%%%%%%%%%%%%%%%%%%%%%%%%%%%%%%%%%%%%%%%%%%%%%%%%%%%%%%%%%%%%%%%%%%%%
\section{Discussion \label{sec_discuss}}
%%%%%%%%%%%%%%%%%%%%%%%%%%%%%%%%%%%%%%%%%%%%%%%%%%%%%%%%%%%%%%%%%%%%

In this work, we  investigated the modular invariance of D3-D1-D(-1)-instanton corrections
to the hypermultiplet moduli space of type IIB string theory compactified on a CY threefold $\CYm$. We focused
on the large volume limit, where D5-NS5-instantons can be consistently
ignored. Using similar arguments as the ones which enter in the proof of modular invariance of the partition
function of D4-D2-D0 black holes in type IIA string theory compactified on the same threefold
$\CYm$, we showed that the DT invariants which govern D3-instanton corrections are expressible
in terms of the Fourier coefficients of the elliptic genus of the MSW superconformal field theory,
which we refer to as MSW invariants. Unlike the DT invariants, the MSW invariants are independent
of the moduli and have simple modular properties.  In the one-instanton approximation,
we found that the D3-instanton corrections to the standard Darboux coordinates
are expressible as Eichler integrals of the MSW elliptic genus, and modular derivatives thereof.
Thus, they suffer from modular anomalies, which can however be absorbed by a local complex
contact transformation. The generating function for this contact
transformation was recognized
as the holomorphic mock theta series of signature $(1,b_2-1)$
introduced in \cite{Zwegers-thesis}. In this physical set-up, the Eichler integral which provides
the modular completion of the mock theta series arises as a Penrose-type integral along the twistor
fiber, or rather its Fourier transform.\footnote{See \cite{Troost:2010ud} for
another example where the modular completion of a mock modular form arises naturally from physics.}
Unlike Darboux coordinates, the contact potential (also known as the four-dimensional dilaton)
transforms covariantly, and is proportional to the (modular derivative
of the) MSW elliptic genus.

While it is very satisfying to see modularity emerge in the large volume, one-instanton approximation,
an outstanding problem is to extend the results of this paper beyond this
regime. This can be in principle addressed by expanding the integral equations \eqref{eqTBA}
to higher order in the MSW invariants, as indicated in \S\ref{sec_msw}.
There are two contributions at next-to-leading order, namely (i) the  two-instanton
correction in the usual iterative expansion of \eqref{eqTBA}, with the DT invariant
$\bar\Omega(\gamma;\bfz)$ replaced by the MSW invariant $\bar\Omega_{\bfp,\bfmu}$,
and (ii) the usual one-instanton correction, with $\bar\Omega(\gamma;\bfz)$
replaced by the  quadratic term in \eqref{multiMSW}.  The first contribution is weighted by
$e^{-2\pi\tau_2 ( |Z(\gamma_1,\bfz)|+|Z(\gamma_2,\bfz)|)}$,  whereas the second
contribution is weighted by $e^{-2\pi\tau_2|Z(\gamma_1+\gamma_2,\bfz)|}$.
By construction, the sum of these contributions
is continuous across walls of marginal stability, although
they are separately discontinuous.  We expect that the D3-instanton correction at
two-instanton order will be modular invariant, and will be related to the two-center
D4-brane partition function in the same fashion as at one-instanton order, with the two-center
Siegel-Narain theta
series  replaced
by an indefinite theta series.
It is encouraging that  the partition function for two-center D4-black
holes, corresponding to the contribution of (i), indeed becomes modular invariant and
continuous after adding a
suitable non-holomorphic completion \cite{Manschot:2009ia}, which should arise
from contributions of type (ii).

More ambitiously, it would be very interesting to show that
S-duality invariance holds at the non-linear level. One way to achieve this
would be to recast the standard type IIA twistorial construction into the manifestly
invariant framework developed in \cite{Alexandrov:2012bu}. A naive attempt using
the standard representation of the MSW elliptic genus as a Poincar\'e
series \cite{Manschot:2007ha} fails, due to
the necessity of regulating the Poincar\'e series (see Appendix \ref{sec_poinca}).
It is conceivable that this obstruction may be
avoided by taking into account the constraints on the polar degeneracies from modular
invariance. Eventually, we hope that it will be possible to formulate all D5-D3-D1-D(-1)
and NS5-brane instanton corrections in a manifestly invariant S-duality fashion, going beyond
the linear order analysis of \cite{Alexandrov:2010ca} and
providing a tight set of constraints on the generalized Donaldson-Thomas invariants on
Calabi-Yau threefolds.

Finally, we expect that the structure uncovered in this paper in the context of the
HM moduli space in type II string theories on a compact CY threefold  will continue to
exist in the rigid limit near singularities of $\CYm$.
Specifically,  the Coulomb branch of $\cN=2$ five-dimensional $SU(2)$ gauge theories
compactified to a two-torus \cite{Haghighat:2011xx}, obtained by
considering type IIA string theory on a non-compact CY threefold given by the anti-canonical bundle $\cO(-K_S)\to S$
over a rational surface $S$ \cite{Morrison:1996xf}, gives a family of \hk manifolds parametrized
by the modular parameter of the torus, which must be invariant under modular transformations.
We expect this family of HK metrics to be described by a similar twistorial
construction as in this paper, using the generalization of the QK/HK correspondence
put forward in \cite{Alexandrov:2012bu}. For such line bundles over rational surfaces,
the DT invariants $\Omega(\gamma;z^a)$ with $p^0=0$ can be computed explicitly \cite{Manschot:2011ym},
and the analogue of the MSW CFT is a (0,4)-supersymmetric sigma model with target
space given by the Atiyah-Hitchin moduli space of monopoles in three
dimensions \cite{Haghighat:2011xx, Haghighat:2012}.
An important difference between the rigid case and the assumptions in this paper is that
the generating function $h_{\bfp,\bfmu}(\tau)$ of DT invariants for
rational surfaces is mock
modular  \cite{Vafa:1994tf, Manschot:2011dj}. It is natural to expect that
its non-holomorphic modular completion will arise from multi-instanton corrections.
Finally, by the general logic of the QK/HK
correspondence, this family of HK metrics will be naturally endowed with
a rank two modular invariant hyperholomorphic line bundle analogous to
the rank one hyperholomorphic line bundle which arises in 4D gauge theories
\cite{Alexandrov:2011ac,Neitzke:2011za}, whose physical role remains to be understood.

\acknowledgments

We would like to thank Daniel Persson for valuable comments on the draft.

\appendix

%%%%%%%%%%%%%%%%%%%%%%%%%%%%%%%%%%%%%%%%%%%%%%%%%%%%%%%%%%%%%%%%%%%%
\section{Indefinite theta series and period integrals \label{app_mock}}
%%%%%%%%%%%%%%%%%%%%%%%%%%%%%%%%%%%%%%%%%%%%%%%%%%%%%%%%%%%%%%%%%%%%

In this Appendix, we review some aspects of indefinite theta series with Lorentzian
signature $(1,n-1)$, which are essential for describing D3-instanton corrections.

\subsection{Vign\'eras' theorem}

Let $Q(\bfx)\equiv \bfx^2$ be a quadratic form on $\IR^n$ with signature $(1,n)$, and
$\Lambda$ an $n$-dimensional lattice such that $Q(\bf k)\in\IZ$ for
${\bf k}\in\Lambda$. Let
 $\mathcal{P}(\bfx)$ be a function on $\IR^n$
 such that $\mathcal{P}(\bfx)\, e^{\pi\bfx^2}$ is integrable. Let
$\bfp\in\Lambda$  be a characteristic vector (such
that ${\bf k}^2+{\bf k}\cdot \bfp\in 2\mathbb{Z}$, $\forall \,{\bf k} \in \Lambda$),
 $\bfmu\in\Lambda^*/\Lambda$ a glue vector, and $\lambda$ an arbitrary integer.
 Following \cite{Vigneras:1977}, with some changes of notations,
we construct the family of theta series
\be
\label{Vignerasth}
\vartheta_{\bfp,\bfmu}
\left( \mathcal{P} ,\lambda; \tau; \bfb,\bfc\right)=\tau_2^{-\lambda/2}
\sum_{\bfk \in \Lambda+\bfmu+\frac12 \bfp}\, (-1)^{\bfk\cdot\bfp}\,
\mathcal{P}(\sqrt{2\tau_2}(\bfk+\bfb))\, \bar q^{- \half (\bfk+\bfb)^2}\,
\expe{- \bfc \cdot (\bfk+\tfrac12\bfb)},
\ee
where $q=e^{2\pi \I \tau}$ and $\bfb, \bfc \in \Lambda \otimes \mathbb{R}$.
Irrespective of the choice of $\mathcal{P}$ and $\lambda$, \eqref{Vignerasth} satisfies
the elliptic transformation properties
\be
\begin{split}
\label{Vigell}
\vartheta_{\bfp,\bfmu}\left( \mathcal{P} ,\lambda; \tau; \bfb+\bf
  k,\bfc\right) =&(-1)^{\bf k\cdot\bfp}\,
\expe{\tfrac12 \bfc\cdot \bf k}\, \vartheta_{\bfp,\bfmu}\left( \mathcal{P} ,\lambda; \tau; \bfb,\bfc\right),
\\
\vartheta_{\bfp,\bfmu}\left( \mathcal{P} ,\lambda; \tau; \bfb,\bfc+\bf
  k\right)=&(-1)^{\bf k\cdot\bfp}\,
\expe{-\tfrac12 \bfb\cdot \bf k}\, \vartheta_{\bfp,\bfmu}\left( \mathcal{P} ,\lambda; \tau; \bfb,\bfc\right),
\end{split}
\ee
for any ${\bf k}\in \Lambda$.

Now, let us assume that $\mathcal{P}(\bfx)\, e^{\pi\bfx^2}$ decays sufficiently
fast\footnote{More precisely, $f(\bfx)\equiv\, \mathcal{P}(\bfx)\, e^{\pi\bfx^2}$ should be such that,
for any $D(\bfx)$ be any differential operator of order $\leq 2$ and
$R(\bfx)$ any polynomial of degree $\leq 2$,
$f(\bfx)$, $D(\bfx)f(\bfx)$ and $R(\bfx)f(\bfx)$ should be both integrable and square integrable on $\IR^n$.}
at infinity,
and that $\mathcal{P}(\bfx)$ is annihilated by the differential operator
\be
\label{Vigdif}
\left[ \Delta   + 2\pi ( \bfx\cdot \pa_{\bfx} -\lambda) \right] \mathcal{P}(\bfx)  = 0,
\ee
where $\Delta=\partial_{\bfx}^2$ is the Laplacian operator on $\IR^n$.
Then, using the same methods
as in \cite{Vigneras:1977}, one may show
that the family of theta series \eqref{Vignerasth}, with $\bfmu$
running over all cosets in  $\Lambda^*/\Lambda$,
 transforms as a vector-valued Jacobi form of weight $(0,\lambda+n/2)$
under the full modular group. This means that it satisfies the modular properties
\bea
\vartheta_{\bfp,\bfmu}\left( \mathcal{P} ,\lambda; -1/\tau; \bfc,-\bfb\right)
&=&\frac{1}{\sqrt{|\Lambda^*/\Lambda|}}(\I\bar \tau)^{\lambda+\tfrac{n}{2}}
\expe{-\frac14 \bfp^2}
 \sum_{\bfnu\in\Lambda^*/\Lambda}
\expe{-\bfmu\cdot \bfnu}
\vartheta_{\bfp,\bfmu}\left( \mathcal{P} ,\lambda; \tau; \bfb,\bfc\right),
\nn\\
\vartheta_{\bfp,\bfmu}\left( \mathcal{P} ,\lambda; \tau+1; \bfb,\bfc+\bfb)\right)
&=&\expe{\tfrac12(\bfmu+\tfrac12 \bfp)^2}\,
\vartheta_{\bfp,\bfmu}\left( \mathcal{P} ,\lambda; \tau; \bfb,\bfc\right) ,
\label{eq:thetatransforms}
\eea
on top of the elliptic properties \eqref{Vigell}.
In the next subsections, we present the relevant instances of this general
construction relevant for the present work.

\subsection{Siegel-Narain Theta series}

The simplest example arises for ${\mathcal P}(\bfx)=e^{-\pi (x_+^{(\bft)})^2}$, $\lambda=-1$,
where  $x_+^{(\bft)}$ is the projection of $\bfx$ along a fixed time-like vector $\bft$ in $\IR^n$
($\bft^2>0$):
\be
x_+^{(\bft)}= \frac{\bfx \cdot \bft}{\sqrt{\bft\cdot \bft}},
\qquad
\bfx_+^{(\bft)} = \frac{\bfx \cdot \bft}{\bft\cdot \bft}\, \bft = \bfx - \bfx_-^{(\bft)}
\ee
so that $(\bfx_+^{(\bft)})^2 =(x_+^{(\bft)})^2\geq 0$,  $(\bfx_-^{(\bft)})^2\leq 0$. In
case no confusion arises, we write  $x_+^{(\bft)}=x_+$.
The corresponding theta series \eqref{Vignerasth} is then, up to a power of $\tau_2$ and
lattice shift, the
standard Siegel-Narain theta series,
\be
\label{thSieg}
\begin{split}
\theta_{\bfp,\bfmu} (\tau,\bft,\bfb,\bfc) \equiv &\tau_2^{-1/2}
\vartheta_{\bfp,\bfmu}\left( \e^{-\pi x_+^2} ,-1; \tau; \bfb,\bfc\right)
\\
=& \sum_{\bfk\in \Lambda+\bfmu+\tfrac12\bfp} \, (-1)^{\bfk\cdot\bfp}\,
q^{\tfrac12(\bfk+\bfb)_+^2}
\, \bar q^{-\tfrac12 (\bfk+\bfb)_-^2} \, \expe{- \bfc \cdot (\bfk+\tfrac12\bfb)}.
\end{split}
\ee
Thus, $\theta_{\bfp,\bfmu}$ transforms as a vector-valued Jacobi form of weight  $(\tfrac12,\tfrac{n-1}{2})$.

\subsection{Zwegers' mock theta series}

The other prime example is Zwegers' indefinite theta series
of weight $(0,\tfrac{n}{2})$  \cite{Zwegers-thesis} (based on earlier works \cite{MR1623706,0961.14022}),
\be
\label{Zwegthetahat}
\begin{split}
\widehat\Theta_{\bfp,\bfmu}(\tau,\bft,\bft',\bfb,\bfc)=& \sum_{\bfk \in \Lambda+\bfmu+\tfrac12 \bfp}
\signkp
\left[ \Erf\left(\sqrt{2\pi \tau_2} \kbp^{(\bft)} \right)-
\Erf\left(\sqrt{2\pi \tau_2} \kbp^{(\bft')}  \right) \right]
\\
&\qquad\times
\bar q^{-\tfrac12(\bfk+\bfb)^2}\,\expe{-\bfc\cdot (\bfk+\tfrac{\bfb}{2})} ,
\end{split}
\ee
where $\bft$ and $\bft'$ are two time-like vectors and $\Erf(\sqrt{\pi}
x)=2\int_{0}^xe^{-\pi\,t^2}\de t$ is the standard error function. This series again belongs
to the class \eqref{Vignerasth} with ${\mathcal P}(\bfx)=\Erf(\sqrt{\pi} \bfx_+^{(\bft)})
-\Erf(\sqrt{\pi} \bfx_+^{(\bft')})$. Each of the two terms in ${\mathcal P}(\bfx)$
satisfies the differential equation \eqref{Vigdif} for $\lambda=0$, and the sum
of the two terms satisfies the decay conditions, although this is not true for
each term separately. Using
\be
\Erf(\sqrt{\pi} x) = \sgn(x)\, \left(1-\Erfc(\sqrt{\pi} |x|) \right),
\label{signtoErf}
\ee
one may decompose \eqref{Zwegthetahat} into a sum of three terms
\be
\label{eq:indefThetadec}
\widehat\Theta_{\bfp,\bfmu}(\tau,\bft,\bft',\bfb,\bfc)=- \Phi_{\bfp,\bfmu}(\tau,\bft,\bfb,\bfc)+
\Theta_{\bfp,\bfmu}(\tau,\bft,\bft',\bfb,\bfc) + \Phi_{\bfp,\bfmu}(\tau,\bft',\bfb,\bfc),
\ee
where the middle term is anti-holomorphic in $\tau$,
\be
\label{eq:indefTheta}
\Theta_{\bfp,\bfmu}(\tau,\bft,\bft',\bfb,\bfc)=\sum_{\bfk \in \Lambda+\bfmu+\tfrac12 \bfp}
 \signkp\[\sgn( \kbp^{(\bft)} )-\sgn( \kbp^{(\bft')} )\]
 \bar q^{-\tfrac12(\bfk+\bfb)^2}\,\expe{- \bfc\cdot (\bfk+\tfrac12\bfb)},
\ee
while the first and last term are given by
\be
\label{eq:completedPhi}
\Phi_{\bfp,\bfmu}(\tau,\bft,\bfb,\bfc)=\sum_{\bfk\in \Lambda+\bfmu+\tfrac12 \bfp}
\signkp\,\sgn(\kbp^{(\bft)})\,
\Erfc \left(\sqrt{2 \pi \tau_2}\, |\kbp^{(\bft)}| \right)\,\bar q^{-\tfrac12(\bfk+\bfb)^2}\,
\expe{- \bfc\cdot (\bfk+\tfrac12\bfb)}.
\ee
It is useful to note that both \eqref{eq:indefTheta} and \eqref{eq:completedPhi} are
of the form \eqref{Vignerasth} with a function $\mathcal{P}(\bfx)$  which is square
integrable\footnote{For the holomorphic theta series \eqref{eq:indefTheta},
this follows from the fact that the difference of sign functions vanishes on all time-like
vectors.} and obeys the differential
equation \eqref{Vigdif} everywhere, except on the locus where $x_+^{(\bft)}$ vanishes.
As a result, the three terms in \eqref{eq:indefThetadec} are not separately modular invariant,
although their sum is.

The modular anomaly of \eqref{eq:completedPhi} may be exposed by
using the identity
\be
\sgn(\kbp^{(\bft)})\, e^{2 \pi \tau_2 [\kbp^{(\bft)}]^2}\,
\Erfc\left( \sqrt{2 \pi \tau_2}| \kbp^{(\bft)}|\right) =-\I\,
\int_{\tau}^{\I \infty}
\, \expe{\frac12 [\kbp^{(\bft)}]^2 (w-\tau)} \frac{\kbp^{(\bft)}\, \de w}
{\sqrt{-\I(w-\bar\tau)}}
\ee
to rewrite \eqref{eq:completedPhi} as a (generalized) Eichler integral,
\be
\label{eq:completedPhip}
\Phi_{\bfp,\bfmu}(\tau,\bft,\bfb,\bfc)=-\I\, \int_{\tau}^{\I\infty}
\sum_{\bfk\in \Lambda+\bfmu+\tfrac12\bfp}\signkp\,
q_w^{\tfrac12(\bfk+\bfb)^2_+} \, \bar q^{\,-\tfrac12(\bfk+\bfb)_-^2}\,
\frac{\kbp\, \de w }{[-\I(w-\bar \tau)]^{1/2}}\,
\expe{- \bfc\cdot (\bfk+\tfrac12 \bfb)},
\ee
where $q_w=\expe{w}$.
Recall that if $F(\tau,\bar\tau)$ is an analytic modular
form of weight $(\wh,\bwh)$, and if $F(w,\bar\tau)$ is its analytic continuation away
from the slice $w=\tau$, the Eichler integral
\be
\label{PhiEich}
\Phi(\tau) =  \int_{\tau}^{\I\infty} \,  \frac{F(w,\bar\tau)\, \de w }{[-\I(w-\bar \tau)]^{2-\wh}}
\ee
transforms with modular weight $(0,\bwh+2-\wh)$, up to modular anomaly given by
a period integral,
\be
\label{Eichmod}
\Phi(\gamma\tau) =  (c\bar\tau+d)^{\bwh+2-\wh}\, \left(
\Phi(\tau) - \int_{-d/c}^{\I\infty} \, \frac{F(w,\bar\tau)\, \de w }{[-\I(w-\bar \tau)]^{2-\wh}} \right).
\ee
The function $F(\tau,\bar\tau)$ is known as the shadow of the Eichler integral \eqref{PhiEich},
and can be extracted from $\Phi(\tau)$ by acting with the operator $-(2\tau_2)^{2-\wh} \pa_\tau$.

Returning to \eqref{eq:completedPhip},  $\Phi_{\bfp,\bfmu}(\tau,\bft,\bfb,\bfc)$ is then identified as
an Eichler integral of weight $(0,\tfrac{n}{2})$,
with shadow proportional to the theta series of weight $(\tfrac32,\tfrac{n-1}{2})$,
\be
\label{defSieg1}
\begin{split}
\tilde\theta_{\bfp,\bfmu} (\tau,\bft,\bfb,\bfc) \equiv &\  \tau_2^{-3/2}
\vartheta_{\bfp,\bfmu}\left( (x_+/\sqrt{2}) \, \e^{-\pi x_+^2} ,-2; \tau; \bfb,\bfc\right)
\\
=& \sum_{\bfk\in \Lambda+\bfmu+\tfrac12\bfp}\signkp\,
\kbp\, q^{\tfrac12(\bfk+\bfb)^2_+} \, \bar q^{\,-\tfrac12(\bfk+\bfb)_-^2}\,
\expe{- \bfc \cdot (\bfk+\tfrac12 \bfb)} .
\end{split}
\ee

Let us now consider what happens with \eqref{eq:indefThetadec} when one of the
time-like vectors, say $\bft'$, approaches the boundary of the K\"ahler cone, i.e. becomes light-like $\bft'^2=0$.
Then $\kbp^{(\bft')} $ diverges and, since  $\Erfc(|x|)$ is exponentially suppressed as $x\to \infty$,
the last term in \eqref{eq:indefThetadec} vanishes, leaving only the first two.
Thus, for any choice of light-like vector $\bft'$, the holomorphic
theta series $\Theta_{\bfp,\bfmu}(\tau,\bft,\bft',\bfb,\bfc)$ in \eqref{eq:indefTheta}
gives a modular completion of the Eichler integral $\Phi_{\bfp,\bfmu}(\tau,\bft,\bfb,\bfc)$ (or conversely,
the Eichler integral gives a modular completion of the holomorphic
theta series).

\subsection{An infinite sequence of mock theta series}

Finally, let us describe an infinite sequence of indefinite theta series similar to \eqref{Zwegthetahat},
obtained by applying the projection $\Db_+^{(\bft)}$ of the covariant derivative $\bfDb$ \eqref{defDw}
along a time-like vector $\bft$.
This derivative acts on any Jacobi form of weight $(\wh,\bwh)$
by changing its modular weight to $(\wh,\bwh+1)$. In particular,
it preserves the class \eqref{Vignerasth} of theta series and acts on them in the following way
\be
\Db_+^{(\bft)}\vartheta_{\bfp,\bfmu}\left( \mathcal{P} ,\lambda; \tau; \bfb,\bfc\right)
=\vartheta_{\bfp,\bfmu}\left( \p_{(\bft)}\mathcal{P} ,\lambda+1; \tau; \bfb,\bfc\right),
\qquad
\p_{(\bft)}\mathcal{P}(\bfx) \equiv \[\I \,  \frac{\bft\cdot ( \pa_{\bfx} + 2\pi \bfx )}{\sqrt{2 \bft^2}} \] \mathcal{P}(\bfx) .
\ee
Using this result as well as the identity
\be
\p_{(\bft)}^m\,\sgn(x_+^{(\bft')}) = (\pi/2)^{m/2} \,
H_{m}(\I \sqrt{\pi} \, x_+^{(\bft)})\, \sgn(x_+^{(\bft')}),
\ee
where $H_m$ are the Hermite polynomials, one finds that the $m$-th derivative of \eqref{eq:indefTheta}
is given by
\bea
\bigl(\Db_+^{(\bft)}\bigr)^m\, \Theta_{\bfp,\bfmu}(\tau,\bft,\bft',\bfb,\bfc)&=&
\(\frac{\pi}{2\tau_2}\)^{m/2}\, \sum_{\bfk\in \Lambda+\bfmu+\tfrac12 \bfp}
\signkp
\[\sgn( \kbp^{(\bft)} ) -\sgn( \kbp^{(\bft')} )\]
\nn\\
&& \times
H_{m}\(\I \sqrt{2\pi\tau_2}\, \kbp^{(\bft)}\)
\bar q^{-\tfrac12(\bfk+\bfb)^2}\,\expe{- \bfc\cdot (\bfk+\tfrac12\bfb)} .
\label{eq:indefThetam}
\eea
Similarly, the $m$-th derivative of
\eqref{eq:completedPhi} is found to be
\be
\label{eq:completedPhim}
\bigl(\Db_+^{(\bft)}\bigr)^m \,\Phi_{\bfp,\bfmu}(\tau,\bft,\bfb,\bfc)=\tau_2^{-m/2}
\sum_{\bfk\in \Lambda+\bfmu+\tfrac12 \bfp} \signkp\,
S_{m}\left( \sqrt{2\tau_2}\, |\kbp^{(\bft)}| \right)\,\bar q^{-\tfrac12(\bfk+\bfb)^2}\,
\expe{-\bfc\cdot (\bfk+\tfrac12\bfb)} ,
\ee
where we introduced the function
\be
\begin{split}
S_m(x) =&\,
\p_{(\bft)}^m\(\sgn(x_+^{(\bft)})\Erfc\(\sqrt{\pi}|x_+|\) \)
\\
=&\,
\frac{2}{\I^{m}\sqrt{\pi}}\,  m!\,
(2\pi)^{m/2}\, [\sgn(x_+)]^{m+1} \, H_{-m-1}(\sqrt{\pi} |x_+|) \, e^{-\pi x_+^2} .
\end{split}
\ee

The modular anomaly of \eqref{eq:completedPhim}  can be exposed by writing
it as an Eichler integral similar to
\eqref{eq:completedPhip} using the integral representation of $S_m$
\be
\begin{split}
e^{2 \pi \tau_2 (\bfk+\bfb)^2_+}  S_m\left(\sqrt{2\tau_2}|\kbp|\right)
=&\, - \I m!\int_{\tau}^{\I \infty}
 \expe{-\frac12 (\bfk+\bfb)^2_+ (\tau-w)}
\\
& \times  \begin{cases}
\frac{1}{\left(\tfrac{m}{2}\right)!}
\left( -\frac{\pi}{2}\, \frac{w-\tau}{w-\bar\tau}\right)^{\tfrac{m}{2}}
\,\frac{\kbp\,\de w}{[-\I(w-\bar\tau)]^{1/2}} ,
&\quad m\ {\rm even}
\\
\frac{1}{\left(\tfrac{m-1}{2}\right)!}
\left( -\frac{\pi}{2}\, \frac{w- \tau}{w-\bar \tau}\right)^{\tfrac{m-1}{2}}
\, \frac{-\I \sqrt{\tau_2}\,\de w}{[-\I(w-\bar\tau)]^{3/2}} .
&\quad m\ {\rm odd}
\end{cases}
\end{split}
\ee
In this way, we find
\be
\label{eq:completedPhipm}
\begin{split}
\bigl(\Db_+^{(\bft)}\bigr)^m \Phi_{\bfp,\bfmu}=\, &
-\I\, \tau_2^{-m/2}\, m!\, \int_{\tau}^{\I\infty}
\sum_{\bfk\in \Lambda+\bfmu+\tfrac12\bfp}
\signkp\,
q_w^{\tfrac12(\bfk+\bfb)^2_+} \, \bar q^{\,-\tfrac12(\bfk+\bfb)_-^2}\,
\expe{- \bfc\cdot (\bfk+\tfrac12 \bfb)}
\\
&\times \,\begin{cases}
\tfrac{1}{\left(\tfrac{m}{2}\right)!}
 \left( -\frac{\pi}{2}\,\frac{w-\tau}{w-\bar\tau}\right)^{\tfrac{m}{2}}\,
\frac{\kbp\, \de w }{[-\I(w-\bar \tau)]^{1/2}},
&\quad m\ {\rm even}
\\
\tfrac{1}{\left(\tfrac{m-1}{2}\right)!}
 \left( -\frac{\pi}{2}\,\frac{w- \tau}{w-\bar\tau}\right)^{\tfrac{m-1}{2}}\,
\frac{-\I \sqrt{\tau_2} \de w }{[-\I(w-\bar \tau)]^{3/2}},
 &\quad m\ {\rm odd}
\end{cases}
\end{split}
\ee
The two theta series \eqref{eq:indefThetam} and \eqref{eq:completedPhim} have anomalous modular transformations,
but their difference, provided the vector $\bft'$ belongs to the boundary of the \kahler cone,
transforms as a modular function of weight $(0,\tfrac{n}{2}+m)$.

The infinite set of derivatives of the theta series constructed above
occurs in the analysis  of instanton corrections to Darboux coordinates in \S\ref{sec_mock}
through their  generating functions
\be
\label{defcG}
\cG^\Theta_{\bfp,\bfmu}(y)=\,
\sum_{m=0}^{\infty}\frac{y^m }{m!}\,
\bigl(\Db_+^{(\bft)}\bigr)^m\, \Theta_{\bfp,\bfmu}\, ,
\qquad
\cG^\Phi_{\bfp,\bfmu}(y)=\,
\sum_{m=0}^{\infty}\frac{y^m }{m!}\,
\bigl(\Db_+^{(\bft)}\bigr)^m\, \Phi_{\bfp,\bfmu}\, .
\ee
Using
\be
e^{2xy-y^2} = \sum_{m=0}^{\infty} H_m(x)\,\frac{y^m}{m!}\, ,
\ee
one finds that the first generating function is given by
\be
\begin{split}
\cG^\Theta_{\bfp,\bfmu}(y)=&\,
\sum_{\bfk\in \Lambda+\bfmu+\tfrac12 \bfp}
\signkp
\[\sgn( \kbp^{(\bft)} ) -\sgn( \kbp^{(\bft')} )\]
\\
&\hspace{1cm} \times
\expe{\kbp^{(\bft)}y+\frac{\I y^2}{4\tau_2} }
\bar q^{-\tfrac12(\bfk+\bfb)^2}\expe{- \bfc\cdot (\bfk+\tfrac12\bfb)},
\end{split}
\label{genf-Theta}
\ee
while the second generating function can be easily found from \eqref{eq:completedPhipm},
\be
\begin{split}
\cG^\Phi_{\bfp,\bfmu}(y)=&\,
-\I \int_{\tau}^{\I\infty}
\frac{\de w }{[-\I(w-\bar \tau)]^{1/2}}\,
\expe{\frac{\I y^2}{4\tau_2}\,\frac{w-\tau}{w-\bar\tau} }
\\
&\ \times
\sum_{\bfk\in \Lambda+\bfmu+\tfrac12\bfp}
\signkp\(\kbp+\frac{y}{w-\bar\tau}\)
q_w^{\tfrac12(\bfk+\bfb)^2_+} \, \bar q^{\,-\tfrac12(\bfk+\bfb)_-^2}\,
\expe{- \bfc\cdot (\bfk+\tfrac12 \bfb)}.
\end{split}
\label{genf-Phi}
\ee
Assuming  that the generating parameter $y$ transforms under $SL(2,\IZ)$ with weight $(0,-1)$,
it follows that the difference
$
\cG^\Phi_{\bfp,\bfmu}(y)-\cG^\Theta_{\bfp,\bfmu}(y)
$
is a non-anomalous modular form, with the same  weight as that of the original theta series \eqref{eq:indefThetadec},
i.e. $(0,\tfrac{n}{2})$.

%%%%%%%%%%%%%%%%%%%%%%%%%%%%%%%%%%%%%%%%%%%%%%%%%%%%%%%%%%%%%%%%%%%%
\section{Instanton corrections in the large volume limit \label{sec_inslve}}
%%%%%%%%%%%%%%%%%%%%%%%%%%%%%%%%%%%%%%%%%%%%%%%%%%%%%%%%%%%%%%%%%%%%

%%%%%%%%%%%%%%%%%%%%%%%%%%%%%%%%%
\subsection{S-duality and mirror map \label{sec_inslve1}}
%%%%%%%%%%%%%%%%%%%%%%%%%%%%%%%%%

To establish the modularity of the type IIA construction of D-instantons, we have to find
instanton corrections to the classical mirror map \eqref{symptobd} such that the
Darboux coordinates on the twistor space expressed in terms of the type IIB fields transform
as in \eqref{SL2Zxi}. For this purpose, it will be useful to borrow some methods from
the work \cite{Alexandrov:2012bu} where a manifestly S-duality invariant
twistorial construction of a QK manifold with two commuting isometries was provided.
This is despite the fact that the Poincar\'e series representation is not immediately
applicable to our framework, as further discussed in appendix \ref{sec_poinca}.

One of the main insights of the construction in \cite{Alexandrov:2012bu}
was that the kernel $\frac{\de t'}{t'}\, \frac{t'+t}{t'-t}$
entering the integral equations \eqref{eqTBA},
which transforms in  complicated way under S-duality,
can be replaced by an invariant kernel
\be
\label{modker}
K(\varpi,\varpi') \frac{\de t'}{t'}\equiv
\frac12\(\frac{\varpi'+\varpi}{\varpi'-\varpi}+\frac{1/\varpi'-\varpi'}{1/\varpi'+\varpi'}\)
\frac{\de \varpi'}{\varpi'}
=\frac12\,\frac{z'+z}{z'-z}\, \frac{\de z'}{z'} .
\ee
It differs from the original one by a $t$-independent term which can be absorbed
into a redefinition of the coordinates parametrizing the HM moduli
space $\zeta^\Lambda, \tzeta_\Lambda, \sigma$.
Essentially this redefinition provides a substantial part of quantum corrections to the mirror map.

However, the kernel \eqref{modker} is still not convenient for the study of the large volume limit
because the additional $t$-independent term gives a divergent contribution in this limit
(recall that the integrals over $t'$ are localized at $t'=\pm\I$).
Fortunately, there is a way to cure this problem. To this end, let us note that the invariant kernel \eqref{modker}
is not unique. In particular,  two other natural choices suggest themselves,
\be
K^+(t,t')\frac{\de t'}{t'}=\frac{\de t'}{t'-t}\,\frac{1+\I t}{1+\I t'}=\frac{\de z'}{z'-z},
\qquad
K^-(t,t')\frac{\de t'}{t'}=\frac{\de t'}{t'-t}\,\frac{1-\I t}{1-\I t'}=\frac{z}{z'}\,\frac{\de z'}{z'-z}.
\label{pm-ker}
\ee
Both are S-duality invariant, differ from the original one by a $t$-independent term, and moreover
are finite in the limit where $t'\to \mp \I$, respectively.
The crucial difference between them and \eqref{modker} is that the latter is invariant under the combined
action of the antipodal map $\antipod: t\mapsto -1/\bar t$ and complex conjugation,
whereas the kernels \eqref{pm-ker} are mapped into each other
\be
\overline{\antipod\[K^+(t,t')\frac{\de t'}{t'}\]}=K^-(t,t')\frac{\de t'}{t'}.
\ee
This plays an important role in ensuring the reality conditions on the Darboux coordinates
\be
\overline{\antipod[\xii{i}^\Lambda]}=\xii{\bi}^\Lambda,
\qquad\overline{\antipod[\txii{i}_\Lambda]}=\txii{\bi}_\Lambda,
\qquad
\overline{\antipod\[\ai{i}\]}=\ai{\bi},
\ee
where $\cU_{\bi}$ is the image of the patch $\cU_i$ under the antipodal map.
Therefore, the new kernels can be used to generalize the construction of
modular invariant QK manifolds  in \cite{Alexandrov:2012bu} as follows:

{\it
Let $\cZ$ be the $2n+1$-complex dimensional contact manifold defined by the infinite covering
\be
\cZ= \cU_+\cup\cU_- \cup\cU_{0}\cup\(\cup_{m,n}' \cU_{m,n}^+\) \cup\(\cup_{m,n}' \cU_{m,n}^-\),
\ee
and transition functions
\be
\label{transfun-gen}
\Hij{+0}=\Fcl(\xii{+}^\Lambda) ,
\qquad
\Hij{-0}=\bFcl(\xii{-}^\Lambda) ,
\qquad
\Hij{(m,n)^\pm 0}= G_{m,n}^\pm(\xi^0,\xii{m,n}^a,\txii{0}_a)\ ,
\ee
where $\Fcl(X)=-\kappa_{abc} \,\frac{X^a X^b X^c}{6 X^0}$ is an arbitrary cubic prepotential,
$\Lambda=(0,a)=0,1,\dots {n-1}$.
Here $\cU_\pm$ are the usual patches around the poles of $\CP$,
$\cU_{m,n}^\pm$ are two sets of patches which are mapped to each other under $SL(2,\IZ)$-transformations,
and $\cU_0$ covers the rest.
They must be chosen so that $\cU_0$ is mapped to itself under the antipodal map
and S-duality, whereas $\cU_{m,n}$ are mapped to each other according to
\be
\antipod\[\cU_{m,n}^\pm\]= \cU^\mp_{-m,-n},
\qquad
\cU_{m,n}^\pm\mapsto \cU^\mp_{m',n'},
\qquad
\( m'\atop n'\) =
\(
\begin{array}{cc}
a & c
\\
b & d
\end{array}
\)
\( m \atop n \).
\ee
The holomorphic functions $G_{m,n}^\pm$ are assumed to transform
as\footnote{We refrain from writing the full non-linear transformation of $G_{m,n}^\pm$,
since we are interested only in the linear approximation in this paper.
The  transformation property required for modular invariance at the non-linear level
can be found in \cite{Alexandrov:2012bu}.}
\be
\overline{\antipod\[G_{m,n}^\pm\]}=G_{m,n}^\mp,
\qquad
G_{m,n}^\pm
\mapsto \frac{G^\pm_{m',n'}}{c\xi^0+d}
\ +\mbox{non-linear terms}\
+\mbox{reg.}
\label{StransG}
\ee
where $+ \mbox{reg.}$  denotes equality up to terms which are regular in $\cU_{m',n'}^\pm$.}

Given these conditions, the associated QK manifold $\cM$ can be shown to carry an isometric action of $SL(2,\IZ)$.
The only difference with \cite{Alexandrov:2012bu} is the presence of two sets of twistorial data, labeled
by $+$ and $-$, which are mapped to each other by the antipodal map.
Physically, they can be interpreted as contributions of branes and anti-branes.
The Darboux coordinates in the patch $\cU_0$ then satisfy the following integral
equations:
\bea
\xi^0 &=& \zeta^0+\frac{\tau_2}{2}\(\varpi^{-1}-\varpi\),
\nn
\\
\xii{0}^a &=& \zeta_{\rm cl}^a +\varpi^{-1}Y^a -\varpi \bY^a+
\sum_{\veps=\pm} {\sum_{m,n}}' \oint_{C^\veps_{m,n}} \frac{\de \varpi'}{2 \pi \I \varpi'} \,
K^\veps(\varpi,\varpi') \, \p_{\txii{0}_a}G_{m,n}^\veps,
\nn
\\
\txii{0}_\Lambda& =& \tzeta^{\rm cl}_\Lambda +\varpi^{-1}\Fcl_\Lambda(Y)-\varpi\bFcl_\Lambda(\bY)
- \sum_{\veps=\pm} {\sum_{m,n}}' \oint_{C^\veps_{m,n}} \frac{\de \varpi'}{2 \pi \I \varpi'} \,
K^\veps(\varpi,\varpi')\, \p_{\xii{m,n}^\Lambda}G_{m,n}^\veps,
\label{twistlines}
\\
\ai{0}&= &  -\hf(\tsigma+\zeta^\Lambda \tzeta_\Lambda)^{\rm cl}
-\(\varpi^{-1}+\varpi\)\(\varpi^{-1} \Fcl(Y)+\varpi \bFcl(\bY)\)
-\zeta_{\rm cl}^\Lambda\( \varpi^{-1}\Fcl_\Lambda(Y)-\varpi\bFcl_\Lambda(\bY)\)
\nn \\
&-&
\sum_{\veps=\pm} {\sum_{m,n}}' \oint_{C^\veps_{m,n}} \frac{\de \varpi'}{2 \pi \I \varpi'} \,
\[ K^\veps(\varpi,\varpi')
 \(1-\xii{m,n}^\Lambda(\varpi') \p_{\xii{m,n}^\Lambda}\)G_{m,n}^\veps
\right.
\nn
\\
&&\left.
\qquad
+\( \varpi^{-1}K^\veps(0,\varpi') \Fcl_a(Y)-\varpi K^\veps(\infty,\varpi') \bFcl_a(\bY) \)\,\p_{\txii{0}_a}G_{m,n}^\veps \],
\nn
\eea
where $C^\pm_{m,n}$ are contours on $\CP$ surrounding $\cU^\pm_{m,n}$ counter-clockwise
and $(\tau_2,Y^a,\zeta^0,\zeta_{\rm cl}^a,\tzeta^{\rm cl}_\Lambda,\tsigma^{\rm cl})$
are coordinates on $\cM$.

The key to prove the presence of S-duality is to find expressions of the coordinates
$(Y^a,\zeta_{\rm cl}^a,\tzeta^{\rm cl}_\Lambda,\tsigma^{\rm cl})$ in terms of $(\tau,b^a,t^a,c^a,\cla,\psi)$
which ensure the proper transformations of the Darboux coordinates. Let us do this explicitly for the coordinates
entering $\xi^a$. To this end, note that the derivative $\p_{\txii{0}_a}G_{m,n}(t')$
transforms, like the transition function \eqref{StransG},
with an overall factor of $(c \xi^0(\varpi')+d)^{-1}$.
Using the property
\be
1-z^2 \ \mapsto\ \frac{c\xi^0+d}{c\tau+d}\,(1-z^2),
\ee
this factor can be converted into $(c \xi^0(\varpi)+d)^{-1}$ by multiplying the kernel by
$\frac{1-z'^2}{1-z^2}$. Put differently, the Darboux coordinate $\xi^a$
would transform as in \eqref{SL2Zxi} if it were equal to
\be
\begin{split}
\xii{0}^a =&\,\tau b^a-c^a +\frac{2z\tau_2}{1-z^2}\(t^a +\I zb^a\)
\\
&\, +
\frac{1}{2 \pi \I } {\sum_{m,n}}'\[ \oint_{C^+_{m,n}} \frac{\de z'}{z'-z} \,\frac{1-z'^2}{1-z^2}
\, \p_{\txi_a}G^+_{m,n}
+\oint_{C^-_{m,n}} \frac{\de z'}{z'-z} \,\frac{1-z'^2}{1-z^2}\, \frac{z^3}{(z')^3}
\, \p_{\txi_a}G^-_{m,n}\].
\end{split}
\ee
It is easy to verify that this equality can be achieved if one chooses
\be
\begin{split}
Y^a
=&\,
\frac{\tau_2}{2}\( b^a+\I t^a\)-{\sum_{m,n}}'\[
\oint_{C^+_{m,n}}\frac{\de z}{8\pi} (1-z)\,\p_{\txii{0}_a}G^+_{m,n}
-\oint_{C^-_{m,n}}\frac{\de z}{8\pi z^3} (1-z)\,\p_{\txii{0}_a}G^-_{m,n}\],
\\
\zeta_{\rm cl}^a
=&\,
- (c^a - \tau_1 b^a)-{\sum_{m,n}}' \[\oint_{C^+_{m,n}} \frac{z\de z}{4 \pi \I}\,\p_{\txii{0}_a}G^+_{m,n}
-\oint_{C^-_{m,n}} \frac{\de z}{4 \pi \I z^3}\,\p_{\txii{0}_a}G^-_{m,n}\].
\end{split}
\label{mir-map-ver2}
\ee
We will leave the other relations undetermined because their derivation is more involved and
they are not of interest for our purposes.

%%%%%%%%%%%%%%%%%%%%%%%%%%%%%%%%%
\subsection{Large volume limit}
%%%%%%%%%%%%%%%%%%%%%%%%%%%%%%%%%

We shall now use the above results, in particular \eqref{pm-ker} and \eqref{mir-map-ver2},
as a guidance for finding the mirror map between type IIA and type IIB fields.
To apply them to our situation, we replace $(C_{m,n}^\pm,G_{m,n}^\pm)$ by $(\ell_{\gammapm},H_{\gammapm})$
where $\gammap=(0,p^a,q'_a,q_0)$ with $\bfp\cdot \bft^2>0$.
After this replacement, the Darboux coordinates are given by the same integral equations \eqref{twistlines}
where $\zeta_{\rm cl}^a,\tzeta^{\rm cl}_\Lambda,\tsigma^{\rm cl}$ differ from the type IIA fields
$\zeta^a,\tzeta_\Lambda,\sigma$ by the terms coming from the redefinition of
the kernel.\footnote{$\tsigma^{\rm cl}$ also contains an additional
classical term $\frac{\tau_2^3}{3}\kappa_{abc} b^a b^b b^c$.}
Our aim here is to study their large volume limit $t^a\to\infty$.

In this limit we concentrate on the infinitely small region of the $\CP$ fiber around the S-duality
invariant point $z=0$ such that the combination $zt^a$ stays finite.
Moreover, all integrals along $\ell_{\gammapm}$ are dominated by a saddle point
at $z'=0$ for the set of charges $\gammap$ and at $z'=\infty$ for the set of charges $\gammam$.
Using this fact and the relations \eqref{mir-map-ver2} (with the above replacements),
the Darboux coordinates of the type IIA construction in the one-instanton,
large volume approximation are given by \eqref{darblv} where the instanton contributions
read as
\be
\begin{split}
\delta\xi^a =&\,p^a\sum_{q_\Lambda} \int_{\ell_{\gammap}} \frac{\de z'}{z'-z} \, H_{\gammap},
\\
\delta\txi'_a = & \, \delta\tzeta'_a
+\sum_{q_\Lambda}\[ \int_{\ell_{\gammap}} \de z'\(\frac{q'_a}{z'-z} -\frac{\I}{2}\,\kappa_{abc}p^c t^b \)H_{\gammap}
+\frac{\I}{2}\,\kappa_{abc}p^c t^b\int_{\ell_{\gammam}} \frac{\de z'}{(z')^2}\,H_{\gammam}\],
\\
\delta\txi'_0 = & \,\delta\tzeta'_0
+\sum_{q_\Lambda}\[ \int_{\ell_{\gammap}} \de z'\(\frac{q'_0}{z'-z} +\frac{1}{4}\,\kappa_{abc}t^a t^b p^c (z'+2z)
+\frac{\I}{2}\, \kappa_{abc} p^a t^b b^c \)H_{\gammap}
\right.
\\
&\left.
-\frac{1}{4}\, \kappa_{abc}t^b p^c
\int_{\ell_{\gammam}}\frac{\de z'}{(z')^3}\(2\I z' b^a +t^a(1+2z z')\)H_{\gammam}\],
\\
\delta\alpha'= & \, -\hf\(\delta\sigma+\zeta^\Lambda\delta\tzeta'_\Lambda\)
+\sum_{q_\Lambda}\[ \int_{\ell_{\gammap}} \de z'\(-\frac{\frac{1}{2\pi\I}
+ q'_0\tau+q'_a(\tau b^a-c^a+2 \tau_2 t^a\, z' )}{z'-z}
\right.\right.
\\
&\left.
-\hf\,\kappa_{abc}p^c\(\I c^a t^b
+2\tau_2 b^a t^b+\(z(\tau_1-3\I\tau_2)+\frac{z'}{2}\(\tau_1-3\I\tau_2\)\)t^at^b\)  \)
H_{\gammap}
\\
& \left.
+\frac{1}{4}\,\kappa_{abc}t^bp^c \int_{\ell_{\gammam}} \frac{\de z'}{(z')^3}
\((2\I c^a-4\tau_2 b^a)z'+t^a \(\tau_1(1+2z z')+\I \tau_2(3+2z z')\) \) H_{\gammam}\].
\end{split}
\label{DClimit}
\ee
Here $\delta\tzeta'_\Lambda,\delta\sigma$ denote instanton corrections to the classical mirror map \eqref{symptobd}
to be found from the condition of modularity.

To study the modular properties of \eqref{DClimit}, it is more convenient to replace the last three
expressions by the combinations introduced in \eqref{defhctJa}.
Using the covariant derivative operators \eqref{defDw}, they can be shown to be given by
\bea
\delta\xi^a &=&2\pi \I p^a \cJ_\bfp
\nn\\
\hat\delta\txi'_a &= &  \delta\tzeta'_a-D_a\cJ_\bfp
+\frac{\I}{2}\,\kappa_{abc}t^b p^c\sum_{q_\Lambda}
\[\int_{\ell_{\gammap}} \de z'\,H_{\gammap}+\int_{\ell_{\gammam}} \frac{\de z'}{(z')^2}\,H_{\gammam}\],
\nn\\
\hat\delta_+\alpha'&= &  \tau\delta\tzeta'_0-\hf\(\delta\sigma+\zeta^\Lambda\delta\tzeta'_\Lambda\)
+\[\frac{\tau_2}{\pi}\kappa^{ab} D_a \bar D_{b} -(b_2+1)\]\cJ_\bfp
\label{instDP}
\\
&+&\frac{1}{2\I}\,\kappa_{abc}t^bp^c\sum_{q_\Lambda} \[\int_{\ell_{\gammap}}\de z'
\( b^a (\tau_1+3\I \tau_2) - c^a + 2 \tau_2 t^a z' \)H_{\gammap}
- \int_{\ell_{\gammam}} \frac{\de z'}{(z')^3}
\((c^a-\bar\tau b^a)z'+ \tau_2 t^a  \) H_{\gammam}\],
\nn
\\
\hat\delta_-\alpha'&= &  \bar\tau\delta\tzeta'_0-\hf\(\delta\sigma+\zeta^\Lambda\delta\tzeta'_\Lambda\)
-  \left[4\pi \tau_2 \cD_{-1}+(c^a-\bar\tau b^a)D_a \right] \cJ_\bfp
+\I \tau_2 z(p\cdot t^2) \sum_{q_\Lambda}\int\frac{\de z'}{(z')^2}H_{\gammam}
\nn\\
&+&
\frac{1}{2\I}\,\kappa_{abc}t^bp^c\sum_{q_\Lambda} \[\int_{\ell_{\gammap}}\de z'
\( b^a \tau - c^a + \tau_2 t^a z' \) H_{\gammap}
-\int_{\ell_{\gammam}} \frac{\de z'}{(z')^3}
\((c^a+(3\I\tau_2-\tau_1) b^a)z'+2 \tau_2 t^a \) H_{\gammam}\],
\nn
\eea
where we used the notation \eqref{defcJ}.
All the terms independent on the $\CP$ variable $z$ can be canceled by choosing appropriately
the mirror map. This fixes the instanton corrections to the mirror map \eqref{symptobd}
in the large volume, one-instanton approximation to be
\be
\begin{split}
\delta z^a=&\,-\frac{1}{4\pi\tau_2}{\sum_{q_\Lambda}}\[
\int_{\ell_{\gammap}}\de z\,(1-z)\,\p_{\txii{0}_a}H_{\gammap}
-\int_{\ell_{\gammam}}\frac{\de z}{z^3} (1-z)\,\p_{\txii{0}_a}H_{\gammam}\]
\\
\delta\zeta^a=&\, 0,
\\
\delta\tzeta'_a =&\,
-\frac{1}{2\pi}\,\kappa_{abc}t^b \sum_{q_\Lambda} \Re\(\int_{\ell_{\gammap}} \de z\,\p_{\txi_c}H_{\gammap}\),
\\
\delta\tzeta'_0=&\,
\frac{1}{2\pi}\, \kappa_{abc}t^b\sum_{q_\Lambda} \Re \int_{\ell_{\gammap}}\de z
\( b^a -\frac{\I}{2}\, t^a z \)\p_{\txi_c} H_{\gammap},
\\
\delta\sigma=&\,
\frac{1}{2\pi}\,\kappa_{abc}t^b\sum_{q_\Lambda} \Re\int_{\ell_{\gammap}}\de z
\( c^a -4\I\tau_2 b^a-\( \frac{\I}{2}\,\tau_1+3\tau_2\) t^a z \)\p_{\txi_c} H_{\gammap}.
\end{split}
\label{inst-mp}
\ee
Plugging these relations into \eqref{instDP}, one arrives at the expressions \eqref{cov-DC}
given in the main text. Note that the integrals over $\ell_{\gamma}$ are Gaussian and
can be easily computed. It is worth stressing that $\delta z^a$ gives a correction to the mirror map between
the type IIA complex structure moduli $z^a$
and the type IIB \kahler moduli $b^a+\I t^a$.
Such corrections do not arise for D1-D(-1)-instantons, but are in general necessary
in the presence of D3-instantons \cite{Alexandrov:2012bu}.

%%%%%%%%%%%%%%%%%%%%%%%%%%%%%%%%%%%%%%%%%%%%%%%%%%%%%%%%%%%%%%%%%%%%
\section{Relation to the Poincar\'e series construction \label{sec_poinca}}
%%%%%%%%%%%%%%%%%%%%%%%%%%%%%%%%%%%%%%%%%%%%%%%%%%%%%%%%%%%%%%%%%%%%

In this section, we make a first attempt at recasting the type IIA twistorial construction
into the manifestly S-duality invariant framework developed in \cite{Alexandrov:2012bu}.
The basic idea is to represent the partition function \eqref{thetaeg} as a Poincar\'e
series, and assign each term in the sum over cosets to a different transition function
$G_{m,n}$.

For this purpose, recall that the partition function of MSW invariants \eqref{defchimu}
is a vector-valued holomorphic modular form of negative weight $\wh=-b_2/2-1$,
therefore the positive frequency Fourier coefficients can be expressed in terms of the polar
(negative frequency) coefficients by the Rademacher formula \cite{Ap-book}
\be
\label{radom}
\bar\Omega_{\bfp,\bfmu}(\hat q_0) = 2\pi \I^{-\wh}\!\!\!\!
 \sum_{0\leq \hat q'_0 \leq r \chi(\cD)/24}\, \bar\Omega_{\bfp,\bfmu}(\hat q_0')\,
 \sum_{c=1}^{\infty} c^{-1}\, K(\hat q_0,\hat q'_0,c)\,
 \left( \frac{-\hat q'_0}{\hat q_0}\right)^{\frac{1-\wh}{2}}\,
 I_{1-\wh}\left(\frac{4\pi}{c}\sqrt{-\hat q_0\,\hat q'_0}\right),
\ee
where $K(\hat q_0,\hat q'_0,c)$ is the Kloosterman sum
\be
K(\hat q_0,\hat q'_0,c) = \sum_{-c\leq d<0, (c,d)=1}\, M^{-1}
\begin{pmatrix} a & b \\ c & d \end{pmatrix}\, \expe{\hat q'_0 \frac{a}{c}+\hat q_0\frac{d}{c}},
\ee
Using this representation one can express the elliptic genus \eqref{thetaeg}
as a Poincar\'e series \cite{Manschot:2007ha}\footnote{For simplicity, we assume
that the constant term of  \eqref{defchimu} vanishes. Moreover, the sum over $c,d$
should be regulated to $1\leq c\leq K$,
$|d|\leq K$, before sending $K$ to infinity.}
\begin{eqnarray}
\label{poincaZ}
 \mathcal{Z}(\tau,\bfy) &=& \hf
  \sum_{0\leq \hat q_0\leq r \chi(\cD)/24}
  \sum_{\bfmu\in \Lambda^*/\Lambda }
  \sum_{\bfk\in\Lambda + \bfmu+ \frac12 \bfp}
  \sum_{\trans\in\Gamma_\infty\backslash \Gamma}  (c\tau+d)^{3/2}(c\bar\tau+d)^{-1/2}\,
  \bar\Omega_{\bfp,\bfmu}(\hat q_0)\,(-1)^{\bfk\cdot \bfp}
\nonumber  \\
&&\times
\expe{
    -\left[ \hat q_0
      +\frac{1}{2} \bfk^2\right]
   \frac{a\tau+b}{c\tau+d} -\frac{\bfk\cdot \bfy}{c\tau+d}}\,
      R_{-1-\frac{b_2}{2}}\(\frac{2\pi \I \hat q_0}{c(c\tau+d)}\),
\end{eqnarray}
where the factor
\be
R_\wh(x)=1-\frac{1}{\Gamma(1-\wh)}\int_x^\infty e^{-z} z^{-\wh}\de z
\label{prefactorR}
\ee
approaches $1$ exponentially fast
at $\Re(x)\to\infty$ and $R_\wh(x)\sim x^{1-\wh}/\Gamma(2-\wh)$ at $x\to 0$.
The group element $\trans={\scriptsize \left( \begin{array}{cc} a & b \\ c & d\end{array}\right)}$
runs over the coset $\Gamma_\infty\backslash SL(2,\IZ)$, i.e.
$(c,d)$ run over pairs of coprime integers and $a,b$ are chosen
such that $ad-bc=1$.

Applying the same strategy to the formal theta series
$H_{(1)}$ \eqref{Hdec} on the twistor space, one arrives at
a formal Poincar\'e series. This gives
\be
H_{(1)}=\sum_{0\leq \hat q_0\leq \frac{r \chi(\cD)}{24}}\,
  \sum_{\bfmu\in \Lambda^*/\Lambda }\,
  \sum_{\bfk\in\Lambda + \bfmu+ \frac12 \bfp}^!\,
  \sum_{\trans\in\Gamma_\infty\backslash \Gamma} G_{c,d;\hgamma}
\label{poinca}
\ee
where $\hgamma=(\bfp,\bfmu,\bfk,\hat q_0)$,
\begin{eqnarray}
\label{ansatzGmn}
G_{c,d;\hgamma}&=&\frac{(-1)^{\bfk\cdot \bfp}}{8\pi^2}\,\bar\Omega_{\bfp,\bfmu}(\hat q_0)\,
(c\xi_0+d)\,
\expe{\bfp\cdot  \trans(\bftxip)
    -\left[ \hat q_0
      +\frac{1}{2} \bfk^2 \right]
      \trans(\xi^0)-\bfk \cdot \trans(\boldsymbol \xi)}
  \nonumber
  \\
&&\times
      R_{-1-\frac{b_2}{2}}\(\frac{2\pi \I \hat q_0}{c(c\xi^0+d)}\),
\end{eqnarray}
and the action of $\trans\in \Gamma_\infty\backslash SL(2,\IZ)$
on $\xi^0$, $\bfxi$ and $\bftxip$ is given by (\ref{SL2Zxi}).
Although the series \eqref{poinca} is divergent due to the sum over D1-brane
charges $\bfk$, it formally
transforms like a holomorphic modular form of weight $-1$.

The representation \eqref{poinca} suggests a natural Ansatz for
the holomorphic transition functions \eqref{transfun-gen} that enter
the manifestly S-duality invariant twistorial construction
outlined in \S \ref{sec_inslve1}, namely
\be
G_{m,n}^{\pm}=G_{m,n;\pm\hgamma},
\ee
and choose the contours $C_{m,n}^\pm$ as the geodesic circles joining
 the two essential singularities of \eqref{ansatzGmn}, namely the two roots $t^\pm_{m,n}$
of  the quadratic polynomial $t(m\xi^0(t)+n)$.
In particular, it is easy to check that the contact potential generated by these twistorial data
in the large volume limit coincides with \eqref{resPhi}.
Unfortunately, the transition functions \eqref{ansatzGmn} do not satisfy the transformation property \eqref{StransG}
due to the presence of the $R$-factor \eqref{prefactorR} which transforms as a period integral.
As a result, the $SL(2,\IZ)$ invariance of the construction is not guaranteed, and
indeed the  Darboux coordinates computed from \eqref{ansatzGmn} acquire a modular anomaly.
This anomaly  should however  be cancelable by  a similar holomorphic contact transformation
as the one described in \S\ref{sec_mock}. We hope to put this construction on a more solid
basis in future work.

%\bibliography{../common/combined}

\begin{thebibliography}{10}

\bibitem{Bagger:1983tt}
J.~Bagger and E.~Witten, ``{M}atter couplings in {${\mathcal N}=2$}
  supergravity,'' {\em Nucl. Phys.} {\bf B222} (1983)
1.
%%CITATION = NUPHA,B222,1;%%.

\bibitem{deWit:1984px}
B.~de~Wit, P.~Lauwers, and A.~Van~Proeyen, ``{Lagrangians of N=2 Supergravity -
  Matter Systems},'' {\em Nucl.Phys.} {\bf B255} (1985)
569.
%%CITATION = NUPHA,B255,569;%%.

\bibitem{Cecotti:1988qn}
S.~Cecotti, S.~Ferrara, and L.~Girardello, ``{G}eometry of type {II}
  superstrings and the moduli of superconformal field theories,'' {\em Int. J.
  Mod. Phys.} {\bf A4} (1989)
2475.
%%CITATION = IMPAE,A4,2475;%%.

\bibitem{Bodner:1989cg}
M.~Bodner and A.~C. Cadavid, ``Dimensional reduction of type {II}b supergravity
  and exceptional quaternionic manifolds,'' {\em Class. Quant. Grav.} {\bf 7}
  (1990)
829.
%%CITATION = CQGRD,7,829;%%.

\bibitem{Ferrara:1989ik}
S.~Ferrara and S.~Sabharwal, ``{Q}uaternionic manifolds for type {II}
  superstring vacua of {C}alabi-{Y}au spaces,'' {\em Nucl. Phys.} {\bf B332}
  (1990)
317.
%%CITATION = NUPHA,B332,317;%%.

\bibitem{Antoniadis:1997eg}
I.~Antoniadis, S.~Ferrara, R.~Minasian, and K.~S. Narain, ``{$R^4$ couplings in
  M- and type II theories on Calabi-Yau spaces},'' {\em Nucl. Phys.} {\bf B507}
  (1997) 571--588,
\href{http://www.arXiv.org/abs/hep-th/9707013}{{\tt hep-th/9707013}}.
%%CITATION = HEP-TH/9707013;%%.

\bibitem{Gunther:1998sc}
H.~G{\"u}nther, C.~Herrmann, and J.~Louis, ``{Quantum corrections in the
  hypermultiplet moduli space},'' {\em Fortsch. Phys.} {\bf 48} (2000)
  119--123,
\href{http://www.arXiv.org/abs/hep-th/9901137}{{\tt hep-th/9901137}}.
%%CITATION = HEP-TH/9901137;%%.

\bibitem{Antoniadis:2003sw}
I.~Antoniadis, R.~Minasian, S.~Theisen, and P.~Vanhove, ``String loop
  corrections to the universal hypermultiplet,'' {\em Class. Quant. Grav.} {\bf
  20} (2003) 5079--5102,
\href{http://www.arXiv.org/abs/hep-th/0307268}{{\tt hep-th/0307268}}.
%%CITATION = HEP-TH/0307268;%%.

\bibitem{Robles-Llana:2006ez}
D.~Robles-Llana, F.~Saueressig, and S.~Vandoren, ``String loop corrected
  hypermultiplet moduli spaces,'' {\em JHEP} {\bf 03} (2006) 081,
\href{http://www.arXiv.org/abs/hep-th/0602164}{{\tt hep-th/0602164}}.
%%CITATION = HEP-TH 0602164;%%.

\bibitem{Alexandrov:2007ec}
S.~Alexandrov, ``{Quantum covariant c-map},'' {\em JHEP} {\bf 05} (2007) 094,
\href{http://www.arXiv.org/abs/hep-th/0702203}{{\tt hep-th/0702203}}.
%%CITATION = HEP-TH/0702203;%%.

\bibitem{Becker:1995kb}
K.~Becker, M.~Becker, and A.~Strominger, ``Five-branes, membranes and
  nonperturbative string theory,'' {\em Nucl. Phys.} {\bf B456} (1995)
  130--152,
\href{http://www.arXiv.org/abs/hep-th/9507158}{{\tt hep-th/9507158}}.
%%CITATION = NUPHA,B456,130;%%.

\bibitem{Becker:1999pb}
K.~Becker and M.~Becker, ``{Instanton action for type II hypermultiplets},''
  {\em Nucl. Phys.} {\bf B551} (1999) 102--116,
\href{http://www.arXiv.org/abs/hep-th/9901126}{{\tt hep-th/9901126}}.
%%CITATION = HEP-TH/9901126;%%.

\bibitem{MR664330}
S.~M. Salamon, ``Quaternionic {K}\"ahler manifolds,'' {\em Invent. Math.} {\bf
  67} (1982), no.~1, 143--171.

\bibitem{Karlhede:1984vr}
A.~Karlhede, U.~Lindstr{\"o}m, and M.~Ro\v{c}ek, ``Selfinteracting tensor
  multiplets in {$\N=2$} superspace,'' {\em Phys. Lett.} {\bf B147} (1984)
297.
%%CITATION = PHLTA,B147,297;%%.

\bibitem{Hitchin:1986ea}
N.~J. Hitchin, A.~Karlhede, U.~Lindstr{\"o}m, and M.~Ro\v{c}ek,
  ``Hyperk{\"a}hler metrics and supersymmetry,'' {\em Commun. Math. Phys.} {\bf
  108} (1987)
535.
%%CITATION = CMPHA,108,535;%%.

\bibitem{MR1001707}
C.~LeBrun, ``Quaternionic-{K}\"ahler manifolds and conformal geometry,'' {\em
  Math. Ann.} {\bf 284} (1989), no.~3, 353--376.

\bibitem{MR1096180}
A.~Swann, ``Hyper-{K}\"ahler and quaternionic {K}\"ahler geometry,'' {\em Math.
  Ann.} {\bf 289} (1991), no.~3, 421--450.

\bibitem{deWit:2001dj}
B.~de~Wit, M.~Ro\v{c}ek, and S.~Vandoren, ``{H}ypermultiplets, hyperk{\"a}hler
  cones and quaternion-{K\"a}hler geometry,'' {\em JHEP} {\bf 02} (2001) 039,
\href{http://www.arXiv.org/abs/hep-th/0101161}{{\tt hep-th/0101161}}.
%%CITATION = HEP-TH 0101161;%%.

\bibitem{Lindstrom:2008gs}
U.~Lindstrom and M.~Rocek, ``{Properties of hyperkahler manifolds and their
  twistor spaces},'' {\em Commun.Math.Phys.} {\bf 293} (2010) 257--278,
  \href{http://www.arXiv.org/abs/0807.1366}{{\tt 0807.1366}}.

\bibitem{Alexandrov:2008nk}
S.~Alexandrov, B.~Pioline, F.~Saueressig, and S.~Vandoren, ``{Linear
  perturbations of quaternionic metrics},'' {\em Commun. Math. Phys.} {\bf 296}
  (2010) 353--403,
\href{http://www.arXiv.org/abs/0810.1675}{{\tt 0810.1675}}.
%%CITATION = 0810.1675;%%.

\bibitem{Alexandrov:2008ds}
S.~Alexandrov, B.~Pioline, F.~Saueressig, and S.~Vandoren, ``{Linear
  perturbations of Hyperkahler metrics},'' {\em Lett. Math. Phys.} {\bf 87}
  (2009) 225--265,
\href{http://www.arXiv.org/abs/0806.4620}{{\tt 0806.4620}}.
%%CITATION = 0806.4620;%%.

\bibitem{RoblesLlana:2006is}
D.~Robles-Llana, M.~Ro\v{c}ek, F.~Saueressig, U.~Theis, and S.~Vandoren,
  ``{Nonperturbative corrections to 4D string theory effective actions from
  SL(2,Z) duality and supersymmetry},'' {\em Phys. Rev. Lett.} {\bf 98} (2007)
  211602,
\href{http://www.arXiv.org/abs/hep-th/0612027}{{\tt hep-th/0612027}}.
%%CITATION = HEP-TH/0612027;%%.

\bibitem{Alexandrov:2006hx}
S.~Alexandrov, F.~Saueressig, and S.~Vandoren, ``{Membrane and fivebrane
  instantons from quaternionic geometry},'' {\em JHEP} {\bf 09} (2006) 040,
\href{http://www.arXiv.org/abs/hep-th/0606259}{{\tt hep-th/0606259}}.
%%CITATION = HEP-TH/0606259;%%.

\bibitem{RoblesLlana:2007ae}
D.~Robles-Llana, F.~Saueressig, U.~Theis, and S.~Vandoren, ``{Membrane
  instantons from mirror symmetry},'' {\em Commun.Num.Theor.Phys.} {\bf 1}
  (2007) 681, \href{http://www.arXiv.org/abs/0707.0838}{{\tt 0707.0838}}.

\bibitem{Saueressig:2007dr}
F.~Saueressig and S.~Vandoren, ``{Conifold singularities, resumming instantons
  and non- perturbative mirror symmetry},'' {\em JHEP} {\bf 07} (2007) 018,
\href{http://www.arXiv.org/abs/arXiv:0704.2229 [hep-th]}{{\tt arXiv:0704.2229
  [hep-th]}}.
%%CITATION = ARXIV:0704.2229;%%.

\bibitem{Gaiotto:2008cd}
D.~Gaiotto, G.~W. Moore, and A.~Neitzke, ``{Four-dimensional wall-crossing via
  three-dimensional field theory},'' {\em Commun.Math.Phys.} {\bf 299} (2010)
  163--224, \href{http://www.arXiv.org/abs/0807.4723}{{\tt 0807.4723}}.

\bibitem{Alexandrov:2008gh}
S.~Alexandrov, B.~Pioline, F.~Saueressig, and S.~Vandoren, ``{D-instantons and
  twistors},'' {\em JHEP} {\bf 03} (2009) 044,
\href{http://www.arXiv.org/abs/0812.4219}{{\tt 0812.4219}}.
%%CITATION = 0812.4219;%%.

\bibitem{Alexandrov:2009zh}
S.~Alexandrov, ``{D-instantons and twistors: some exact results},'' {\em J.
  Phys.} {\bf A42} (2009) 335402,
\href{http://www.arXiv.org/abs/0902.2761}{{\tt 0902.2761}}.
%%CITATION = 0902.2761;%%.

\bibitem{Haydys}
A.~Haydys, ``{Hyper-K\"ahler and quaternionic K\"ahler manifolds with
  $S^{1}$-symmetries},'' {\em J. Geom. Phys.} {\bf 58} (2008), no.~3, 293--306.

\bibitem{Alexandrov:2011ac}
S.~Alexandrov, D.~Persson, and B.~Pioline, ``{Wall-crossing, Rogers
  dilogarithm, and the QK/HK correspondence},'' {\em JHEP} {\bf 1112} (2011)
  027,
\href{http://www.arXiv.org/abs/1110.0466}{{\tt 1110.0466}}.
%%CITATION = ARXIV:1110.0466;%%.

\bibitem{Alexandrov:2010ca}
S.~Alexandrov, D.~Persson, and B.~Pioline, ``{Fivebrane instantons, topological
  wave functions and hypermultiplet moduli spaces},'' {\em JHEP} {\bf 1103}
  (2011) 111, \href{http://www.arXiv.org/abs/1010.5792}{{\tt 1010.5792}}.

\bibitem{Pioline:2009qt}
B.~Pioline and D.~Persson, ``{The automorphic NS5-brane},'' {\em Commun. Num.
  Th. Phys.} {\bf 3} (2009), no.~4, 697--754,
\href{http://www.arXiv.org/abs/0902.3274}{{\tt 0902.3274}}.
%%CITATION = 0902.3274;%%.

\bibitem{Bao:2009fg}
L.~Bao, A.~Kleinschmidt, B.~E.~W. Nilsson, D.~Persson, and B.~Pioline,
  ``{Instanton Corrections to the Universal Hypermultiplet and Automorphic
  Forms on SU(2,1)},'' {\em Commun. Num. Theor. Phys.} {\bf 4} (2010) 187--266,
\href{http://www.arXiv.org/abs/0909.4299}{{\tt 0909.4299}}.
%%CITATION = 0909.4299;%%.

\bibitem{Alexandrov:2010np}
S.~Alexandrov, D.~Persson, and B.~Pioline, ``{On the topology of the
  hypermultiplet moduli space in type II/CY string vacua},'' {\em Phys.Rev.}
  {\bf D83} (2011) 026001, \href{http://www.arXiv.org/abs/1009.3026}{{\tt
  1009.3026}}.

\bibitem{Alexandrov:2009qq}
S.~Alexandrov and F.~Saueressig, ``{Quantum mirror symmetry and twistors},''
  {\em JHEP} {\bf 09} (2009) 108,
\href{http://www.arXiv.org/abs/0906.3743}{{\tt 0906.3743}}.
%%CITATION = 0906.3743;%%.

\bibitem{Alexandrov:2012bu}
S.~Alexandrov and B.~Pioline, ``{S-duality in Twistor Space},''
\href{http://www.arXiv.org/abs/1206.1341}{{\tt 1206.1341}}.
%%CITATION = ARXIV:1206.1341;%%.

\bibitem{Maldacena:1997de}
J.~M. Maldacena, A.~Strominger, and E.~Witten, ``{B}lack hole entropy in
  {M}-theory,'' {\em JHEP} {\bf 12} (1997) 002,
\href{http://www.arXiv.org/abs/hep-th/9711053}{{\tt hep-th/9711053}}.
%%CITATION = HEP-TH 9711053;%%.

\bibitem{Manschot:2009ia}
J.~Manschot, ``{Stability and duality in N=2 supergravity},'' {\em
  Commun.Math.Phys.} {\bf 299} (2010) 651--676,
  \href{http://www.arXiv.org/abs/0906.1767}{{\tt 0906.1767}}.

\bibitem{Manschot:2010xp}
J.~Manschot, ``{Wall-crossing of D4-branes using flow trees},'' {\em
  Adv.Theor.Math.Phys.} {\bf 15} (2011) 1--42,
\href{http://www.arXiv.org/abs/1003.1570}{{\tt 1003.1570}}.
%%CITATION = ARXIV:1003.1570;%%.

\bibitem{Zwegers-thesis}
S.~Zwegers, ``Mock theta functions.'' PhD dissertation, 2002, Utrecht.

\bibitem{Dijkgraaf:2000fq}
R.~Dijkgraaf, J.~M. Maldacena, G.~W. Moore, and E.~P. Verlinde, ``{A black hole
  Farey tail},''
\href{http://www.arXiv.org/abs/hep-th/0005003}{{\tt hep-th/0005003}}.
%%CITATION = HEP-TH/0005003;%%.

\bibitem{Manschot:2007ha}
J.~Manschot and G.~W. Moore, ``{A Modern Fareytail},'' {\em Commun. Num. Theor.
  Phys.} {\bf 4} (2010) 103--159,
\href{http://www.arXiv.org/abs/0712.0573}{{\tt 0712.0573}}.
%%CITATION = 0712.0573;%%.

\bibitem{Hosono:1993qy}
S.~Hosono, A.~Klemm, S.~Theisen, and S.-T. Yau, ``{Mirror symmetry, mirror map
  and applications to Calabi-Yau hypersurfaces},'' {\em Commun. Math. Phys.}
  {\bf 167} (1995) 301--350,
\href{http://www.arXiv.org/abs/hep-th/9308122}{{\tt hep-th/9308122}}.
%%CITATION = HEP-TH/9308122;%%.

\bibitem{Bohm:1999uk}
R.~B{\"o}hm, H.~G{\"u}nther, C.~Herrmann, and J.~Louis, ``{Compactification of
  type IIB string theory on Calabi-Yau threefolds},'' {\em Nucl. Phys.} {\bf
  B569} (2000) 229--246,
\href{http://www.arXiv.org/abs/hep-th/9908007}{{\tt hep-th/9908007}}.
%%CITATION = HEP-TH/9908007;%%.

\bibitem{Haghighat:2011xx}
B.~Haghighat and S.~Vandoren, ``{Five-dimensional gauge theory and
  compactification on a torus},'' {\em JHEP} {\bf 09} (2011) 060,
\href{http://www.arXiv.org/abs/1107.2847}{{\tt 1107.2847}}.
%%CITATION = 1107.2847;%%.

\bibitem{Alexandrov:2011va}
S.~Alexandrov, ``{Twistor Approach to String Compactifications: a Review},''
\href{http://www.arXiv.org/abs/1111.2892}{{\tt 1111.2892}}.
%%CITATION = ARXIV:1111.2892;%%.

\bibitem{Neitzke:2007ke}
A.~Neitzke, B.~Pioline, and S.~Vandoren, ``{Twistors and Black Holes},'' {\em
  JHEP} {\bf 04} (2007) 038,
\href{http://www.arXiv.org/abs/hep-th/0701214}{{\tt hep-th/0701214}}.
%%CITATION = HEP-TH/0701214;%%.

\bibitem{ks}
M.~Kontsevich and Y.~Soibelman, ``{Stability structures, motivic
  Donaldson-Thomas invariants and cluster transformations},''
  \href{http://www.arXiv.org/abs/0811.2435}{{\tt 0811.2435}}.

\bibitem{Alexandrov:2010pp}
S.~Alexandrov and P.~Roche, ``{TBA for non-perturbative moduli spaces},'' {\em
  JHEP} {\bf 1006} (2010) 066, \href{http://www.arXiv.org/abs/1003.3964}{{\tt
  1003.3964}}.

\bibitem{Gaiotto:2006ns}
D.~Gaiotto, A.~Strominger, and X.~Yin, ``From {A}d{S}(3)/{CFT}(2) to black
  holes / topological strings,''
\href{http://www.arXiv.org/abs/hep-th/0602046}{{\tt hep-th/0602046}}.
%%CITATION = HEP-TH 0602046;%%.

\bibitem{Gaiotto:2006wm}
D.~Gaiotto, A.~Strominger, and X.~Yin, ``{The M5-brane elliptic genus:
  Modularity and BPS states},'' {\em JHEP} {\bf 08} (2007) 070,
\href{http://www.arXiv.org/abs/hep-th/0607010}{{\tt hep-th/0607010}}.
%%CITATION = HEP-TH/0607010;%%.

\bibitem{deBoer:2006vg}
J.~de~Boer, M.~C.~N. Cheng, R.~Dijkgraaf, J.~Manschot, and E.~Verlinde, ``{A
  farey tail for attractor black holes},'' {\em JHEP} {\bf 11} (2006) 024,
\href{http://www.arXiv.org/abs/hep-th/0608059}{{\tt hep-th/0608059}}.
%%CITATION = HEP-TH/0608059;%%.

\bibitem{Denef:2007vg}
F.~Denef and G.~W. Moore, ``Split states, entropy enigmas, holes and halos,''
\href{http://www.arXiv.org/abs/hep-th/0702146}{{\tt hep-th/0702146}}.
%%CITATION = HEP-TH/0702146;%%.

\bibitem{Douglas:2000gi}
M.~R. Douglas, ``{D-branes, categories and N = 1 supersymmetry},'' {\em J.
  Math. Phys.} {\bf 42} (2001) 2818--2843,
\href{http://www.arXiv.org/abs/hep-th/0011017}{{\tt hep-th/0011017}}.
%%CITATION = HEP-TH/0011017;%%.

\bibitem{Aspinwall:2004jr}
P.~S. Aspinwall, ``{D-branes on Calabi-Yau manifolds},''
\href{http://www.arXiv.org/abs/hep-th/0403166}{{\tt hep-th/0403166}}.
%%CITATION = HEP-TH/0403166;%%.

\bibitem{Dabholkar:2005dt}
A.~Dabholkar, F.~Denef, G.~W. Moore, and B.~Pioline, ``Precision counting of
  small black holes,'' {\em JHEP} {\bf 10} (2005) 096,
\href{http://www.arXiv.org/abs/hep-th/0507014}{{\tt hep-th/0507014}}.
%%CITATION = HEP-TH/0507014;%%.

\bibitem{Sharpe:1999qz}
E.~R. Sharpe, ``{D-branes, derived categories, and Grothendieck groups},'' {\em
  Nucl.Phys.} {\bf B561} (1999) 433--450,
\href{http://www.arXiv.org/abs/hep-th/9902116}{{\tt hep-th/9902116}}.
%%CITATION = HEP-TH/9902116;%%.

\bibitem{Minasian:1997mm}
R.~Minasian and G.~W. Moore, ``{K-theory and Ramond-Ramond charge},'' {\em
  JHEP} {\bf 11} (1997) 002,
\href{http://www.arXiv.org/abs/hep-th/9710230}{{\tt hep-th/9710230}}.
%%CITATION = HEP-TH/9710230;%%.

\bibitem{Douglas:2006jp}
M.~R. Douglas, R.~Reinbacher, and S.-T. Yau, ``{Branes, bundles and attractors:
  Bogomolov and beyond},''
\href{http://www.arXiv.org/abs/math/0604597}{{\tt math/0604597}}.
%%CITATION = MATH/0604597;%%.

\bibitem{Diaconescu:2007bf}
E.~Diaconescu and G.~W. Moore, ``{Crossing the Wall: Branes vs. Bundles},''
\href{http://www.arXiv.org/abs/0706.3193}{{\tt 0706.3193}}.
%%CITATION = 0706.3193;%%.

\bibitem{Joyce:2008pc}
D.~Joyce and Y.~Song, ``{A theory of generalized Donaldson-Thomas
  invariants},''
\href{http://www.arXiv.org/abs/0810.5645}{{\tt 0810.5645}}.
%%CITATION = 0810.5645;%%.

\bibitem{Manschot:2010qz}
J.~Manschot, B.~Pioline, and A.~Sen, ``{Wall Crossing from Boltzmann Black Hole
  Halos},'' {\em JHEP} {\bf 1107} (2011) 059,
\href{http://www.arXiv.org/abs/1011.1258}{{\tt 1011.1258}}.
%%CITATION = ARXIV:1011.1258;%%.

\bibitem{deBoer:2008ss}
J.~de~Boer, J.~Manschot, K.~Papadodimas, and E.~Verlinde, ``{The Chiral ring of
  AdS(3)/CFT(2) and the attractor mechanism},'' {\em JHEP} {\bf 0903} (2009)
  030,
\href{http://www.arXiv.org/abs/0809.0507}{{\tt 0809.0507}}.
%%CITATION = ARXIV:0809.0507;%%.

\bibitem{deBoer:2008fk}
J.~de~Boer, F.~Denef, S.~El-Showk, I.~Messamah, and D.~Van~den Bleeken,
  ``{Black hole bound states in AdS(3) x S**2},'' {\em JHEP} {\bf 0811} (2008)
  050,
\href{http://www.arXiv.org/abs/0802.2257}{{\tt 0802.2257}}.
%%CITATION = ARXIV:0802.2257;%%.

\bibitem{Troost:2010ud}
J.~Troost, ``{The non-compact elliptic genus: mock or modular},'' {\em JHEP}
  {\bf 1006} (2010) 104,
\href{http://www.arXiv.org/abs/1004.3649}{{\tt 1004.3649}}.
%%CITATION = ARXIV:1004.3649;%%.

\bibitem{Vafa:1994tf}
C.~Vafa and E.~Witten, ``{A Strong coupling test of S duality},'' {\em
  Nucl.Phys.} {\bf B431} (1994) 3--77,
\href{http://www.arXiv.org/abs/hep-th/9408074}{{\tt hep-th/9408074}}.
%%CITATION = HEP-TH/9408074;%%.

\bibitem{Morrison:1996xf}
D.~R. Morrison and N.~Seiberg, ``{Extremal transitions and five-dimensional
  supersymmetric field theories},'' {\em Nucl.Phys.} {\bf B483} (1997)
  229--247,
\href{http://www.arXiv.org/abs/hep-th/9609070}{{\tt hep-th/9609070}}.
%%CITATION = HEP-TH/9609070;%%.

\bibitem{Manschot:2011ym}
J.~Manschot, ``{BPS invariants of semi-stable sheaves on rational surfaces},''
\href{http://www.arXiv.org/abs/1109.4861}{{\tt 1109.4861}}.
%%CITATION = ARXIV:1109.4861;%%.

\bibitem{Haghighat:2012}
B.~Haghighat, J.~Manschot, and S.~Vandoren, ``work in progress,''.

\bibitem{Manschot:2011dj}
J.~Manschot, ``{BPS invariants of N=4 gauge theory on a surface},''
\href{http://www.arXiv.org/abs/1103.0012}{{\tt 1103.0012}}.
%%CITATION = ARXIV:1103.0012;%%.

\bibitem{Neitzke:2011za}
A.~Neitzke, ``{On a hyperholomorphic line bundle over the Coulomb branch},''
\href{http://www.arXiv.org/abs/1110.1619}{{\tt 1110.1619}}.
%%CITATION = 1110.1619;%%.

\bibitem{Vigneras:1977}
M.-F. Vign\'eras, ``{S\'eries th\^eta des formes quadratiques ind\'efinies},''
  {\em Springer Lecture Notes} {\bf 627} (1977) 227 -- 239.

\bibitem{MR1623706}
L.~G{\"o}ttsche and D.~Zagier, ``Jacobi forms and the structure of {D}onaldson
  invariants for {$4$}-manifolds with {$b_+=1$},'' {\em Selecta Math. (N.S.)}
  {\bf 4} (1998), no.~1, 69--115.

\bibitem{0961.14022}
L.~G{\"o}ttsche, ``{Theta functions and Hodge numbers of moduli spaces of
  sheaves on rational surfaces.},'' {\em Commun. Math. Phys.} {\bf 206} (1999),
  no.~1, 105--136.

\bibitem{Ap-book}
T.~Apostol, {\em Modular Functions and Dirichlet Series in Number Theory}.
\newblock Graduate Texts in Mathematics. Springer-Verlag, 1976.

\end{thebibliography}
%\bibliographystyle{../common/utphys}

\providecommand{\href}[2]{#2}\begingroup\raggedright\endgroup

\end{document}